\documentclass[aps,prx,twocolumn,superscriptaddress,showpacs]{revtex4-1}
\usepackage[colorlinks,bookmarks=false,citecolor=blue,linkcolor=blue,urlcolor=bl
ue]{hyperref}
\usepackage{graphicx}
\usepackage{amsmath}
\usepackage{amsfonts}
\usepackage{amssymb}
\usepackage{natbib}
\usepackage{xspace}
\usepackage{braket}
\usepackage{bm}

\makeatletter

\setcitestyle{numbers,square}

\makeatother

\begin{document}

\title{Investigating field-induced magnetic order in Han Purple by neutron 
scattering up to 25.9 T}

\author{S. Allenspach}
\affiliation{Quantum Criticality and Dynamics Group, Paul Scherrer Institute, 
CH-5232 Villigen-PSI, Switzerland}
\affiliation{Department of Quantum Matter Physics, University of Geneva, 
CH-1211 Geneva, Switzerland}

\author{A. Madsen}
\affiliation{Quantum Criticality and Dynamics Group, Paul Scherrer Institute, 
CH-5232 Villigen-PSI, Switzerland}
\affiliation{Institute of Computational Science, Universit\`{a} della Svizzera 
italiana, CH-6900 Lugano, Switzerland}

\author{A. Biffin}
\affiliation{Laboratory for Neutron Scattering and Imaging, Paul Scherrer 
Institute, CH-5232 Villigen, Switzerland}

\author{M. Bartkowiak}
\affiliation{Helmholtz-Zentrum Berlin f\"{u}r Materialien und Energie, 
Hahn-Meitner-Platz 1, 14109 Berlin, Germany}

\author{O. Prokhnenko}
\affiliation{Helmholtz-Zentrum Berlin f\"{u}r Materialien und Energie, 
Hahn-Meitner-Platz 1, 14109 Berlin, Germany}

\author{A. Gazizulina}
\affiliation{Karlsruhe Institute of Technology, Institute for Quantum 
Materials and Technologies, 76021 Karlsruhe, Germany}

\author{X. Liu}
\affiliation{School of Physics, Sun Yat-sen University, Guangzhou 510275, 
China}

\author{R. Wahle}
\affiliation{Helmholtz-Zentrum Berlin f\"{u}r Materialien und Energie, 
Hahn-Meitner-Platz 1, 14109 Berlin, Germany}

\author{S. Gerischer}
\affiliation{Helmholtz-Zentrum Berlin f\"{u}r Materialien und Energie, 
Hahn-Meitner-Platz 1, 14109 Berlin, Germany}

\author{S. Kempfer}
\affiliation{Helmholtz-Zentrum Berlin f\"{u}r Materialien und Energie, 
Hahn-Meitner-Platz 1, 14109 Berlin, Germany}

\author{P. Heller}
\affiliation{Helmholtz-Zentrum Berlin f\"{u}r Materialien und Energie, 
Hahn-Meitner-Platz 1, 14109 Berlin, Germany}

\author{P. Smeibidl}
\affiliation{Helmholtz-Zentrum Berlin f\"{u}r Materialien und Energie, 
Hahn-Meitner-Platz 1, 14109 Berlin, Germany}

\author{A. Mira}
\affiliation{Data Science Laboratory, Universit\`{a} della Svizzera 
italiana, CH-6900 Lugano, Switzerland}
\affiliation{Dipartimento di Scienza e Alta Tecnologia, Universit\`{a} degli 
Studi dell'Insubria, 2210 Como, Italy}

\author{N. Laflorencie}
\affiliation{Laboratoire de Physique Th\'{e}orique, CNRS and Universit\'{e} 
de Toulouse, 31062 Toulouse, France}

\author{F. Mila}
\affiliation{Institute of Physics, \'{E}cole Polytechnique F\'{e}d\'{e}rale 
de Lausanne (EPFL), CH-1015 Lausanne, Switzerland}

\author{B. Normand}
\affiliation{Quantum Criticality and Dynamics Group, Paul Scherrer Institute, 
CH-5232 Villigen-PSI, Switzerland}
\affiliation{Institute of Physics, \'{E}cole Polytechnique F\'{e}d\'{e}rale 
de Lausanne (EPFL), CH-1015 Lausanne, Switzerland}

\author{Ch. R\"{u}egg}
\affiliation{Quantum Criticality and Dynamics Group, Paul Scherrer Institute, 
CH-5232 Villigen-PSI, Switzerland}
\affiliation{Department of Quantum Matter Physics, University of Geneva, 
CH-1211 Geneva, Switzerland}
\affiliation{Institute of Physics, \'{E}cole Polytechnique F\'{e}d\'{e}rale 
de Lausanne (EPFL), CH-1015 Lausanne, Switzerland}
\affiliation{Institute for Quantum Electronics, ETH Z\"{u}rich, 
CH-8093 H\"{o}nggerberg, Switzerland}

\begin{abstract}
BaCuSi$_2$O$_6$ is a quasi-two-dimensional (2D) quantum antiferromagnet 
containing three different types of stacked, square-lattice bilayer hosting 
spin-1/2 dimers. Although this compound has been studied extensively over the 
last two decades, the critical applied magnetic field required to close the 
dimer spin gap and induce magnetic order, which exceeds 23 T, has to date 
precluded any kind of neutron scattering investigation. However, the HFM/EXED 
instrument at the Helmholtz-Zentrum Berlin made this possible at magnetic 
fields up to 25.9 T. Thus we have used HFM/EXED to investigate the field-induced 
ordered phase, in particular to look for quasi-2D physics arising from the 
layered structure and from the different bilayer types. From neutron diffraction 
data, we determined the global dependence of the magnetic order parameter on 
both magnetic field and temperature, finding a form consistent with 3D quantum 
critical scaling; from this we deduce that the quasi-2D interactions and 
nonuniform layering of BaCuSi$_2$O$_6$ are not anisotropic enough to induce 
hallmarks of 2D physics. From neutron spectroscopy data, we measured the 
dispersion of the strongly Zeeman-split magnetic excitations, finding good 
agreement with the zero-field interaction parameters of BaCuSi$_2$O$_6$. We 
conclude that HFM/EXED allowed a significant extension in the application of 
neutron scattering techniques to the field range above 20 T and in particular 
opened new horizons in the study of field-induced magnetic quantum phase 
transitions. 
\end{abstract}

\maketitle

\section{Introduction}
One of the basic concepts of statistical physics is that any continuous 
classical or quantum phase transition (QPT) can be assigned to a certain 
universality class \cite{Zinn-Justin2002}. Because the characteristic 
energy scale of a system vanishes and the correlation length diverges at 
such a transition, the microscopic details become irrelevant and the critical 
properties of the system are dictated only by global and scale-invariant 
characteristics such as the dimensionality, symmetry, and in special cases 
also the topology. Reducing the effective dimensionality of a system enhances 
the role of quantum fluctuations, leading to the emergence of exotic 
``low-dimensional'' behavior.

Quantum magnetic systems are ideal testbeds for the study of phase transitions, 
criticality, and low-dimensional physics. One family of quantum magnets well
suited for this task is spin-dimer systems, which consist of interacting 
spin-1/2 pairs with internal antiferromagnetic (AF) coupling, resulting in a 
global singlet ground state whose excitations are ``triplons,'' propagating 
triplet quasiparticles. When a sufficiently strong magnetic field is applied, 
a QPT takes place from the dimer-singlet phase, which magnetically is quantum 
disordered to a field-induced ordered phase resembling an effective XY spin 
model \cite{Sachdev2011}. The magnetization of the dimers in this ordered phase 
can be separated into a longitudinal component, $m_{\parallel}$, along the 
field direction and a transverse component, $m_{\perp}$, perpendicular to 
the field; $m_{\parallel}$ is of necessity ferromagnetic (FM) and local in 
nature, whereas $m_{\perp}$ is AF as a result of the interactions both 
within and between the dimers \cite{Matsumoto2004}, and its magnitude 
represents the order parameter.

If the interaction network has dimensionality $d = 3$, the phase transition 
is in the 3D-XY universality class \cite{Giamarchi2008,Zapf2014} and the 
field-induced ground state has the critical behavior of a Bose-Einstein 
Condensate (BEC) \cite{Bose1924,Einstein1924}, as observed experimentally 
for example in the 3D spin-dimer material TlCuCl$_3$ \cite{Nikuni2000,
Tanaka2001,Oosawa2001,Rueegg2003,Glazkov2004}. In systems with $d \leq$ 2, 
the field-induced phase cannot possess long-range order (LRO) at finite 
temperatures due to the Mermin-Wagner theorem \cite{Mermin1966}, but 
quasi-ordered phases may nevertheless be found in systems with weak 
interactions between low-dimensional substructures (e.g.~chains, ladders, 
or planes) when the temperature exceeds the weak energy scale. 
Spin-dimer systems with $d = 1$ exhibit a 
field-induced phase that resembles a Tomonaga-Luttinger liquid (TLL), with 
algebraically decaying spin-spin correlations in the ground state and 
fractional (spin-1/2) excitations in the spectrum \cite{Sachdev1994,
Giamarchi1999,Giamarchi2003}. For spin-dimer systems with $d = 2$, the 
field-induced phase is equivalent to the 2D-XY model, which is special 
because of the Berezinskii-Kosterlitz-Thouless (BKT) transition 
\cite{Berezinskii1971,Kosterlitz1973} from a quasi-ordered phase with 
algebraically decaying spin correlations at lower temperatures to a disordered 
phase with exponentially decaying correlations. Quasi-low-dimensional 
spin-dimer materials consist of low-dimensional substructures such as chains, 
ladders, or (bi)layers that are strongly coupled internally but weakly coupled 
to each other. Such systems display 3D coherence up to a certain temperature, 
above which thermal fluctuations destroy 3D LRO to produce different forms 
of low-dimensional behavior.

\begin{figure}[p]
\begin{center}
\includegraphics[width=0.96\columnwidth]{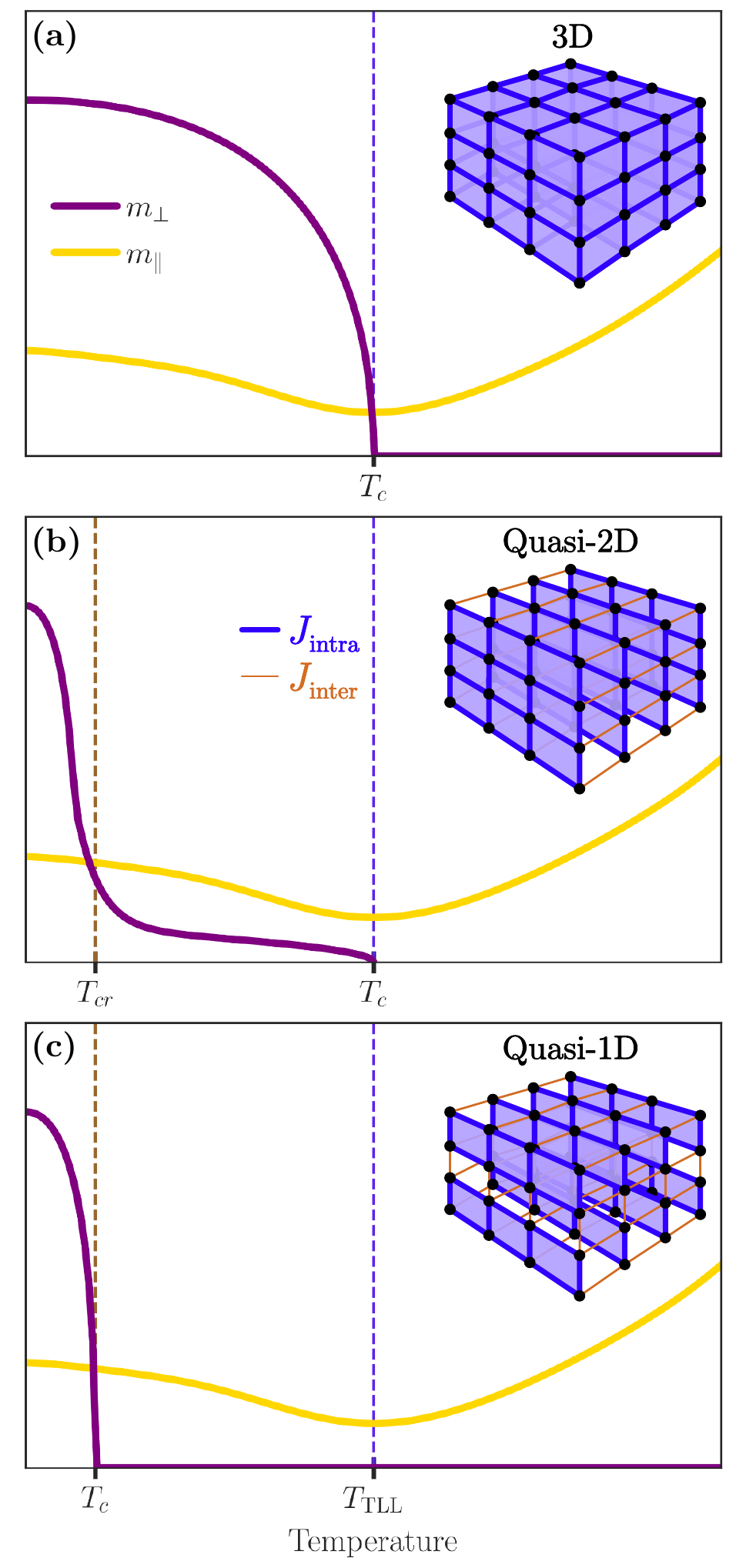}
\caption[]{Schematic temperature-dependence of the longitudinal magnetization, 
$m_{\parallel}$, and the transverse magnetization, $m_{\perp}$, in the 
field-induced ordered phases of spin-dimer systems whose substructures, with 
characteristic energy scale $J_{\rm intra}$, have different dimensionalities 
and weak mutual coupling $J_{\rm inter}$. (a) 3D. (b) 2D, showing an example 
in which $m_{\perp}$ undergoes particularly strong thermal suppression at 
$T_{\rm cr}$, but nevertheless remains finite up to $T_c$ \cite{Furuya2016}. 
(c) 1D, where $T_c$ is determined by $J_{\rm inter}$ whereas $J_{\rm intra}$ 
governs the behavior only of the substructure, setting the characteristic 
temperature $T_{\rm TLL}$.} 
\label{fig:magnetization_case}
\end{center}
\end{figure}

To see how certain hallmarks of low-dimensional physics manifest themselves in 
the temperature-dependence of $m_{\perp}$ and $m_{\parallel}$ in the field-induced 
ordered phase, Fig.~\ref{fig:magnetization_case} represents both quantities as 
functions of temperature for spin-dimer systems with different substructure 
dimensionalities. In 3D, $m_{\parallel}$ displays a minimum where magnetic order 
is lost, before increasing as the rising temperature causes a preferential 
population of the lowest of the Zeeman-split triplet states 
[Fig.~\ref{fig:magnetization_case}(a)], and this property was used to define 
the phase boundary in TlCuCl$_3$ \cite{Nikuni2000}. This minimum marks the 
point where thermal fluctuations destroy coherence within the low-dimensional 
substructure, and hence it still corresponds to $T_c$ in a quasi-2D system 
[Fig.~\ref{fig:magnetization_case}(b)]. By contrast, in a quasi-1D system 
[Fig.~\ref{fig:magnetization_case}(c)] it indicates only a crossover 
temperature, $T_{\rm TLL}$, out of the quasi-ordered TLL phase \cite{Maeda2007}, 
and this physics has been observed in the two spin-ladder compounds BPCB 
[(C$_5$H$_{12}$N])$_2$CuBr$_4$] \cite{Klanjsek2008,Lorenz2008,Thielemann2009} 
and DIMPY [(C$_7$H$_{10}$N)$_2$CuBr$_4$] \cite{Schmidiger2012,Ninios2012}. 
Turning to the order parameter, in 3D and quasi-1D systems 
[Figs.~\ref{fig:magnetization_case}(a) and \ref{fig:magnetization_case}(c)] 
$m_{\perp}$ shows a conventional temperature-dependence, but it has been 
proposed \cite{Furuya2016} that a quite different form could be observed 
in sufficiently 2D systems [Fig.~\ref{fig:magnetization_case}(b)]. In this 
scenario, $m_{\perp}$ falls rapidly at low temperatures as 3D coherence is 
lost, before attaining a steady but strongly suppressed value beyond a 
crossover temperature, $T_{\rm cr}$, and remaining finite up to a critical 
temperature, $T_c$; although the magnitude of $T_c$ is characteristic of 
the in-plane (2D) energy scale, the critical properties around it may be 
3D nature. 

While the forms of $m_{\parallel}$ and $m_{\perp}$ depicted in 
Fig.~\ref{fig:magnetization_case} have been confirmed by neutron diffraction 
and nuclear magnetic resonance (NMR) in 3D \cite{Nikuni2000,Tanaka2001} and 
in quasi-1D spin-dimer materials \cite{Thielemann2009,Ninios2012}, the special 
quasi-2D form of $m_{\perp}$ has not so far been observed. This raises the 
question of which quasi-2D spin-dimer material may offer a suitable candidate 
to search for such hallmarks of 2D physics. As reviewed recently in 
Ref.~\cite{Allenspach2021}, quasi-2D materials including the 
Shastry-Sutherland compound SrCu$_2$(BO$_3$)$_2$ \cite{Kageyama1999}, 
the ``triplon-breakdown'' compound (C$_4$H$_{12}$N$_2$)Cu$_2$Cl$_6$ (PHCC) 
\cite{Stone2001,Stone2006}, and the triangular-dimer-lattice chromate 
compounds Ba$_3$Cr$_2$O$_8$ \cite{Nakajima2006,Kofu2009,Aczel2009a} and 
Sr$_3$Cr$_2$O$_8$ \cite{Aczel2009b,Islam2010,Nomura2020} all have additional 
physics that appear to remove them from consideration in this context. 
Although a possible BKT phase has been reported in the metal-organic material 
TK91 [C$_{36}$H$_{48}$Cu$_2$F$_6$N$_8$O$_{12}$S$_2$], which consists of 
stacked and distorted honeycomb planes \cite{Tutsch2014}, the maximal 
temperature for quasi-LRO of 50~mK poses a serious challenge to a systematic 
experimental investigation by neutron diffraction or NMR.

BaCuSi$_2$O$_6$ is both a purple pigment known in ancient China 
\cite{FitzHugh1992} and a quasi-2D quantum magnet composed of $S = 1/2$ 
Cu$^{2+}$ dimers arranged in a square-lattice geometry with offset bilayer 
stacking. Early experiments by torque magnetometry were interpreted as 
indicating a counterintuitive reduction of the effective system dimension 
from 3D to 2D as the temperature was reduced towards the field-induced 
critical point \cite{Sebastian2006}, which was presumed to be a consequence 
of frustrated inter-bilayer interactions. However, intensive subsequent 
investigation revealed that the low-temperature phase contains three different 
types of structurally \cite{Samulon2006,Sheptyakov2012} and magnetically 
\cite{Rueegg2007,Kraemer2007} inequivalent bilayer, and as we discuss 
in Sec.~\ref{sec:bacusio} this both explains the presence of an anomalous 
critical scaling regime, previously misinterpreted as dimensional reduction 
\cite{Allenspach2020}, and offers a qualitatively different route to 
realizing the quasi-2D physics of Fig.~\ref{fig:magnetization_case}(b) 
\cite{Furuya2016}. 

Although neutron scattering is the method of choice for characterizing 
the structure and excitations of magnetic states, its application to the 
field-induced ordered phase of BaCuSi$_2$O$_6$ has to date been impossible 
due the fact that the critical magnetic field is 23.35~T. Here we report a 
neutron scattering study of BaCuSi$_2$O$_6$ at fields up to 25.9~T, made 
possible by the HFM/EXED facility, which was installed and operated at the 
Helmholtz-Zentrum Berlin from 2015 until 2019. Working in diffraction mode, 
we measure the intensities of a number of Bragg peaks of the ordered phase 
as functions of magnetic field and temperature, and perform a thorough 
statistical analysis to identify the evolution of the magnetic order 
parameter. In spectroscopy mode, we measure the magnetic excitations at 
fields both below and within the regime of field-induced order, and 
compare these to the modes expected on the basis of the interactions 
determined at zero field.

The structure of this article is as follows. In Sec.~\ref{sec:bacusio} we 
summarize the collected body of knowledge concerning BaCuSi$_2$O$_6$ and 
introduce the possible consequences of its inequivalent layering. In 
Sec.~\ref{sec:preparation} we introduce the instrument HFM/EXED and the 
experimental possibilities it allowed. Section \ref{sec:neutron_diffraction} 
presents the results of our diffraction measurements and a systematic global 
analysis of these data. In Sec.~\ref{sec:ins_measurements} we discuss the 
evolution of the magnetic excitation spectrum as a function of the applied 
field. Section \ref{sec:dc} contains a brief discussion and conclusion. 

\begin{figure*}[t]
\begin{center}
\includegraphics[width=0.96\linewidth]{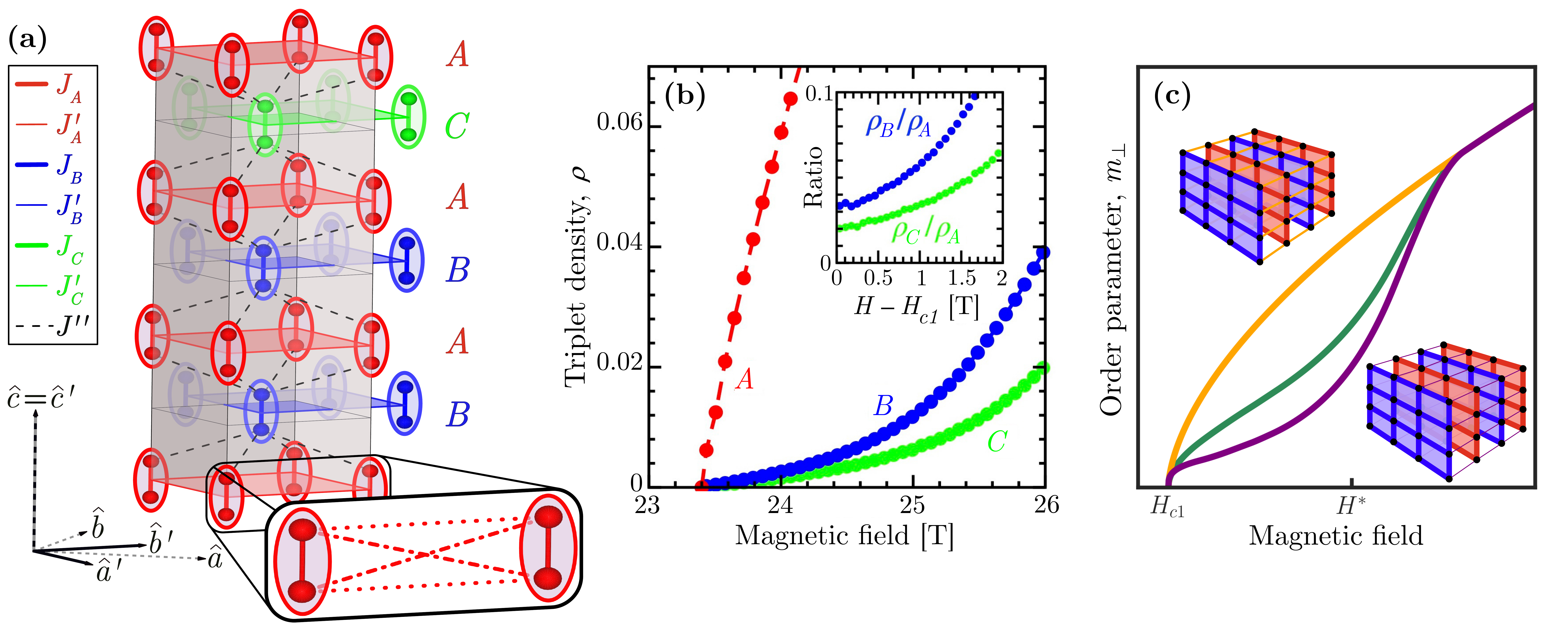}
\caption[]{(a) Representation of a primitive unit cell of BaCuSi$_2$O$_6$.
The three distinct bilayer types are labelled A, B, and C, and their 
ABABAC stacking sequence gives them population ratios 3:2:1. The minimal 
magnetic model requires the intra-dimer, inter-dimer, and inter-bilayer 
Heisenberg interactions $\{J_\sigma, J_\sigma^\prime, J^{\prime\prime} \}$; the 
effective inter-dimer interaction parameters within each bilayer 
($J_\sigma^\prime$, edges of colored squares) are sums of four pairwise
ionic interactions (inset). 
(b) Triplon occupations in the three types of bilayer, computed by quantum 
Monte Carlo for unrenormalized $\{J\}$ parameters and shown as a function 
of field at an effective temperature $T = 100$ mK. Inset: occupation ratios.
(Both panels from Ref.~\cite{Allenspach2021}).
(c) Schematic representation of the field-induced magnetic order parameter, 
$m_\perp$, at zero temperature for a magnetic model of inequivalent stacked 
bilayers. For an approximately uniform model, meaning with near-equivalent 
layers (yellow), one expects a conventional growth of $m_\perp$ with a 
conventionally broad regime of 3D critical scaling. For an extremely 
non-uniform model, with $J_B - J_A \gg J^{\prime\prime}$ (violet), one expects 
$m_\perp$ to become finite once $H$ exceeds the gap of the A bilayers, but to 
remain strongly suppressed over the regime $H_{c1} < H < H^*$, where $H^* - 
H_{\rm c1} = (J_B - J_A)/g \mu_B \mu_0$, until the field establishes a 
significant triplon occupation on all bilayers. For a model of intermediate 
uniformity (dark green), one may expect a partial suppression of $m_\perp$ 
above $H_{c1}$ that acts to reduce the regime of 3D critical scaling very 
strongly.}
\label{fig:bacusio_model}
\end{center}
\end{figure*}

\section{B\lowercase{a}C\lowercase{u}S\lowercase{i}$_2$O$_6$}
\label{sec:bacusio}

The spin-dimer compound BaCuSi$_2$O$_6$ forms large, purple-colored single 
crystals. The Cu$^{2+}$ ions provide $S = 1/2$ quantum spins, which are 
coupled first into dimer units, which in turn form square-lattice bilayers, 
and finally these bilayers are stacked with a relative [1/2 1/2] offset to 
form the 3D structure. At room temperature there is only one type of bilayer 
\cite{Sparta2006,Samulon2006}, but BaCuSi$_2$O$_6$ undergoes a structural phase 
transition below 90 K to a weakly orthorhombic structure \cite{Samulon2006}, 
in which the adjacent bilayers become structurally inequivalent 
\cite{Sheptyakov2012}. At zero field, the ground state is a global singlet, 
whose triplon excitations have a gap of 3.15 meV, and early neutron scattering 
\cite{Rueegg2007} and NMR investigations \cite{Kraemer2007,Kraemer2013} 
demonstrated from the presence of multiple triplon modes that the different 
bilayer types also become magnetically inequivalent. 

While the critical field required to close the spin gap, $\mu_0 H_{c1} = 23.35$ 
T for ${\hat H} \parallel {\hat c}$, has excluded neutron scattering as a probe 
of the field-induced ordered phase, NMR and a number of other high-field 
experimental techniques \cite{Jaime2004,Sebastian2005,Sebastian2006} have been 
applied to measure the field-temperature phase boundary. An effort to analyze 
the critical scaling of this phase boundary, as extracted from detailed torque 
magnetometry measurements \cite{Sebastian2006}, led the authors to propose an 
unusual dimensional reduction, from 3D to 2D scaling, on approaching the 
quantum critical point. The origin of this behavior was proposed to lie in the 
exact frustration of the presumed AF interactions between the offset bilayers, 
and was supported by some subsequent theoretical studies \cite{Batista2007,
Schmalian2008} but contested by others \cite{Maltseva2005,Roesch2007a}, 
notably those including the inequivalent bilayers \cite{Roesch2007b}. 

The situation was resolved for BaCuSi$_2$O$_6$ by the theoretical observation
\cite{Mazurenko2014} that the inter-dimer interactions within each bilayer 
should in fact be effectively FM, as a consequence of the relative couplings 
between ion pairs represented in the inset of Fig.~\ref{fig:bacusio_model}(a). 
This would mean that the inter-bilayer interactions are entirely unfrustrated, 
obviating the dimensional reduction scenario. Inelastic neutron scattering 
(INS) studies at zero magnetic field verified this situation experimentally 
\cite{Allenspach2020} by confirming the presence of three triplon modes, 
previously observed in Ref.~\cite{Rueegg2007}, establishing an ABABAC 
$c$-axis stacking of the three corresponding magnetic bilayer types (labelled 
in ascending order of triplon energy as A, B, and C), and determining that the 
spin Hamiltonian, depicted in Fig.~\ref{fig:bacusio_model}(a), has effective 
FM intra-bilayer and unfrustrated AF inter-bilayer interactions. It is clear 
from Fig.~\ref{fig:bacusio_model}(a) that the bilayer stacking introduces an 
energy scale $J_B - J_A$, and with it a regime of magnetic field directly 
above $H_{c1}$ where the response of the different bilayers to the applied 
field will not be the same, a result confirmed \cite{Allenspach2020} by 
quantum Monte Carlo (QMC) simulations of the bilayer triplon densities 
[Fig.~\ref{fig:bacusio_model}(b)]. For the 3D critical scaling of the 
ABABAC system, the important quantity is the energy scale ${\tilde 
J}^{\prime\prime} = (J^{\prime\prime})^2 / [J_B - J_A]$, which is approximately 
0.04~K in BaCuSi$_2$O$_6$ \cite{Allenspach2020}, because thermal fluctuations 
above this scale may cause an effective decoupling of the triplon-condensed 
$A$ bilayers. This converts the true 3D scaling into an anomalous effective 
scaling regime, which arises from the bilayer inequivalence and was originally 
misinterpreted as dimensional reduction.

The concept of non-uniform layering raises the prospect of a dramatic 
strengthening, encapsulated in the renormalization of $J^{\prime\prime}$ to 
${\tilde J}^{\prime\prime}$, of the quasi-2D nature of a system, and hence of 
the possibility that the physics of Fig.~\ref{fig:magnetization_case}(b) 
could be observed in experiment. Now that neutron scattering measurements 
have become possible at the low edge of the field-induced magnetically 
ordered phase in BaCuSi$_2$O$_6$, precisely where the different bilayers 
respond differently [Fig.~\ref{fig:bacusio_model}(b)], one may anticipate 
how the order parameter could evolve for different degrees of non-uniformity 
in the bilayer properties. Figure \ref{fig:bacusio_model}(c) depicts the 
contrast between the conventional growth of $m_\perp$ with $H$ for a uniform 
system ($J_B - J_A < J^{\prime\prime}$), and a situation where the extremely 
non-uniform bilayer properties ($J_B - J_A \gg J^{\prime\prime}$) ensure a 
wide regime of applied field in which the triplon gap would be closed on an 
isolated A bilayer but not on a B bilayer. In this region, the B bilayer has 
only weak, proximity-induced magnetic order arising from the A-bilayer 
condensation and the interlayer coupling, and one may expect the order 
parameter to remain suppressed until the field is large enough to create a 
significant B-bilayer triplon condensation. As the bilayers are made more 
similar, first the suppression would become less pronounced and then one may 
not observe a non-monotonic first derivative in $m_\perp (H)$, but it is clear 
that any finite $J_B - J_A > J^{\prime\prime}$ will act to curtail the regime 
of true 3D critical scaling (as observed in Ref.~\cite{Allenspach2020}). We 
stress that the regime of suppressed $m_\perp (H)$ for the non-uniform 
bilayer stack in Fig.~\ref{fig:bacusio_model}(c) has no direct correspondence 
with the finite-temperature plateau in $m_\perp (T)$ shown in 
Fig.~\ref{fig:magnetization_case}(b); the latter is proposed 
\cite{Furuya2016} to be a property even of a uniform stack (depicted in 
the figure inset), and its field response is that of the yellow curve in 
Fig.~\ref{fig:bacusio_model}(c). The non-uniform stack should not only 
improve the prospects for observing such extreme 2D physics in $m_\perp (T)$ 
but also offer a less stringent possibility, in $m_\perp (H)$, for observing 
a response function reflecting 2D substructures behaving in a quasi-isolated 
manner. 

Next we comment that reducing the non-uniformity of the bilayer stacking 
in BaCuSi$_2$O$_6$ has already been achieved. Stoichiometric substitution 
of Sr for Ba ions acts to suppress the 90~K structural phase transition, 
with 5\% Sr being sufficient to stabilize the tetragonal structure down 
to the lowest temperatures \cite{Puphal2016}. INS measurements of 
Ba$_{0.9}$Sr$_{0.1}$CuSi$_2$O$_6$ at zero magnetic field confirmed that there 
is only one type of bilayer at 1.5~K and determined the spin Hamiltonian 
\cite{Allenspach2021}. NMR measurements were performed at high magnetic 
fields and low temperatures to determine the phase boundary of field-induced 
order, and a detailed analysis revealed 3D quantum critical scaling with no 
hallmarks of anomalous behavior. Similarly, the relative order parameter 
extracted from the NMR spectra showed no deviation from the conventional 
form \cite{Allenspach2021}. Because the ratio of intra- to inter-bilayer 
interactions is 25 in Ba$_{0.9}$Sr$_{0.1}$CuSi$_2$O$_6$ (a value double that 
of the unsubstituted system \cite{Allenspach2020}), it is clear that, 
for a uniform bilayer stack, this value of the interaction ratio is not 
sufficiently close to the quasi-2D regime to observe any of the fingerprints 
of 2D physics proposed in Fig.~\ref{fig:magnetization_case}(b). 

Based on this result we observe that, even if samples were found that realize 
interaction ratios orders of magnitude beyond the value of 25, the regime 
where the order parameter exhibits quantum critical scaling would be pushed 
below mK temperatures and would become impossible to investigate by present 
methods. By contrast, creating non-uniform layered structures falls within 
present technological capabilities for engineering atomically thin magnetic 
materials \cite{Gibertini2019,Klein2019,Ubrig2019,Cao2019,Wang2019,Kim2019}. 
While still relying on the non-uniform layering occurring naturally in 
BaCuSi$_2$O$_6$, we will use our ability to perform direct measurements of 
the order parameter to investigate whether this degree of non-uniformity is 
sufficient to provide one of the unconventional curves shown in 
Fig.~\ref{fig:bacusio_model}(c).

\section{HFM/EXED}
\label{sec:preparation}

The benchmark value for the critical field of BaCuSi$_2$O$_6$ is $\mu_0 H_{c1}
 = 23.35$~T, determined by $^{29}$Si NMR measurements with the field aligned 
along the sample $c$ axis \cite{Kraemer2007}. Thus the field-induced ordered 
phase has remained inaccessible to the conventional superconducting magnets 
used in neutron scattering experiments, which are limited to maximal static 
magnetic fields of 17~T. However, neutron scattering measurements were 
possible at higher fields on the Extreme Environment Diffractometer (EXED) 
within the High Field Magnet (HFM) facility \cite{Prokhnenko2017} at the 
Helmholtz-Zentrum Berlin, which operated from 2015 to 2019. 

HFM was a horizontal, series-connected hybrid magnet consisting of three 
concentric solenoids, an outer superconducting coil and a combination of two 
inner, normal conducting coils \cite{Smeibidl2016}. Magnetic fields up to 26~T 
could be reached with a 4 MW inner coil at an operating current of 20~kA. These 
fields were maintained across the 50 mm diameter of the warm bore, in which the 
cryostat was installed, with a field homogeneity of 0.5\% across a cryogenic 
sample volume of order (15 mm)$^3$. EXED was a time-of-flight (TOF) neutron 
instrument built at HFM that had three different modes of operation, namely 
diffraction, spectroscopy, and low-$Q$ \cite{Bartkowiak2015,Prokhnenko2015,
Prokhnenko2017}. We refer henceforth to the combination of HFM and EXED as 
HFM/EXED. 

The primary targets for a neutron scattering characterization of the 
field-induced ordered phase are the magnetic order parameter, $m_\perp$, and the 
dispersion of the magnetic excitations throughout the Brillouin zone. Concerning 
the order parameter, NMR is sensitive to the local magnetization of the 
Cu$^{2+}$ ions through the positions of the peaks in the spectrum, and thus a 
relative $m_\perp$ can in principle be extracted as a function of field and 
temperature from the frequency splitting of these peaks \cite{Berthier2017,
Allenspach2021}. However, this turned out not to be possible in BaCuSi$_2$O$_6$ 
because of the broad and complex line shapes caused by the incommensurability 
present in the low-temperature structure \cite{Samulon2006,Sheptyakov2012,
Stern2014}. By contrast, neutron diffraction allows a direct determination of 
$m_\perp$, from the intensity of the magnetic Bragg peaks, whose location can 
be well separated from the nuclear Bragg peaks. In fact the magnetic Bragg 
peaks accessible in BaCuSi$_2$O$_6$ with the alignment we used on HFM/EXED do 
coincide with the locations of the nuclear Bragg peaks, a situation that is 
readily dealt with by including the nuclear contribution within the global 
intensity model used to fit the diffraction data 
(Sec.~\ref{sec:neutron_diffraction}C). Regarding the dispersion relation of 
magnetic excitations, there is no alternative to inelastic neutron scattering.

\begin{figure*}[t]
\begin{center}
\includegraphics[width=1.99\columnwidth]{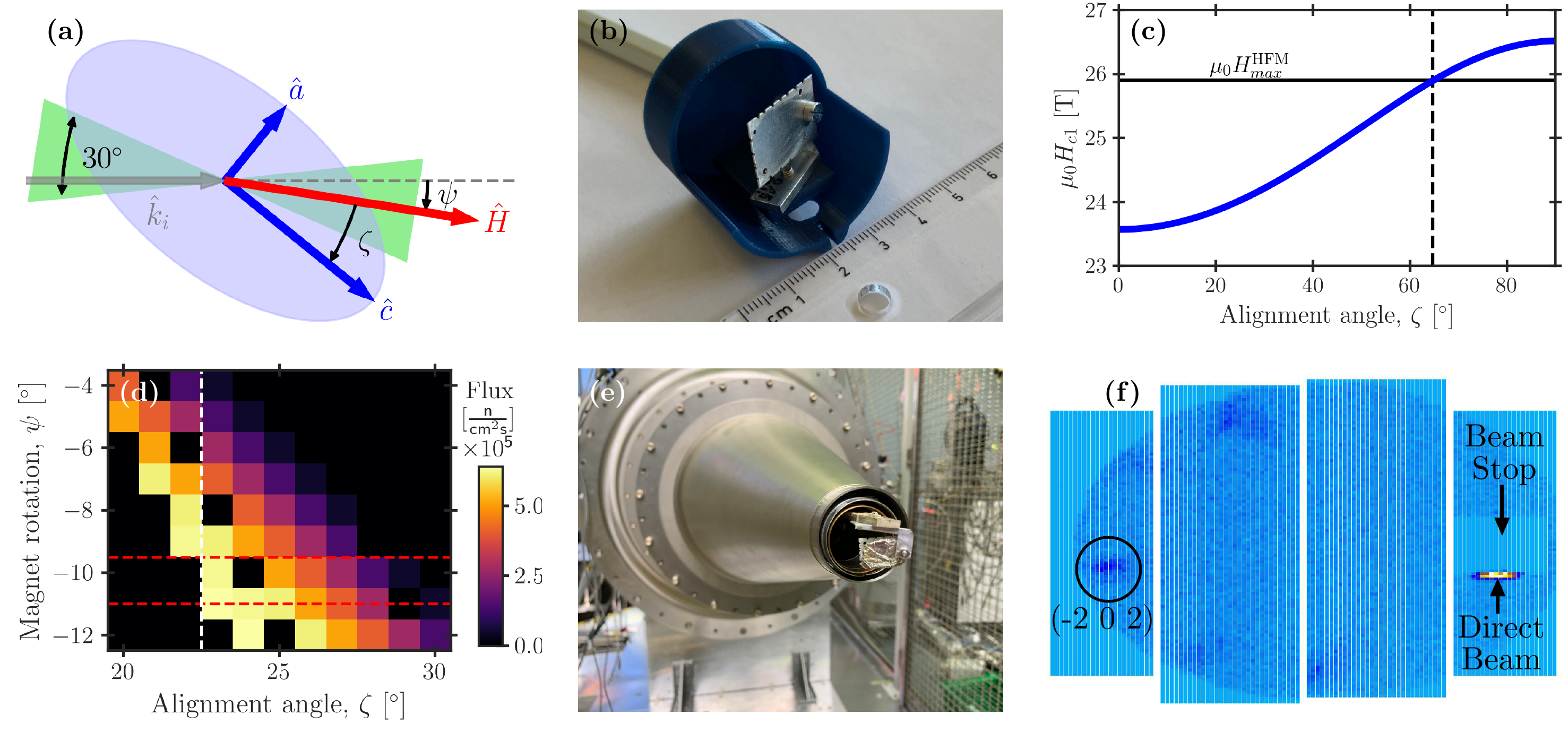}
\caption[]{Summary of the experimental conditions on HFM/EXED. (a) Top view 
of the scattering plane, illustrating the magnet rotation angle, $\psi$, 
and sample alignment angle, $\zeta$. The $g$-factor \cite{Zvyagin2006} is 
represented as the (exaggerated) blue ellipse and its values in the field of 
the experiment therefore depend on $\zeta$. The 30$^{\circ}$ conical opening 
allowed by the magnet is indicated by the green area. 
(b) 3D-printed sample chamber and aluminum sample holder. 
(c) Critical field, $\mu_0 H_{c1}$, shown as a function of $\zeta$. The 
horizontal line corresponds to the maximum magnetic field achievable on 
HFM/EXED (25.9~T) and imposes an upper limit for $\zeta$ (dashed vertical line).
(d) Neutron flux on HFM/EXED modelled for the ($- 2$ 0 2) Bragg peak of 
BaCuSi$_2$O$_6$; pixels with zero flux within the region of detector coverage 
are a consequence of neutrons scattered into gaps between detector elements. 
The dashed white line indicates the chosen $\zeta$ (22.5$^{\circ}$) and the 
red dashed lines the chosen $\psi$ values ($- 9.5^{\circ}$ and $- 11^{\circ}$). 
(e) Sample installed below the cold finger of the cryostat. 
(f) Forward-scattering detector panels of HFM/EXED, displaying an example of 
the neutron intensity of BaCuSi$_2$O$_6$ collected in diffraction mode.}
\label{fig:preparation_overview}
\end{center}
\end{figure*}

HFM/EXED offered a solution to both problems, in that the measurements 
of magnetic order presented in Sec.~\ref{sec:neutron_diffraction} were 
performed in diffraction mode and the inelastic data discussed in 
Sec.~\ref{sec:ins_measurements} were collected in spectroscopy mode. 
Figure~\ref{fig:preparation_overview}(a) shows the scattering geometry 
of the experiments. The magnetic field was applied horizontally and the 
magnet could be rotated by an angle, $\psi$, of up to 12$^{\circ}$ relative 
to the beam of incoming neutrons, which had wave vectors $\vec{k}_{i}$. The 
sample was aligned with ($h$ 0 $l$) in the horizontal scattering plane and 
the angle between the magnetic field and the $c$ axis of the sample, $\zeta$, 
is referred to henceforth as the alignment angle. Because the alignment could 
not be changed once the sample was installed inside the cryostats used 
(Sec.~\ref{sec:experimental_conditions}), $\zeta$ had to be selected before 
the experiments. Incoming neutrons entering the sample chamber were scattered 
to the detector through cone-shaped openings, which subtended an angle of 
30$^{\circ}$ and are shown as the green areas in 
Fig.~\ref{fig:preparation_overview}(a). 

There were three factors to be taken into account for the choice of $\zeta$. 
(i) The sample had to be aligned to fit into the sample chamber of the 
cryostat. To ensure this, a 1:1 model of the sample chamber was produced by 
3D printing, and is shown in Fig.~\ref{fig:preparation_overview}(b) together 
with the sample holder. (ii) The critical field varies with the direction 
of the applied magnetic field, because $\mu_0 H_{c1} = \Delta/(g\mu_B)$, 
where $\Delta$ is the spin gap, and the effectively temperature-independent 
$g$-factor of BaCuSi$_2$O$_6$ has an anisotropy of approximately 12\%, with 
$g_{cc} = 2.306(3)$ and $g_{aa} = g_{bb} = 2.050(3)$ \cite{Zvyagin2006}. 
Thus we estimated that $\mu_0 H_{c1} = 23.57$ T when the field direction is 
parallel to the $c$ axis 
($\zeta = 0^{\circ}$) and 26.15~T when the field is within the $ab$ plane 
($\zeta$ = 90$^{\circ}$) [Fig.~\ref{fig:preparation_overview}(c)] (the latter 
case making the field-induced phase inaccessible on HFM/EXED). The alignment 
angle therefore had to be made as small as possible to maximize access to 
the field-induced phase. (iii) It was necessary to choose magnetic Bragg peaks 
that are both strong, to obtain sufficient counting statistics, and maximize 
the ratio of the magnetic to the nuclear signal. Because the 30$^\circ$ opening 
angle of the magnet restricted the $\vec{Q}$-range accessible on HFM/EXED, the 
choice of Bragg peak placed a further constraint on the possible value of $\zeta$.

Based on these three factors, we decided to align the sample with $\zeta = 
22.5^{\circ}$. This alignment allows access to the ($- 2$ 0 2) Bragg peak, which 
we estimated by structure-factor calculations to have a sufficient ratio 
between its magnetic and nuclear intensities to optimize the extraction of 
the magnetic signal. The specific choice of 22.5$^{\circ}$ was based on neutron 
flux calculations performed for the ($- 2$ 0 2) Bragg peak with the beam-line 
software EXEQ \cite{Bartkowiak2020}, whose results are shown in 
Fig.~\ref{fig:preparation_overview}(d). All of our experiments were performed 
with a single-crystal sample of BaCuSi$_2$O$_6$ of weight 1.01~g, which had 
already been used for the zero-field INS measurements reported in 
Ref.~\cite{Allenspach2020}. 

\begin{figure*}[t]
\begin{center}
\includegraphics[width=1.99\columnwidth]{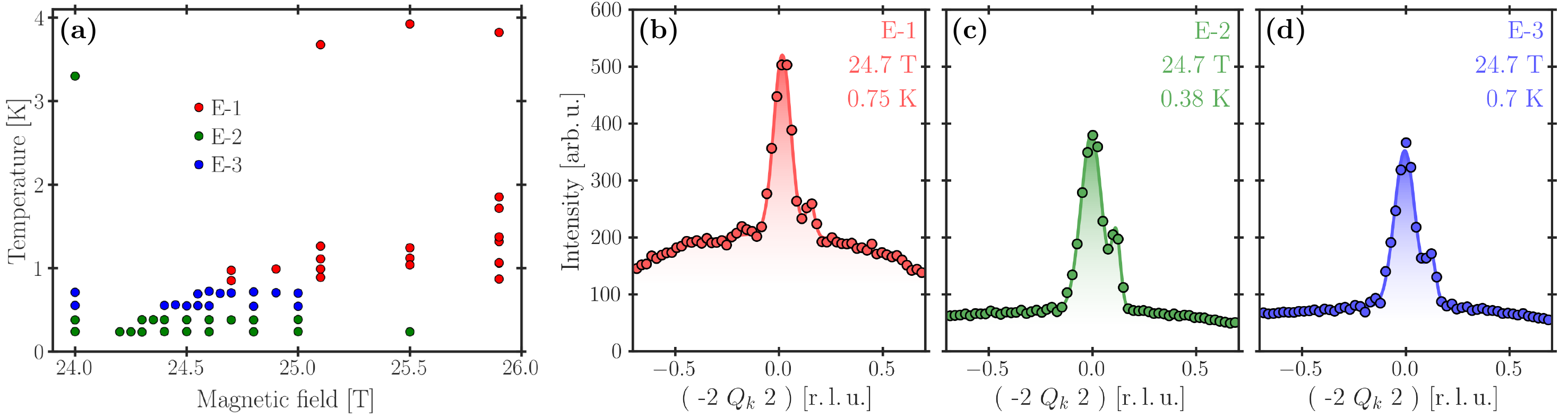}
\caption[]{(a) Overview of the different points in the ($\mu_0 H,T$) phase 
diagram at which the ($- 2$ 0 2) Bragg peak was measured during the three 
experiments E-1, E-2, and E-3. (b-d) Intensity measured along ($- 2$ $Q_k$ 
2). Symbols show the normalized intensity obtained by integrating the data 
over $h$ and $l$. Solid lines display the results of individual fits using 
an intensity model of two Gaussians and a polynomial background, as described 
in the text.}
\label{fig:diffraction}
\end{center}
\end{figure*}

\section{Neutron Diffraction Measurements}
\label{sec:neutron_diffraction}

\subsection{Experimental Conditions}
\label{sec:experimental_conditions}

Diffraction data were collected during three different experiments, which in 
the following we label E-1, E-2, and E-3. In E-1, a $^{3}$He cryostat was used 
to reach temperatures down to 0.6~K, while in E-2 and E-3 the sample was 
installed in a dilution cryostat with the same geometry, but which allowed 
us to extend the measurements down to 0.25~K. 
Figure~\ref{fig:preparation_overview}(e) shows the sample mounted 
on the cold finger of the cryostat. The rotation of the magnet relative to the 
incoming neutron beam, $\psi$, was selected as $- 9.5^{\circ}$ for E-1 and $- 
11^{\circ}$ for E-2 and E-3. These values were chosen not only to maximize the 
incoming neutron flux for the ($- 2$ 0 2) Bragg peak but also to avoid having 
parts of this peak cut off by the edges of or gaps in the detector. The 
chopper settings were adjusted for each experiment so that the wavelength 
band of the incoming neutrons was centered on the ($\psi$-dependent) wavelength 
of the ($- 2$ 0 2) peak and had a width of 1~\AA. Although the sample was 
removed from the cryostats between the experiments, the alignment angle was 
set to the same initial value for all three experiments. Nevertheless, even 
with $\zeta$ aligned to 22.5$^{\circ}$, the cold fingers of the two cryostats 
were not identical, causing a field- and temperature-dependent misalignment 
that affected both the in-plane and out-of-plane positions of the 
($- 2$ 0 2) Bragg peak. The out-of-plane misalignment is visible in 
Fig.~\ref{fig:preparation_overview}(f) as a vertical offset of this peak 
on the forward-scattering detector panel. The additional bend in the cold 
finger when applying a magnetic field was found to be $\delta \zeta \leq 
0.3^{\circ}$ between 0 and 25.9~T, and this further offset was the same for 
both cryostats and thus in all experiments.

The elastic signal of BaCuSi$_2$O$_6$ was measured for the combination 
of magnetic fields and temperatures, ($\mu_0 H$, $T$), shown in 
Fig.~\ref{fig:diffraction}(a). Data were collected by scanning either the 
temperature or the field, while keeping the other constant. These scans 
started either at the lowest temperature or the highest field value, where 
the magnetic order was strongest in the field-induced phase. Either the 
temperature was then increased or the magnetic field was decreased until 
the magnetic order had vanished. Afterwards the temperature was increased 
to 3.5~K before changing the field for the next scan, or the magnetic field 
was set to 24~T before changing the temperature. The goal of this procedure 
was to minimize potential hysteresis effects in our measured order parameter. 
Due to the limited amount of measurement time, we included only a small set 
of ($\mu_0 H, T)$ points in the disordered phase, and chose these to cover 
a wide range of fields and temperatures. To obtain sufficient statistics 
for the magnetic signal on top of the nuclear signal, the data-acquisition 
time was adjusted for each ($\mu_0 H, T$) point within the ordered phase 
based on its distance to the phase boundary, where $m_\perp \rightarrow 0$, 
and ranged from 3 to 16 hours.

A temperature sensor was attached to the cold fingers of the cryostats and 
measured a time series of temperature data for each ($\mu_0 H, T$) point. The 
sample temperatures and their uncertainties at every point were estimated 
based on the mean and standard deviation of each time series. The magnetic 
field of HFM/EXED was calibrated in lower fields and extrapolated to the 
highest field values. From other experiments performed on HFM/EXED 
\cite{Prokes2017,Fogh2020,Prokes2020}, this extrapolation is known to result 
in highly accurate estimates of the field values and thus the uncertainties 
in the fields were assumed to be zero throughout the analysis to follow.

\subsection{Extraction of the Peak Intensities}

The measured data were preprocessed and transformed into $\vec{Q}$-space 
using the software Mantid \cite{Arnold2014}. During the preprocessing 
step, the neutron counts were integrated over the wavelength band of the 
scattered neutrons, normalized by the monitor and a vanadium standard, 
and rescaled by the Lorentz polarization factor \cite{Debye1913} and 
by taking the sample misalignment into account. One-dimensional cuts 
were extracted from the preprocessed data along ($- 2$ $Q_k$ 2) 
by integrating over [$- 2.15$,$- 1.85$] in $h$ and [1.85,2.15] in $l$ 
for each ($\mu_0 H,T$) point, resulting in the normalized intensities 
$I(Q_k,\mu_0 H,T)$. In addition to the ($- 2$ 0 2) peak, the 
($ -2$ $\pm 1$ 2) Bragg peaks are also included in these cuts, but 
have only a weak magnetic signal. These peaks were used to verify the 
absence of any magnetostriction effects, by which the nuclear intensity 
is modified due to field-induced changes of the crystal structure. 
Although magnetostriction does not affect all Bragg peaks equally, 
structural changes affecting the ($- 2$ 0 2) peak would also be visible 
in the ($- 2$ $\pm$1 2) peaks.

Figures \ref{fig:diffraction}(b)-(d) display the cuts along ($- 2$ $Q_k$ 2) 
obtained in each of the three experimental runs, E-1, E-2, and E-3. In addition 
to the main peak centered at $Q_k = 0$, a smaller side peak is visible at $Q_k 
\approx 0.15$. This second peak has the same field- and temperature-dependence 
as the main peak and appears in the same location in the forward- and 
backscattering detector panels, from which we deduce that it is most probably 
a consequence of divergence in the beam profile across the sample volume. 
While the background is almost flat and comparable in intensity for the 
cuts obtained in E-2 [Fig.~\ref{fig:diffraction}(c)] and E-3 
[Fig.~\ref{fig:diffraction}(d)], the background contribution is much larger 
for the E-1 cut [Fig.~\ref{fig:diffraction}(b)] because a different cryostat 
was used. These one-dimensional cuts were fitted individually by two Gaussians 
(for the main and side peaks), while the background was approximated by a 
polynomial in $Q_k$. Thus the peak intensity of ($- 2$ 0 2), $I(H,T)$, is 
determined from the sum of the integrated weights of these two Gaussians.

An additional contribution to the magnetic scattering intensity not 
included in the fit function is the critical scattering that results from 
short-ranged critical fluctuations. Critical scattering would manifest 
itself as an additional broadened peak centered at ($- 2$ 0 2). Such an 
intensity contribution would be detectable due to changes in the background 
and peak width, but because the fitting parameters did not change significantly 
for different ($\mu_0 H,T$) points, this contribution was neglected in the 
peak-intensity analysis to follow. 

\begin{figure}[t]
\begin{center} 
\includegraphics[width=0.99\columnwidth]{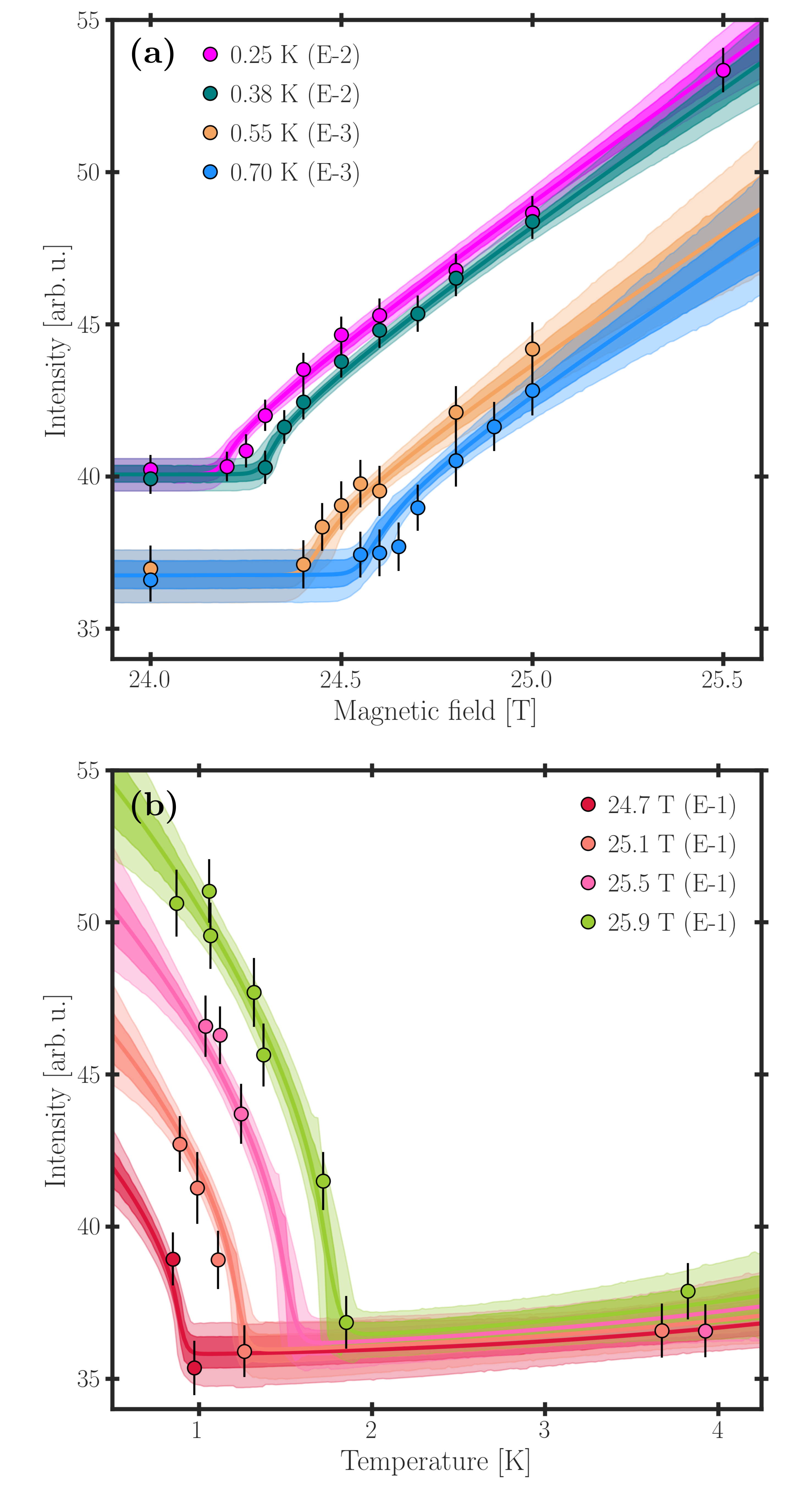}
\caption[]{Peak intensities (symbols) extracted from the diffraction 
data. (a) Dependence of selected peak intensities on the applied 
magnetic field, shown for different constant temperatures. (b) 
Dependence on temperature displayed for different constant magnetic 
fields. Lines and shading show the mean (solid lines), 68\% (light 
shading) and 95\% (dark shading) credible intervals (CI) of the 
posterior distribution obtained from the global fit presented in 
Sec.~\ref{sec:rbia}.}
\label{fig:pi}
\end{center}
\end{figure}

Figure \ref{fig:pi} displays the peak intensities extracted from the data 
(symbols) at constant temperatures [Fig.~\ref{fig:pi}(a)] and at constant 
magnetic fields [Fig.~\ref{fig:pi}(b)]. The temperatures quoted in 
Fig.~\ref{fig:pi}(a) are the set values, whereas in Fig.~\ref{fig:pi}(b) 
they are the actual sample temperatures taken from measurements 
made by the temperature sensor. In both panels, a large constant contribution 
is present due to nuclear scattering. In Fig.~\ref{fig:pi}(a), the intensities 
increase sharply at a temperature-dependent value of the applied field where 
the ordered phase is entered. For fields just above the transition, this 
contribution is dominated by the transverse magnetization ($m_\perp$), while 
for higher fields the contribution of the longitudinal component ($m_\parallel$) 
becomes visible at 0.25~K and 0.38~K, causing the approximately linear 
dependence. In Fig.~\ref{fig:pi}(b), the intensities decrease up to a 
field-dependent value of the temperature at which the ordered phase is left, 
with $m_\perp$ dominating the magnetic signal. In the thermally disordered 
phase, the rise in intensity with temperature is due to the contribution of 
thermally occupied $\ket{t^{+}}$ states to $m_\parallel$; at still higher 
temperatures, where the triplet states $\ket{t^{0}}$ and $\ket{t^{-}}$ also 
become thermally occupied, $m_\parallel$ would decrease again. 

Clearly the functional form of the Bragg-peak intensity, and hence of the 
order parameter ($m_\perp$), does not follow a conventional form across the 
entirety of both panels in Fig.~\ref{fig:pi}. However, before one could 
ascribe this situation to hallmarks of the unconventional physics depicted 
in Figs.~\ref{fig:magnetization_case}(b) or \ref{fig:bacusio_model}(c), 
it is necessary to consider both the contribution of the longitudinal 
magnetization to the measured intensity and the likely width of the  
quantum critical scaling regime. To investigate this situation in a 
fully quantitative manner, while simultaneously making optimal use of
the small number of data points at any given $\mu_0 H$ or $T$, we 
construct a global model of the Bragg-peak intensities extracted 
from the diffraction data and compare its results to the conventional 
forms for the evolution of the order parameter. 

\subsection{Peak-Intensity Model}

Quite generally, the peak intensity can be modelled as
\begin{equation}
I^{\rm model}_s(H,T) = \mathcal{A}_s \Big\{ n^{2}_{\perp} (H,T) + 
\mathcal{B} n^2_{\parallel}(H,T) \Big\} + \mathcal{C}_s,
\label{eq:intensity_model}
\end{equation}
where $\mathcal{A}_s$ are experiment-dependent global scale factors and 
$\mathcal{C}_s$ are constant (spin-independent) contributions to the 
intensity, primarily due to nuclear scattering; here $s$ labels the three 
experiments (E-1, E-2, and E-3). $n_{\perp}$ is proportional to the transverse 
magnetization, $m_{\perp}$, and $n_{\parallel}$ is the triplon density, simulated 
by QMC in Ref.~\cite{Allenspach2020}, which is proportional to the longitudinal 
magnetization, $m_{\parallel}$. $\mathcal{B}$ is an unknown factor governing the 
relative sizes of the two magnetic contributions, which depends on the specific 
magnetic Bragg peak under investigation and on the direction of the staggered 
transverse order.

The observed dependence of the peak intensities on both field and temperature 
[Figs.~\ref{fig:pi}(a) and \ref{fig:pi}(b)] motivates an effective model for 
the $m_\perp$ contribution with the same scaling form as a typical order 
parameter, 
\begin{align}
\!\!\!\! n_{\perp}(H, T) = & \enspace \Theta(g_s \mu_B \mu_0 H - \Delta) 
\Theta(T_c (H) - T) \nonumber \\ & \enspace \times \bigg[ 
\frac{g_s \mu_B \mu_0 H \! - \! \Delta}{\Delta} \bigg]^{\kappa_H }
\bigg[ \frac{T_c(H) \! - \! T}{T_c(H)} \bigg]^{\kappa_T} \!\!\!,
\label{eq:n_perp_model}
\end{align}
where $\Theta(x)$ is the Heavyside function and $\Delta = 3.15(3)$~meV 
is the spin gap determined from INS at zero field \cite{Allenspach2020}.
$g_s$ are the experiment-dependent g-factors that can be estimated from 
the alignment angles, $\zeta_s$, of the three experiments using 
\begin{equation}
g_s = \sqrt{g_{cc}^2 \cos^2(\zeta_s) + g_{aa}^2 \sin^2(\zeta_s)}
\label{eq:g_factor}
\end{equation}
with $g_{cc}$ = 2.306 and $g_{aa}$ = $g_{bb}$ = 2.050 \cite{Zvyagin2006}.

The shape of the phase boundary, $T_c(H) \propto T_c^{\rm QMC}(h)$, is 
known from QMC simulations based on the spin Hamiltonian determined by 
INS \cite{Allenspach2020}. These simulations were performed by mapping 
the spin system to a system of hard-core bosons and with the field $h$ 
expressed in the same units as the interaction parameters (meV). To 
compensate for the $g$-tensor and for the discrepancy between the triplon 
and hard-core-boson dispersions, both the field and the temperature should 
be rescaled. We rescaled the field so that the effective gap of the hard-core 
bosons matched the INS gap, $\Delta$, and included one global scale factor 
for the temperature, $\alpha$, as an unknown model parameter, with which we 
modelled the phase boundary appearing in Eq.~\eqref{eq:n_perp_model} as
\begin{equation}
T_c(H) = \alpha T_c^{\rm QMC}(g_s \mu_B \mu_0 H).
\label{eq:pb_model}
\end{equation}
Finally, the sample temperatures of the ($\mu_0 H, T$) points, 
$\vec{T} \equiv \{T_j\}$ where $j$ labels all the points in 
Fig.~\ref{fig:diffraction}(a), are included as parameters constrained 
by the mean and standard deviation of the temperature-sensor measurements.  

To optimize the parameters of the peak-intensity model based on all of 
the measured data, we used a procedure of Bayesian inference (BI) to 
obtain the joint probability distribution, $p(\bm{\theta}|\mathcal{D})$, 
of these parameters, $\bm{\theta} \equiv (\kappa_H,\kappa_T,\alpha,
\mathcal{B},\mathcal{A}_s,\mathcal{C}_s,\zeta_s,\vec{T})$, where 
$\mathcal{D}$ denotes the data \cite{Sivia1996,Bishop2006}. In BI, 
$p(\bm{\theta}|\mathcal{D})$ is referred to as the posterior 
distribution, and details of its definition and determination are 
in presented in Appendix~\ref{sec:adpd}.

\subsection{Results of the Bayesian Inference Analysis}
\label{sec:rbia}

The parameter set $\bm{\theta}$ clearly contains a significant number 
of heterogeneous and interdependent variables, whose only common 
feature is that their effects within the model are well defined 
[Eq.~\eqref{eq:intensity_model}]. BI offers a powerful and systematic 
statistical procedure for solving this class of problem, i.e.~for 
estimating the values of and quantifying the uncertainties in all of 
these interlinked parameters simultaneously, using the constraints 
set by experimental observation. By contrast, a piecewise 
or sequential method of separating out the contributions to 
Eq.~\eqref{eq:intensity_model} would risk introducing bias, 
propagating significant errors, and increasing the statistical 
uncertainties, particularly in a situation with limited experimental 
data. In the present case, one may summarize the goal of applying the 
BI procedure with such a global and simultaneous model as being to 
separate out the extrinsic uncertainties, which arose unavoidably in 
performing the experiment, from the intrinsic uncertainties arising 
from the imperfectly determined model parameters. In 
Appendix~\ref{sec:aipd} we provide a detailed analysis to illustrate 
the dependence of the posterior distribution on the multiple parameters 
in the peak-intensity model, primarily by means of projections onto 
different parameter pairs. 

Focusing first on the experimental parameters, the fact that 
the alignment angles in the three experimental runs may differ 
(Sec.~\ref{sec:experimental_conditions}) has an effect on the 
$g$-factor and as a result on the position of the phase boundary in 
($\mu_0 H, T$) space. From the posterior distribution we deduce that 
$\zeta_{\rm E-1} = 22.3^{+0.9 \, \circ}_{-0.9}$, $\zeta_{\rm E-2} = 24.7^{+0.7 \, 
\circ}_{-0.5}$, and $\zeta_{\rm E-3} = 24.3^{+0.8 \, \circ}_{-0.9}$. From 
these values we can conclude that our original alignment of the 
sample (22.5$^{\circ}$) was accurate, but that either it was 
misaligned slightly during removal from the $^{3}$He cryostat and 
mounting on the dilution cryostat (E-2, E-3), or the alignment 
changed due to minor differences in the cryostats and their cold 
fingers. The change in $\zeta$ from E-2 to E-3 is likely the result
of a strong quench of the magnet system that terminated E-2 and 
required maintenance of the entire system (before E-3). From these 
$\zeta$ values, the $g$-factors of the three experiments can be 
determined using Eq.~\eqref{eq:g_factor} as $g_{\rm E-1} = 2.271
^{+0.003}_{-0.003}$, $g_{\rm E-2} = 2.263^{+0.002}_{-0.002}$, and 
$g_{\rm E-3} = 2.265^{+0.003}_{-0.003}$. Concerning the $\mathcal{A}_s$ 
and $\mathcal{C}_s$ parameters, we find broad consistency across 
each of the runs, indicating the internal consistency of the 
assumptions made in the global model.

Turning to the parameters of the physical model, we comment first 
that the maximum gain in accuracy compared to our previous study of 
criticality in Sr-doped BaCuSi$_2$O$_6$ \cite{Allenspach2021} stems 
from the fact that $\Delta$ is fixed in Eq.~\eqref{eq:n_perp_model}, 
because this functional form makes all the other parameters 
extremely sensitive to the value of $\Delta$ \cite{Allenspach2020}. 
In the present case this fixes the critical field, $H_{c1}$, up to 
the measurement accuracy of the angles, $\zeta_s$, determining $g$ 
in Eq.~\eqref{eq:g_factor}. For the exponents 
of $n_{\perp}(H, T)$, we find systematic fits to a single value 
$\kappa_T = 0.26^{+0.05}_{-0.06}$ describing the $T$-dependence and 
a single value $\kappa_H = 0.45^{+0.05}_{-0.05}$ describing the 
$H$-dependence. These results may be viewed as a slight surprise, 
given that the widths of our fitting regimes in both $T$ and $H$ far 
exceed the widths over which one might expect to identify critical 
behavior with the same functional forms. In the event that sufficient 
data were available within the critical regimes, one would expect 
these exponents to show an asymptotic approach to the critical values 
for order-parameter scaling of the 3D-XY universality class, $\beta_T
 = 0.3689(3)$ \cite{Campostrini2002} and $\beta_H = 0.5$ (mean-field 
scaling) \cite{Matsumoto2004}, which correspond respectively to 
classical and quantum criticality. Although the widths of the 
associated critical regimes are non-universal and not known {\it a 
priori}, it is clear from the values we obtain for $\kappa_T$ and 
$\kappa_H$ that the data we have included in the analysis extend 
well beyond them. However, the long data-acquisition times per 
($\mu_0 H, T$) point in our experiments made it impossible to 
increase the point density in these regimes to attempt an 
experimental determination of $\beta_H$ and $\beta_T$. Nevertheless, 
our results do provide a clear picture of the requirements for 
overcoming this experimental challenge and the peak-intensity model 
of Eq.~\eqref{eq:n_perp_model} could be used directly in any future 
investigation that seeks to determine these critical exponents.

\begin{figure*}[t]
\begin{center}
\includegraphics[width=1.99\columnwidth]{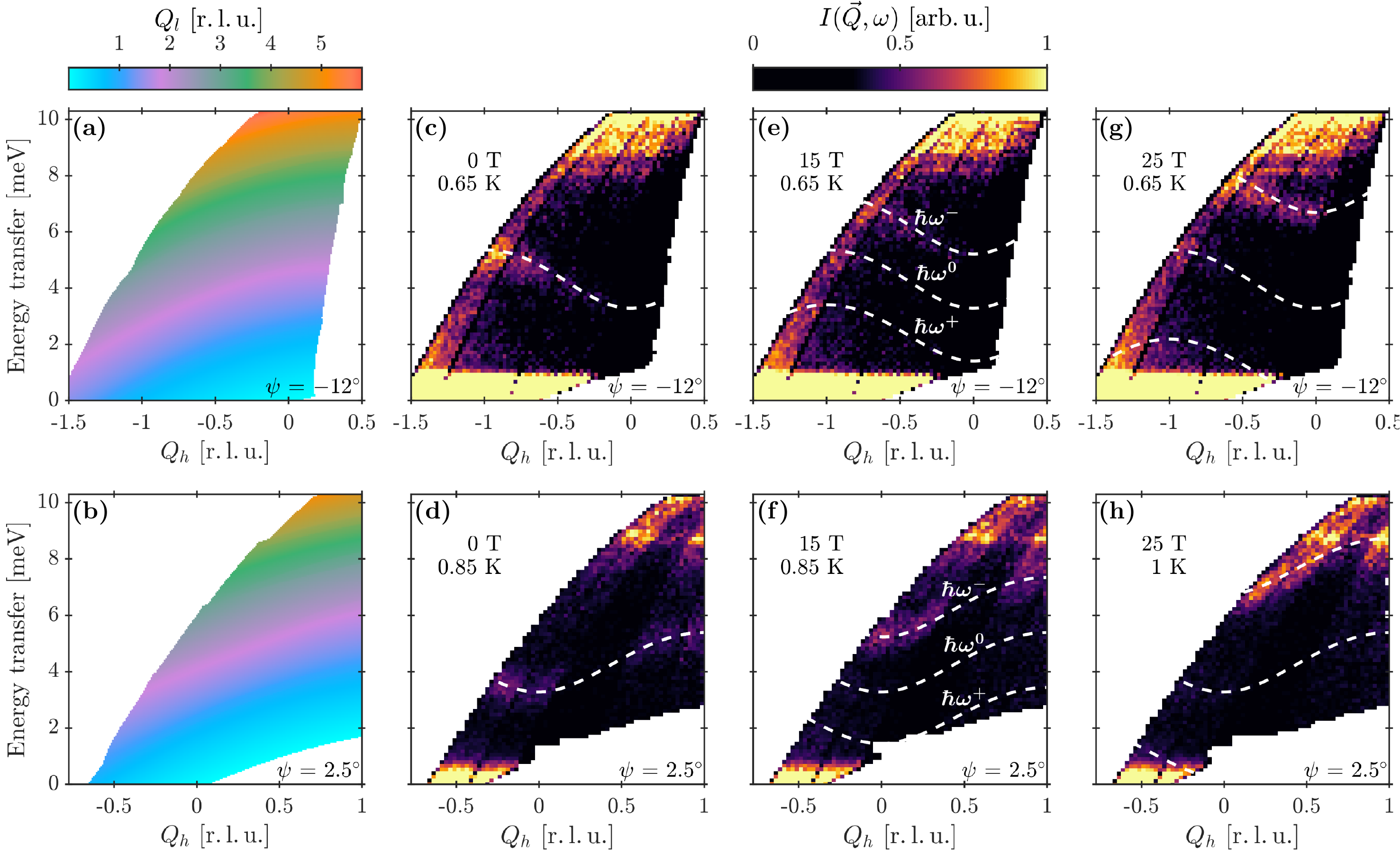}
\caption[]{Dependence of the INS spectrum on magnetic field. (a,b) 
Representation of the dynamic range of scattering processes available 
on HFM/EXED in the space of $Q_h$, $Q_l$, and energy transfer ($\hbar 
\omega$) for sample alignment angles $\psi = - 12^{\circ}$ (a) and 
$2.5^{\circ}$ (b). (c-h) TOF spectra measured on HFM/EXED for both 
values of $\psi$ at three different magnetic fields, reduced to the 
space of $Q_h$ and $\hbar\omega$ by integrating $Q_l$ over the range 
[0,8]; the integration range in $Q_k$ was [$- 0.2$, 0.2] for all 
panels. Dashed white lines represent the average positions of the 
three Zeeman-split triplon modes, whose modelling is described in 
Sec.~\ref{sec:mfd}.}
\label{fig:sfd}
\end{center}
\end{figure*}

For a quantitative discussion, the fit provided by the global peak-intensity 
model, using the parameters optimized by constructing the posterior 
distribution of BI, is represented in Fig.~\ref{fig:pi} by showing the mean 
intensities as the solid lines and the uncertainties as the 68\% confidence 
interval (CI, dark shading) and the 95\% CI (light shading). A small 
discrepancy visible at 0.7~K in Fig.~\ref{fig:pi}(a) may be caused purely 
by statistical fluctuations, because the mean of the posterior is within 
the 68\% CIs of all the peak intensities extracted at 0.7~K except for the 
one at 24.65~T. The temperature-dependence of the extracted peak intensities 
shown in Fig.~\ref{fig:pi}(b) is described rather well below $T_c(H)$ by a 
single critical exponent [Eq.~\eqref{eq:n_perp_model}], and thus within the 
sensitivity and counting statistics of the experiment it is not possible to 
observe any hallmarks of the special quasi-2D form of $m_\perp(T)$ 
illustrated in Fig.~\ref{fig:magnetization_case}(b) \cite{Furuya2016}. 

Turning to $m_\perp(H)$, although its low-temperature evolution at the higher 
fields in Fig.~\ref{fig:pi}(a) is faintly suggestive of the unconventional 
forms illustrated in Fig.~\ref{fig:bacusio_model}(c), it is not possible 
to exclude some more mundane reasons for these observations. Nevertheless, 
we note that the field $H^*$ determined by the B-bilayer interactions in 
BaCuSi$_2$O$_6$ \cite{Allenspach2020} is 25.8 T, and thus it is possible that 
the maximum field available on HFM/EXED falls just short of the value required 
to reveal this type of behavior. While we can conclude that these effects are 
weak for the parameters (two-dimensionality and bilayer stacking) of 
BaCuSi$_2$O$_6$, we cannot exclude that they may be observable to the next 
generation of high-field neutron diffraction experiments, and thus they pose 
an open challenge to future facilities. 

\section{Inelastic Neutron Scattering Measurements}
\label{sec:ins_measurements}

\subsection{Experimental Conditions}

The magnetic excitations of BaCuSi$_2$O$_6$ have been measured by 
INS in the disordered phase at zero magnetic field \cite{Sasago1997,
Allenspach2020} and at fields up to 4~T \cite{Rueegg2007}. Here we 
present INS measurements of the magnetic excitation spectrum up to 
the far higher field of 25~T made possible by using HFM/EXED in its 
spectroscopy mode. These were performed only during E-1, where the base 
temperature of 0.65 K was set by the $^{3}$He cryostat. An incoming 
neutron energy of $E_i = 12$~meV and the associated chopper settings 
were selected to maximize the neutron flux in the energy-transfer 
range of the magnetic excitations. TOF spectra were measured for 
different magnetic fields from 0~T to 25~T using two different 
magnet rotations, $\psi = - 12^{\circ}$ and $2.5^{\circ}$, which 
correspond to the minimal and maximal values of $\psi$ 
[Fig.~\ref{fig:preparation_overview}(d)]. Measurements were performed 
at 0.65~K for 6~hrs at $\psi = - 12^{\circ}$ for five magnetic fields 
(0, 10, 15, 20, and 25~T), but due to time constraints only three TOF 
spectra (at 0, 15, and 25~T) could be measured, each for 3~hrs, with 
$\psi = 2.5^{\circ}$. At this $\psi$ value our measurements were 
performed at 0.85~K (0 and 15~T) and 1~K (25~T), which had 
no effect on the physics of the gapped spin system.

\subsection{Measured INS Spectra}

The measured INS data were preprocessed and transformed into 
energy-momentum space using the software Mantid \cite{Arnold2014}, 
normalized by the monitor and a vanadium standard, and the result was 
scaled by $k_i/k_f$ to provide the normalized intensities $I(\vec{Q},
\omega)$. To improve the statistics, the data were integrated over 
$[-0.2, 0.2]$ in $Q_k$, which corresponds to the direction perpendicular 
to the scattering plane.

In a conventional TOF measurement, the sample may be rotated through a 
wide angular range. Because of the need to fix the value of $\psi$ on 
HFM/EXED, the dynamic range of the $Q_k$-integrated data constitutes 
not a dense 3D subset of energy-momentum space but two 2D subsets that 
cannot be aligned with any high-symmetry directions. Thus accessing 
the full range of $\omega$ values requires sampling over a broad range 
of both $Q_h$ and $Q_l$. The $Q_h$- and $Q_l$-dependence of $\omega$ 
for the dynamic range of our measured TOF spectra is shown for the two 
$\psi$ values in Figs.~\ref{fig:sfd}(a) and \ref{fig:sfd}(b). In 
BaCuSi$_2$O$_6$ it is known \cite{Allenspach2020} that the important 
dispersion information is contained in the $(Q_h,\omega)$ plane, 
with only minimal dependence on $Q_l$, and hence it is clear that 
visualizing the TOF spectrum as a function of $Q_h$ over an 
energy-transfer range covering the Zeeman-split triplon excitations 
up to 25 T requires integration over a very wide range of $Q_l$.

\begin{figure}[p]
\begin{center}
\includegraphics[width=0.96\columnwidth]{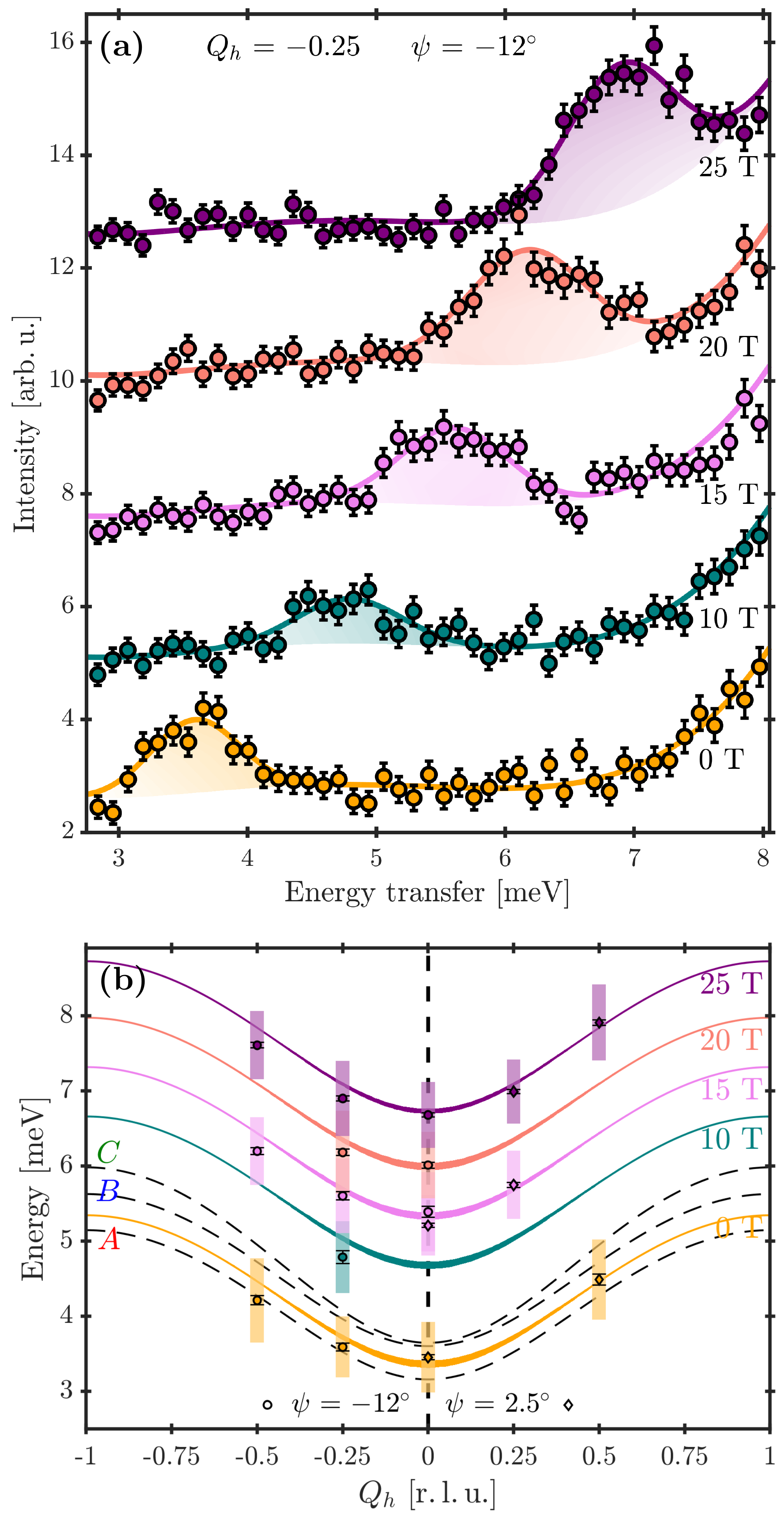}
\caption[]{Field-induced evolution of the triplon spectrum.
(a) Scattered intensity as a function of energy transfer, as extracted 
from the TOF spectra at fixed $Q_h$ and $\psi$, at all five applied 
magnetic fields; respective curves are offset by 2.5 [arb.~u.] for clarity. 
Solid lines and shading show the results of the fitting procedure described 
in the text.
(b) Measured average excitation energy of the upper triplon branch, shown 
as a function of $Q_h$ for the same five fields. Vertical colored bars 
indicate the widths of the Gaussian functions deduced from panel (a), 
expressed as their full width at half-maximum height. Black symbols 
indicate the mode energies, with error bars, extracted from the Gaussian 
center positions. Solid lines show the position of the composite (A+B+C) 
$\hbar\omega^{-}$ triplon mode, modelled as a weighted average intensity 
for each different field. The thicknesses of these lines indicate the effect 
of $Q_l$ on the dispersion. The dashed lines at $H = 0$ show the dispersions 
of the A, B, and C triplons, whose separation explains the widths in energy 
of the observed intensity peaks. All data at $Q_h < 0$ are for $\psi =
 - 12^{\circ}$ and at $Q_h > 0$ for $\psi = 2.5^{\circ}$.}
\label{fig:ins_modes}
\end{center}
\end{figure}

Figures~\ref{fig:sfd}(c)-\ref{fig:sfd}(h) show the spectrum in the plane of 
$Q_h$ and $\omega$ obtained by integrating the TOF data over the range [0,8] 
in $Q_l$. At zero magnetic field [Figs.~\ref{fig:sfd}(c) and \ref{fig:sfd}(d)], 
only one excitation is visible because the inequivalent three triplon modes 
(A, B, C) observed in Refs.~\cite{Rueegg2007,Allenspach2020} cannot be resolved 
individually in the high-flux spectroscopy mode on HFM/EXED. The dashed white 
lines are taken from a model for a single composite mode, A+B+C, as detailed 
below. The spectra measured for $\psi = - 12^{\circ}$ have better statistics 
than those measured at $\psi = 2.5^{\circ}$ due to the longer data-acquisition 
time. An area of zero detector intensity, appearing close to $Q_h = 0$ and 
extending to low but finite energies, is caused by the beam-stop, which blocks 
neutrons scattered in a certain angular range around the direct beam. Above 
8~meV, non-dispersive and field-independent scattering can be observed for 
both $\psi$ values. This signal was not present in previous INS measurements 
performed on BaCuSi$_2$O$_6$ using other TOF spectrometers at zero field 
\cite{Allenspach2020}, but has been observed previously on HFM/EXED, and so 
we ascribe this known background feature to neutrons scattered by the cryostat, 
magnet, or sample holder. The finite scattered intensities close to the lower 
$Q_h$ boundaries of the dynamic range at all energies for the TOF spectra 
measured at $\psi = - 12^{\circ}$ were also not observed in previous experiments 
and are thought to be artifacts of the integration over $Q_l$, which risks 
picking up the tails of a Bragg peak such as ($- 2$ 0 2) or ($- 2$ 0 4).

As an initial model for our spectral data, we will assume that the effect of 
the applied field is only to cause a Zeeman splitting of the three excited 
triplet states, $\ket{t^{-,0,+}}$. However, it has been shown for the 
quasi-1D spin dimer system BPCB that additional bound states can form between 
excited triplets and the condensed triplets that are present above $H_{c1}$ 
\cite{Nayak2020}, and below we comment on the possible applicability of such 
a scenario in BaCuSi$_2$O$_6$. For the triplets $\ket{t^{-,0,+}}$, the ratio of 
the transition probabilities, $p_s^{-,0,+}$, from the singlet ground state is 
given by $p_s^{-} \,$:$\, p_s^{0} \,$:$\, p_s^{+}$ = 1:2:1 \cite{Rueegg2007}. 
Thus the Zeeman-split branch at energy $\hbar\omega^{0}$ is expected to have 
double the intensity of branches at $\hbar\omega^{-,+}$. The intensities of 
each of these modes are also proportional to the dimer structure factor, which 
in BaCuSi$_2$O$_6$ has the form $[1 - \cos (2\pi Q_l d/c)]$, where $c$ is the 
lattice constant in the stacking direction [Fig.~\ref{fig:bacusio_model}(a)] 
and $d$ the separation between the Cu$^{2+}$ ions of the dimers 
\cite{Allenspach2020}. One consequence of the dynamic range of HFM/EXED 
shown in Figs.~\ref{fig:sfd}(a) and \ref{fig:sfd}(b) is that the upward 
shift of the $\hbar\omega^{-}$ branch caused by increasing the field also 
causes its intensity to increase, whereas the $\hbar\omega^{+}$ branch loses 
intensity as its energy decreases. Because the experimental statistics are 
not sufficient to distinguish the $\hbar\omega^{+}$ and $\hbar\omega^{0}$ 
branches from the $\omega$-dependent background at finite fields, in the 
following we focus on the analysis of the $\hbar\omega^{-}$ branch.

\subsection{Field-Dependence of the Triplon Energy}
\label{sec:mfd}

Cuts from the TOF spectra of Figs.~\ref{fig:sfd}(c)-\ref{fig:sfd}(h) were 
extracted to deduce the scattered intensity as a function of $\omega$ by 
using an integration range of $[Q_h - 0.1, Q_h + 0.1]$ about each fixed $Q_h$ 
value. Figure~\ref{fig:ins_modes}(a) shows equivalent cuts for all five 
different magnetic-field values at fixed $Q_h$ and $\psi$. These cuts were 
fitted simultaneously over all $Q_h$ and both $\psi$ values by using an 
individual Gaussian for the excitation branch visible in each cut and a 
polynomial background that was shared for all the cuts at different magnetic 
fields; the resulting fits are represented by the solid lines and shading in 
Fig.~\ref{fig:ins_modes}(a).

The extracted mode energies, corresponding to the center positions 
of the fitted Gaussians, are displayed as black symbols in 
Fig.~\ref{fig:ins_modes}(b). The colored bar indicates the width 
of each Gaussian, which as expected from Fig.~\ref{fig:ins_modes}(a) is 
quite large, and the black symbols within each bar indicate the center 
position and the actual statistical error in its location. To interpret 
these fits, we have modelled the composite mode we observe, A+B+C, 
by taking an intensity-weighted sum of the individual triplon modes, with 
a Zeeman splitting appropriate to the applied field, using the interaction 
parameters determined at $H = 0$ \cite{Allenspach2020}. In this modelling 
procedure, the dependence of each individual mode ($\gamma \in {\rm A,B,C}$) 
on the field is described by the Zeeman term, $\pm g \mu_B \mu_0 H$, for 
fields $H < H_{c1}$. Above $H_{c1}$, the new ground state is predominantly 
\cite{Matsumoto2004} a superposition of $\ket{s_\gamma}$ and $\ket{t^{+}_\gamma}$ 
states on dimers in bilayers of type $\gamma$, so that modes $\hbar \omega^{0}
_\gamma$ increase linearly in energy by $g \mu_B \mu_0 (H - H_{c1})$ and modes 
$\hbar\omega^{-}_\gamma$ by $2 g \mu_B \mu_0 (H - H_{c1})$. Because all our 
inelastic measurements were performed directly after the diffraction 
experiments E-1, without any changes to the sample alignments, the $g$-factor 
determined in the diffraction analysis (Sec.~\ref{sec:neutron_diffraction}), 
$g_{\rm E-1} = 2.271^{+0.003}_{-0.003}$, was used to model the field-dependence of 
the triplon modes. $H_{c1} = \Delta/(g \mu_0 \mu_{B})$ was obtained directly 
from the measured zero-field spin gap, $\Delta = 3.15$ meV 
\cite{Allenspach2020}. 

In Fig.~\ref{fig:ins_modes}(b) we apply this modelling process to the 
zero-field case in order to illustrate the extent of the differences 
between the modes A, B, and C. It is clear that the splitting of the 
three inequivalent triplon modes \cite{Allenspach2020} is almost entirely 
responsible for the width of the intensity distributions appearing in 
Fig.~\ref{fig:ins_modes}(a). We comment that the effect of the wide 
integration over $Q_l$ in the quasi-2D BaCuSi$_2$O$_6$ system is extremely 
small by comparison, causing a broadening of the modelled dispersions that 
is at most 0.1 meV at $Q_h = 0$. At finite fields, we observe that there 
are no significant changes either in the shape of the dispersive $\hbar 
\omega^{-}$ branch or in the widths of the fitting Gaussians and in the 
statistical uncertainties. We note that the dashed lines in 
Figs.~\ref{fig:sfd}(c)-\ref{fig:sfd}(h) show the modelled positions of the 
weighted average intensity peaks for the three branches $\hbar\omega^{-,0,+}$. 

Thus the summary from our INS experiments is that the observed intensities 
are fully consistent with theoretical modelling based on Zeeman-split triplet 
modes over the full field range of the measurements up to 25 T, with no 
significant deterioration of data quality or fitting quality. Returning to 
the question of whether additional modes might appear in the spectrum around 
$H_{c1}$ as a result of bound-state formation, we comment that this type of 
physics is strongly favored by inter-dimer frustration \cite{Nayak2020}. 
Recalling from the inset of Fig.~\ref{fig:bacusio_model}(a) that the full 
magnetic Hamiltonian does contain moderately frustrated pairwise ionic 
interactions, it is certainly possible that bound states could be found in 
the spectrum of BaCuSi$_2$O$_6$ at $H > H_{c1}$. Such states are expected to 
appear as a splitting of the $\ket{t^{0}}$ and $\ket{t^{-}}$ modes at their 
upper band edge ($Q_h = \pm 1$ in Figs.~\ref{fig:sfd} and \ref{fig:ins_modes}), 
where unfortunately the absence of spectral weight at energy $\hbar \omega^0$ 
and the presence of background scattering at $\hbar \omega^-$ preclude any 
reliable identification. The possibility of observing field-induced 
bound-state formation in BaCuSi$_2$O$_6$ therefore remains as a further 
challenge to future high-field spectrometers. 

\section{Discussion and Conclusion}
\label{sec:dc}

We have performed neutron diffraction and spectroscopy experiments on the 
quasi-2D spin-dimer material BaCuSi$_2$O$_6$ at magnetic fields up to 25.9~T 
using the TOF neutron scattering instrument HFM/EXED at the Helmholtz-Zentrum 
Berlin. With these applied fields we were able to access the phase of 
field-induced magnetic long-range order, in which we measured the nuclear 
and magnetic intensities of the ($-2$ 0 2) Bragg peak and the dispersion  
of the uppermost Zeeman branch of the triplon excitation spectrum. The 
diffraction data are well described by a global peak intensity model that 
is fully consistent with a conventional shape of the magnetic order parameter. 
The inelastic data, not previously obtainable by any technique, are also
consistent with Zeeman-split triplon and magnon spectra modelled in both 
the quantum disordered and field-induced ordered phases on the basis of the 
interaction parameters extracted from zero-field INS \cite{Allenspach2020}. 

Technically, the fact that INS measurements on BaCuSi$_2$O$_6$ had previously 
been performed up to only 4~T \cite{Rueegg2007} means that HFM/EXED 
allowed an enormous breakthrough in neutron scattering capabilities. This 
improvement is both quantitative, in vastly increasing the splitting of the 
Zeeman branches of the spectrum, and qualitative, in accessing the order 
parameter of the field-induced magnetic phase. These results showcase the 
capabilities of HFM/EXED, particularly when combined with modern statistical 
methods for the analysis of limited experimental datasets. On this note, our 
results also exemplify the challenges intrinsic to performing measurements at 
such high magnetic fields, and offers some routes to overcoming these. As one 
example, the restricted scattering geometry set by the magnet design meant 
that rather long data-acquisition times were required in our experiments to 
obtain sufficient statistics both for the Bragg-peak intensities at each 
($\mu_0 H, T$) point and for the magnetic excitations at a specific magnet 
rotation. Nevertheless, these trials present invaluable input for assessing 
the factors generating maximum impact on the capabilities of next-generation
high-field magnets at neutron sources. 

Scientifically, BaCuSi$_2$O$_6$ remains a valuable target material for its 
potential to exhibit exotic behavior of both ground and excited states 
under high fields. As a system of stacked, inequivalent bilayer units whose 
non-uniform nature offers a strong enhancement of quasi-two-dimensionality, 
it presents the possibility to observe unconventional behavior of the order 
parameter as a function of either temperature or field. However, even with the 
1:12 inter- to intra-bilayer interaction ratio determined in BaCuSi$_2$O$_6$, 
and with an enhancement factor of $J^{\prime\prime}/[J_B - J_A] \approx 1/10$ 
induced by the non-uniform stacking, the 3D coupling remains sufficiently 
strong that we could not detect an unconventional modification of the 
order parameter, of the types shown in Figs.~\ref{fig:magnetization_case}(b) 
\cite{Furuya2016} or \ref{fig:bacusio_model}(c). Although this spatial 
anisotropy may seem large, it provides a reference point for future 
investigations seeking to confirm the proposal of Ref.~\cite{Furuya2016} 
for quasi-2D magnetic systems. Our investigation broadens the scope of this 
search by underlining the importance of achieving non-uniform layered 
structures with larger energetic mismatches between the interactions in 
the different layers.

\section*{Acknowledgments}

We are grateful to M. Horvati{\'{c}}, P. Naumov, S. Nikitin, and R. Stern for 
helpful discussions and to I. Fisher and S. Sebastian for sample growth. We 
thank Ch.~K\"{a}gi, D.~Sheptyakov, and J.~Stahn for their support in the 
preparation and alignment of the sample, which was performed at the Swiss 
Spallation Neutron Source, SINQ, at the Paul Scherrer Institute. This work 
is based on neutron scattering experiments performed at the Helmholtz-Zentrum 
Berlin. We thank the Swiss National Science Foundation and the ERC grant Hyper 
Quantum Criticality (HyperQC) for financial support, and also acknowledge 
support from the Swiss Data Science Centre (SDSC) through project BISTOM 
C17-12. Work in Toulouse was supported by the French National Research Agency 
(ANR) under projects THERMOLOC ANR-16-CE30-0023-02 and GLADYS ANR-19-CE30-0013, 
and by the use of HPC resources from CALMIP (Grant No.~2020-P0677) and GENCI 
(Grant No.~x2020050225). 

\begin{appendix}

\section{Definition of the Posterior Distribution}
\label{sec:adpd}

The posterior distribution, 
\begin{equation}
p(\bm{\theta}|\mathcal{D}) \propto p(\mathcal{D}|\bm{\theta}) p(\bm{\theta})
\end{equation}
is constructed from the prior distributions, $p(\bm{\theta})$, of all the 
model parameters and the likelihood function, $p(\mathcal{D}|\bm{\theta})$. 
We start our definition of the prior distributions with the 
experiment-dependent rotations, $\zeta_s$, of the sample within the scattering 
plane, which were estimated from the positions of the Bragg peaks on the 
detector. These were gauged in turn using calibration measurements previously 
performed on HFM/EXED, yielding the estimates 22.7$^{\circ}$ for E-1 and 
24.1$^{\circ}$ for E-2 and E-3, which were used as mean values for the prior 
probability distributions 
\begin{align}
p(\zeta_{1}) &= \mathcal{N}(\zeta_{E-1}|\mu=22.7^{\circ},\sigma=1^{\circ})\\
p(\zeta_{2}) &= \mathcal{N}(\zeta_{E-2}|\mu=24.1^{\circ},\sigma=1^{\circ})\\
p(\zeta_{3}) &= \mathcal{N}(\zeta_{E-3}|\mu=24.1^{\circ},\sigma=1^{\circ}).
\end{align}
Here $\mathcal{N}(x|\mu,\sigma)$ denotes the normal distribution of $x$ with 
mean $\mu$ and standard deviation $\sigma$. We chose a standard deviation of 
1$^{\circ}$ to account for uncertainties arising from the out-of-plane 
misalignment, the field-dependent bend of the cryostat sample holder, and a 
possible offset in the magnetic-field direction of the magnet.

The constant intensity offsets, $\mathcal{C}_{s}$, do not include any magnetic 
contributions and are therefore independent of the magnetic field (in the 
absence of magnetostriction), as well as approximately temperature-independent 
for the temperature ranges measured on HFM/EXED. Thus we used the intensity 
extracted from the ($- 2$ 0 2) peak at 15~T and 1~K measured in experiment E-2 
as the mean of a normal prior distribution for $\mathcal{C}_{s}$,
\begin{equation}
p(\mathcal{C}_{s}) = \mathcal{N} (\mathcal{C}_{s} | \mu = 38.2, \sigma = 10)
\enspace \forall \,s.
\end{equation}
The standard deviation of 10 was taken to allow for the fact that this mean 
intensity value was extracted from only one experimental ($\mu_0 H$, $T$) 
point and because it includes a minor contribution due to the longitudinal 
magnetization, which is weak but not zero at 15~T and 1~K.
For the other model parameters, we used a very broad normal distribution 
centered at 0 in order to effect an uninformative (i.e.~unbiased) prior 
distribution,
\begin{equation}
p(\theta_i) = \mathcal{N} (\theta_i|0,\sigma = 1000),
\label{eq:uninformative_prior}
\end{equation}
where $\theta_i \in \{\kappa_H,\kappa_T,\alpha,\mathcal{B},\mathcal{A}_s\}$ 
and the units of $\sigma$ are the same as those of the parameters.

Regarding the uncertainties in the observables, as noted in 
Sec.~\ref{sec:experimental_conditions} the accurate calibration of the 
magnetic-field values allowed us to neglect their uncertainties. The sample 
temperatures were estimated from the temperature-sensor measurements by 
taking the mean, $T^{\rm s}_j$, and standard deviation, $\delta T^{\rm s}_j$, 
of the time series obtained at each ($\mu_0 H, T$) point, $j$. The actual 
sample temperatures, $T_j$, were then treated as unknown parameters (latent 
variables) using the prior distributions
\begin{equation}
p(T_j) = \mathcal{N} (T_j | \mu = T^{\rm s}_j, \sigma = \delta T^{\rm s}_j)
\enspace \forall \, j.
\end{equation}
An alternative approach is to include the uncertainty in the sample 
temperatures as an extra factor in the likelihood function 
\cite{Allenspach2021}, which is mathematically equivalent to the 
procedure presented here. Assuming that the prior distributions of the 
individual peak-intensity model parameters, $\bm{\theta} = (\kappa_H, 
\kappa_T, \alpha, \mathcal{B}, \mathcal{A}_s, \mathcal{C}_s, \zeta_s, 
\vec{T})$, are independent, the joint prior distribution is given by
\begin{equation}
p(\bm{\theta}) = \prod_i p(\theta_i).
\end{equation}

The definition of the likelihood function is much more succinct. Using the 
peak intensities extracted from the data, $I^{\rm data}_j$, their error 
bars, $\delta I^{\rm data}_j$, and the peak-intensity model specified in 
Eq.~\eqref{eq:intensity_model}], 
\begin{equation}
p(\mathcal{D}|\bm{\theta}) = \prod_j \mathcal{N} (I^{\rm data}_j | \mu
 = I^{\rm model}_j (H_j,T_j), \sigma = \delta I^{\rm data}_j)
\end{equation}
where $j$ labels the $(\mu_0 H, T)$ points shown in 
Fig.~\ref{fig:diffraction}(a), each of which was obtained from an 
experiment $s \in \{$E-1, E-2, E-3$\}$. 

The posterior distribution was constructed by sampling from $p(\mathcal{D} 
|\bm{\theta}) p(\bm{\theta})$. Because the peak-intensity model is 
continuously differentiable, Hamiltonian Monte Carlo (HMC) methods  
\cite{Gelman2004,Bishop2006} can be applied and the specific method we 
used to obtain the posterior distribution was a Markov Chain Monte Carlo 
(MCMC) sampling scheme using a No U-Turn Sampler (NUTS) \cite{Hoffman2014} 
implemented in the probabilistic programming Python package PyMC3 
\cite{Salvatier2016}. All the parameters of the peak-intensity model are 
positive quantities, but the performance of HMC methods is improved when 
sampling in unbounded parameter spaces, so instead of encoding this 
information by truncating the prior distributions at zero (for example 
with half-normal or truncated normal distributions), we used normal 
distributions and the absolute values of all the parameters. To assess the 
quality of the sampling, three independent chains were sampled and compared. 
After 5\,000 tuning steps, we recorded 20\,000 further steps and discarded 
the first 5\,000 of these as "burn-in" steps for each chain, thereby 
defining $p(\bm{\theta}|\mathcal{D})$ from a total of 45\,000 samples. 

\begin{figure*}[t]
\begin{center}
\includegraphics[width=1.99\columnwidth]{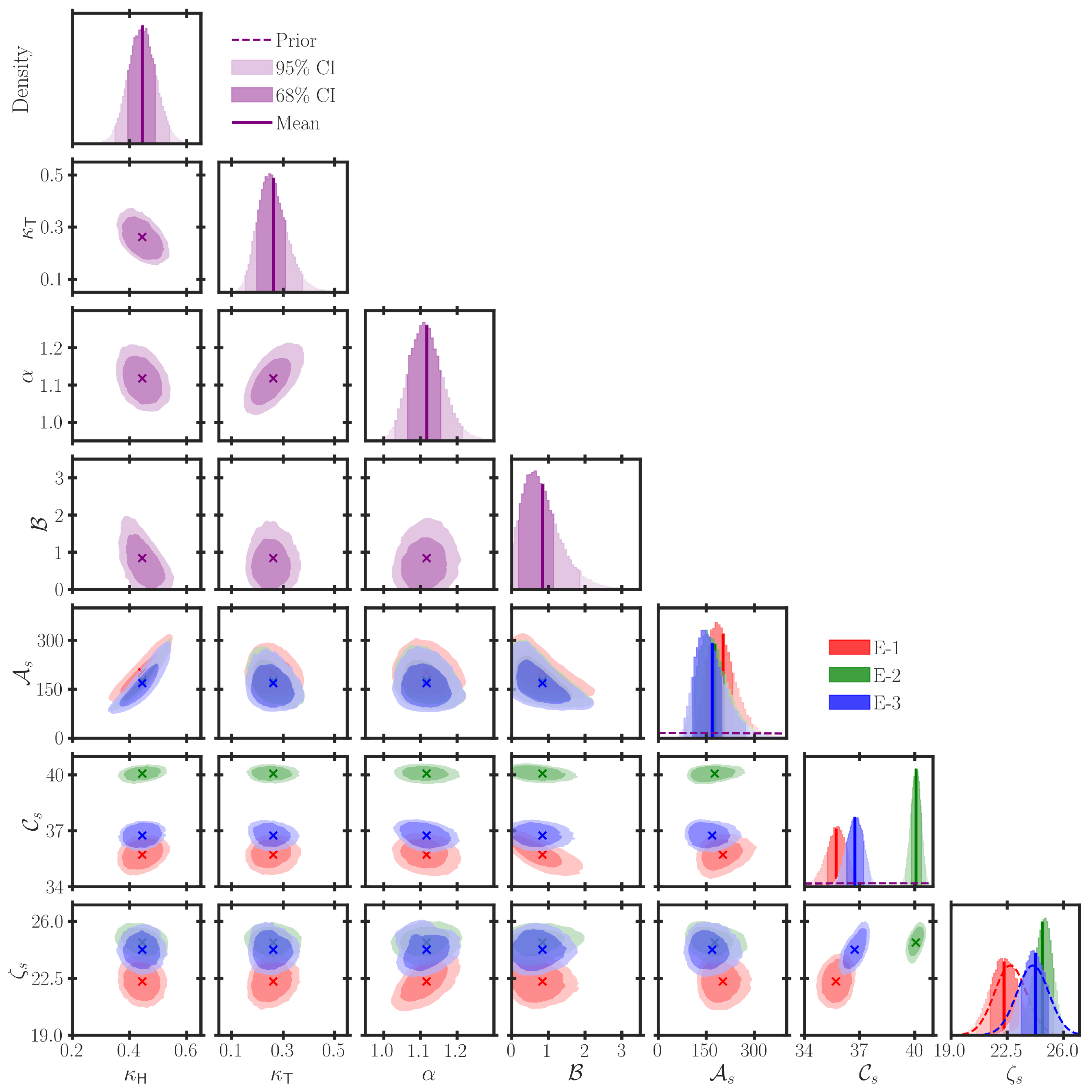}
\caption[]{Projections of the joint posterior distribution of the 
peak-intensity model parameters [Eq.~\eqref{eq:intensity_model}]. 
Solid lines in the diagonal panels and crosses in the off-diagonal 
panels indicate the mean and shaded areas the 68\% (dark) and 95\% 
(light shading) CIs. The quantities dependent on each experiment, 
labelled using $s \in \{$E-1, E-2, E-3$\}$, are displayed in the 
three primary colors. Prior distributions of the model parameters 
(Appendix \ref{sec:adpd}) are shown as dashed lines, but are behind 
or below the horizontal axis in all cases other than $\mathcal{A}_s$, 
$\mathcal{C}_s$, and $\zeta_s$. For $\zeta_s$, the prior distribution 
is shown with the color code for each experiment, although the 
distributions for E-2 and E-3 coincide.}
\label{fig:posterior_distribution}
\end{center}
\end{figure*}

\begin{table}[t]
\begin{center}
\begin{tabular}{ c | c c c c}
&  Global & E-1 & E-2 & E-3 \\
\hline
$\kappa_H$      & 0.45$^{+0.05}_{-0.05}$  & & & \\
$\kappa_T$      & 0.26$^{+0.05}_{-0.06}$  & & & \\
$\alpha$        & 1.12$^{+0.04}_{-0.05}$  & & & \\
$\mathcal{B}$   & 0.8$^{+0.3}_{-0.7}$     & & & \\
$\mathcal{A}_s$ [$\times 10^2$]  & & 2.0$^{+0.3}_{-0.5}$ 
& 1.8$^{+0.3}_{-0.6}$    & 1.7$^{+0.3}_{-0.6}$    \\
$\mathcal{C}_s$ & & 35.7$^{+0.6}_{-0.5}$ & 40.1$^{+0.3}_{-0.3}$ 
& 36.7$^{+0.5}_{-0.4}$ \\
$\zeta_s$ [$^{\circ}$]           & & 22.3$^{+0.9}_{-0.9}$ 
& 24.7$^{+0.7}_{-0.5}$ & 24.3$^{+0.8}_{-0.9}$ \\
\end{tabular}
\caption{Parameters of the peak-intensity model 
[Eq.~\eqref{eq:intensity_model}] inferred using Bayesian inference.
Estimates are taken from the mean of the posterior distribution and 
their uncertainties from the differences between the mean and the 
68\% CI boundaries.}
\label{tab:parameter_estimates}
\end{center}
\end{table}

\section{Investigation of the Posterior Distribution}
\label{sec:aipd}

To characterize the multi-variable joint posterior distribution, in the 
diagonal panels of Fig.~\ref{fig:posterior_distribution} we show its 
projections onto the space of a single model parameter, known as the 
marginal distributions, and in the off-diagonal panels we show projections 
onto pairs of parameters. The parameter estimates and uncertainties are 
determined from the mean and 68\% CI boundaries of the marginal distributions 
and are listed in Table \ref{tab:parameter_estimates}. Because the marginal 
distributions are unimodal, the Highest Posterior Density intervals 
\cite{Gelman2004} are used as the CIs. 

For all three of the physical model parameters, $\kappa_H$, $\kappa_T$, 
and $\alpha$, the marginal distributions show a well-defined single 
peak, with one characteristic width, and are only weakly skewed. For 
$\mathcal{B}$, the marginal distribution extends to very small values, 
with the lower 95\% CI boundary at only 0.00025, but also displays 
a heavy tail towards larger values. To distinguish the experimental 
parameters unique to each run (E-1, E-2, or E-3), the marginal 
distributions of $\mathcal{A}_s$, $\mathcal{C}_s$, and $\zeta_s$ are 
shown using different colors. For $\mathcal{A}_s$, the entire marginal 
distributions of E-2 and E-3 overlap closely, but deviate slightly from 
E-1; these minor differences are most likely a result of different 
levels of neutron absorption in the two different cryostats. The 
marginal distributions of $\mathcal{C}_s$ are similar for E-1 and E-3 
but differ for E-2, a 9\% discrepancy (Table \ref{tab:parameter_estimates}) 
already visible in the extracted peak intensity (Fig.~\ref{fig:pi}).
The coefficients $\mathcal{C}_s$ include not only the nuclear scattering 
contribution but also any other contributions that are constant in 
temperature and magnetic field, including any remnants of the background 
that might be added to the peak intensity in error when extracting it 
from fits to the ($- 2$ $Q_k$ 2) cuts of Fig.~\ref{fig:diffraction}. 
Finally, the marginal distributions of $\zeta_s$ are similar for E-2 and 
E-3 but differ from E-1, as discussed in Sec.~\ref{sec:rbia}. As one 
measure of the effectiveness of the BI procedure, the prior uncertainties 
of the model parameters are clearly reduced in every case other than the 
$\zeta_s$, because their prior distributions were chosen based on strict 
constraints imposed by the refinement of the peak position on the detector. 
Although these prior distributions are as a result informative, the
(marginal) posterior distributions of $\zeta_s$ become narrower and their 
mean positions are nevertheless shifted due to the additional information 
provided by the extracted peak intensities.

Finally, the pair projections shown in the off-diagonal panels of 
Fig.~\ref{fig:posterior_distribution} reveal any correlations between 
the parameters of the model. $\kappa_H$ and $\kappa_T$ show a weak 
negative correlation, while $\kappa_H$ is negatively correlated with 
$\alpha$ and $\kappa_T$ has a stronger positive correlation. 
$\mathcal{B}$ has a weak negative correlation with $\kappa_H$ but is 
almost independent of $\kappa_T$ and $\alpha$. The correlations of any 
of these parameters with an $s$-dependent parameter have the same form 
for all $s$. Other than a strong positive correlation of $\mathcal{A}_s$
with $\kappa_H$ and a negative correlation with $\mathcal{B}$, the 
remaining pair projections reveal no remarkable trends, from which one 
may surmize that the peak-intensity model both provides a meaningful 
account of the underlying physics and is robust against possible artifacts 
arising from the treatment of the experiments and data. 

\end{appendix}


\begin{thebibliography}{81}%
\makeatletter
\providecommand \@ifxundefined [1]{%
 \@ifx{#1\undefined}
}%
\providecommand \@ifnum [1]{%
 \ifnum #1\expandafter \@firstoftwo
 \else \expandafter \@secondoftwo
 \fi
}%
\providecommand \@ifx [1]{%
 \ifx #1\expandafter \@firstoftwo
 \else \expandafter \@secondoftwo
 \fi
}%
\providecommand \natexlab [1]{#1}%
\providecommand \enquote  [1]{``#1''}%
\providecommand \bibnamefont  [1]{#1}%
\providecommand \bibfnamefont [1]{#1}%
\providecommand \citenamefont [1]{#1}%
\providecommand \href@noop [0]{\@secondoftwo}%
\providecommand \href [0]{\begingroup \@sanitize@url \@href}%
\providecommand \@href[1]{\@@startlink{#1}\@@href}%
\providecommand \@@href[1]{\endgroup#1\@@endlink}%
\providecommand \@sanitize@url [0]{\catcode `\\12\catcode `\$12\catcode
  `\&12\catcode `\#12\catcode `\^12\catcode `\_12\catcode `\%12\relax}%
\providecommand \@@startlink[1]{}%
\providecommand \@@endlink[0]{}%
\providecommand \url  [0]{\begingroup\@sanitize@url \@url }%
\providecommand \@url [1]{\endgroup\@href {#1}{\urlprefix }}%
\providecommand \urlprefix  [0]{URL }%
\providecommand \Eprint [0]{\href }%
\providecommand \doibase [0]{http://dx.doi.org/}%
\providecommand \selectlanguage [0]{\@gobble}%
\providecommand \bibinfo  [0]{\@secondoftwo}%
\providecommand \bibfield  [0]{\@secondoftwo}%
\providecommand \translation [1]{[#1]}%
\providecommand \BibitemOpen [0]{}%
\providecommand \bibitemStop [0]{}%
\providecommand \bibitemNoStop [0]{.\EOS\space}%
\providecommand \EOS [0]{\spacefactor3000\relax}%
\providecommand \BibitemShut  [1]{\csname bibitem#1\endcsname}%
\let\auto@bib@innerbib\@empty
\bibitem [{\citenamefont {Zinn-Justin}(2002)}]{Zinn-Justin2002}%
  \BibitemOpen
  \bibfield  {author} {\bibinfo {author} {\bibfnamefont {J.}~\bibnamefont
  {Zinn-Justin}},\ }\href@noop {} {\emph {\bibinfo {title} {{Quantum Field
  Theory and Critical Phenomena}}}}\ (\bibinfo  {publisher} {Oxford University
  Press, Oxford},\ \bibinfo {year} {2002})\BibitemShut {NoStop}%
\bibitem [{\citenamefont {Sachdev}(2011)}]{Sachdev2011}%
  \BibitemOpen
  \bibfield  {author} {\bibinfo {author} {\bibfnamefont {S.}~\bibnamefont
  {Sachdev}},\ }\href@noop {} {\emph {\bibinfo {title} {Quantum Phase
  Transitions, Second Edition}}}\ (\bibinfo  {publisher} {Cambridge University
  Press, Cambridge},\ \bibinfo {year} {2011})\BibitemShut {NoStop}%
\bibitem [{\citenamefont {Matsumoto}\ \emph {et~al.}(2004)\citenamefont
  {Matsumoto}, \citenamefont {Normand}, \citenamefont {Rice},\ and\
  \citenamefont {Sigrist}}]{Matsumoto2004}%
  \BibitemOpen
  \bibfield  {author} {\bibinfo {author} {\bibfnamefont {M.}~\bibnamefont
  {Matsumoto}}, \bibinfo {author} {\bibfnamefont {B.}~\bibnamefont {Normand}},
  \bibinfo {author} {\bibfnamefont {T.~M.}\ \bibnamefont {Rice}}, \ and\
  \bibinfo {author} {\bibfnamefont {M.}~\bibnamefont {Sigrist}},\ }\bibinfo
  {title} {{Field- and pressure-induced magnetic quantum phase transitions in
  {TlCuCl$_3$}}},\ \href {\doibase 10.1103/PhysRevB.69.054423} {\bibfield
  {journal} {\bibinfo  {journal} {Phys. Rev. B}\ }\textbf {\bibinfo {volume}
  {69}},\ \bibinfo {pages} {054423} (\bibinfo {year} {2004})}\BibitemShut
  {NoStop}%
\bibitem [{\citenamefont {Giamarchi}\ \emph {et~al.}(2008)\citenamefont
  {Giamarchi}, \citenamefont {R\"{u}egg},\ and\ \citenamefont
  {Tchernyshyov}}]{Giamarchi2008}%
  \BibitemOpen
  \bibfield  {author} {\bibinfo {author} {\bibfnamefont {T.}~\bibnamefont
  {Giamarchi}}, \bibinfo {author} {\bibfnamefont {C.}~\bibnamefont
  {R\"{u}egg}}, \ and\ \bibinfo {author} {\bibfnamefont {O.}~\bibnamefont
  {Tchernyshyov}},\ }\bibinfo {title} {{Bose{\textendash}Einstein condensation
  in magnetic~insulators}},\ \href {\doibase 10.1038/nphys893} {\bibfield
  {journal} {\bibinfo  {journal} {Nat. Phys.}\ }\textbf {\bibinfo {volume}
  {4}},\ \bibinfo {pages} {198} (\bibinfo {year} {2008})}\BibitemShut {NoStop}%
\bibitem [{\citenamefont {Zapf}\ \emph {et~al.}(2014)\citenamefont {Zapf},
  \citenamefont {Jaime},\ and\ \citenamefont {Batista}}]{Zapf2014}%
  \BibitemOpen
  \bibfield  {author} {\bibinfo {author} {\bibfnamefont {V.}~\bibnamefont
  {Zapf}}, \bibinfo {author} {\bibfnamefont {M.}~\bibnamefont {Jaime}}, \ and\
  \bibinfo {author} {\bibfnamefont {C.~D.}\ \bibnamefont {Batista}},\ }\bibinfo
  {title} {{Bose-Einstein condensation in quantum magnets}},\ \href {\doibase
  10.1103/RevModPhys.86.563} {\bibfield  {journal} {\bibinfo  {journal} {Rev.
  Mod. Phys.}\ }\textbf {\bibinfo {volume} {86}},\ \bibinfo {pages} {563}
  (\bibinfo {year} {2014})}\BibitemShut {NoStop}%
\bibitem [{\citenamefont {Bose}(1924)}]{Bose1924}%
  \BibitemOpen
  \bibfield  {author} {\bibinfo {author} {\bibfnamefont {S.~N.}\ \bibnamefont
  {Bose}},\ }\bibinfo {title} {{Plancks Gesetz und Lichtquantenhypothese}},\
  \href {\doibase 10.1007/BF01327326} {\bibfield  {journal} {\bibinfo
  {journal} {Z. Phys.}\ }\textbf {\bibinfo {volume} {26}},\ \bibinfo {pages}
  {178} (\bibinfo {year} {1924})}\BibitemShut {NoStop}%
\bibitem [{\citenamefont {Einstein}(1924)}]{Einstein1924}%
  \BibitemOpen
  \bibfield  {author} {\bibinfo {author} {\bibfnamefont {A.}~\bibnamefont
  {Einstein}},\ }\bibinfo {title} {{Quantentheorie des einatomigen idealen
  Gases}},\ \href {\doibase 10.1002/3527608958.ch27} {\bibfield  {journal}
  {\bibinfo  {journal} {Sitz. Ber. Kgl. Preuss. Akad. Wiss.}\ }\textbf
  {\bibinfo {volume} {3}},\ \bibinfo {pages} {261} (\bibinfo {year}
  {1924})}\BibitemShut {NoStop}%
\bibitem [{\citenamefont {Nikuni}\ \emph {et~al.}(2000)\citenamefont {Nikuni},
  \citenamefont {Oshikawa}, \citenamefont {Oosawa},\ and\ \citenamefont
  {Tanaka}}]{Nikuni2000}%
  \BibitemOpen
  \bibfield  {author} {\bibinfo {author} {\bibfnamefont {T.}~\bibnamefont
  {Nikuni}}, \bibinfo {author} {\bibfnamefont {M.}~\bibnamefont {Oshikawa}},
  \bibinfo {author} {\bibfnamefont {A.}~\bibnamefont {Oosawa}}, \ and\ \bibinfo
  {author} {\bibfnamefont {H.}~\bibnamefont {Tanaka}},\ }\bibinfo {title}
  {{Bose-Einstein Condensation of Dilute Magnons in TlCuCl$_3$}},\ \href
  {\doibase 10.1103/PhysRevLett.84.5868} {\bibfield  {journal} {\bibinfo
  {journal} {Phys. Rev. Lett.}\ }\textbf {\bibinfo {volume} {84}},\ \bibinfo
  {pages} {5868} (\bibinfo {year} {2000})}\BibitemShut {NoStop}%
\bibitem [{\citenamefont {Tanaka}\ \emph {et~al.}(2001)\citenamefont {Tanaka},
  \citenamefont {Oosawa}, \citenamefont {Kato}, \citenamefont {Uekusa},
  \citenamefont {Ohashi}, \citenamefont {Kakurai},\ and\ \citenamefont
  {Hoser}}]{Tanaka2001}%
  \BibitemOpen
  \bibfield  {author} {\bibinfo {author} {\bibfnamefont {H.}~\bibnamefont
  {Tanaka}}, \bibinfo {author} {\bibfnamefont {A.}~\bibnamefont {Oosawa}},
  \bibinfo {author} {\bibfnamefont {T.}~\bibnamefont {Kato}}, \bibinfo {author}
  {\bibfnamefont {H.}~\bibnamefont {Uekusa}}, \bibinfo {author} {\bibfnamefont
  {Y.}~\bibnamefont {Ohashi}}, \bibinfo {author} {\bibfnamefont
  {K.}~\bibnamefont {Kakurai}}, \ and\ \bibinfo {author} {\bibfnamefont
  {A.}~\bibnamefont {Hoser}},\ }\bibinfo {title} {{Observation of Field-Induced
  Transverse N\'eel Ordering in the Spin Gap System TlCuCl$_3$}},\ \href
  {\doibase 10.1143/JPSJ.70.939} {\bibfield  {journal} {\bibinfo  {journal} {J.
  Phys. Soc. Jpn.}\ }\textbf {\bibinfo {volume} {70}},\ \bibinfo {pages} {939}
  (\bibinfo {year} {2001})}\BibitemShut {NoStop}%
\bibitem [{\citenamefont {Oosawa}\ \emph {et~al.}(2001)\citenamefont {Oosawa},
  \citenamefont {Aruga~Katori},\ and\ \citenamefont {Tanaka}}]{Oosawa2001}%
  \BibitemOpen
  \bibfield  {author} {\bibinfo {author} {\bibfnamefont {A.}~\bibnamefont
  {Oosawa}}, \bibinfo {author} {\bibfnamefont {H.}~\bibnamefont
  {Aruga~Katori}}, \ and\ \bibinfo {author} {\bibfnamefont {H.}~\bibnamefont
  {Tanaka}},\ }\bibinfo {title} {{Specific heat study of the field-induced
  magnetic ordering in the spin-gap system TlCuCl$_3$}},\ \href {\doibase
  10.1103/PhysRevB.63.134416} {\bibfield  {journal} {\bibinfo  {journal} {Phys.
  Rev. B}\ }\textbf {\bibinfo {volume} {63}},\ \bibinfo {pages} {134416}
  (\bibinfo {year} {2001})}\BibitemShut {NoStop}%
\bibitem [{\citenamefont {R{\"u}egg}\ \emph {et~al.}(2003)\citenamefont
  {R{\"u}egg}, \citenamefont {Cavadini}, \citenamefont {Furrer}, \citenamefont
  {G{\"u}del}, \citenamefont {Kr{\"a}mer}, \citenamefont {Mutka}, \citenamefont
  {Wildes}, \citenamefont {Habicht},\ and\ \citenamefont
  {Vorderwisch}}]{Rueegg2003}%
  \BibitemOpen
  \bibfield  {author} {\bibinfo {author} {\bibfnamefont {C.}~\bibnamefont
  {R{\"u}egg}}, \bibinfo {author} {\bibfnamefont {N.}~\bibnamefont {Cavadini}},
  \bibinfo {author} {\bibfnamefont {A.}~\bibnamefont {Furrer}}, \bibinfo
  {author} {\bibfnamefont {H.~U.}\ \bibnamefont {G{\"u}del}}, \bibinfo {author}
  {\bibfnamefont {K.}~\bibnamefont {Kr{\"a}mer}}, \bibinfo {author}
  {\bibfnamefont {H.}~\bibnamefont {Mutka}}, \bibinfo {author} {\bibfnamefont
  {A.}~\bibnamefont {Wildes}}, \bibinfo {author} {\bibfnamefont
  {K.}~\bibnamefont {Habicht}}, \ and\ \bibinfo {author} {\bibfnamefont
  {P.}~\bibnamefont {Vorderwisch}},\ }\bibinfo {title} {{Bose-Einstein
  condensation of the triplet states in the magnetic insulator TlCuCl$_3$}},\
  \href {\doibase 10.1038/nature01617} {\bibfield  {journal} {\bibinfo
  {journal} {Nature}\ }\textbf {\bibinfo {volume} {423}},\ \bibinfo {pages}
  {62} (\bibinfo {year} {2003})}\BibitemShut {NoStop}%
\bibitem [{\citenamefont {Glazkov}\ \emph {et~al.}(2004)\citenamefont
  {Glazkov}, \citenamefont {Smirnov}, \citenamefont {Tanaka},\ and\
  \citenamefont {Oosawa}}]{Glazkov2004}%
  \BibitemOpen
  \bibfield  {author} {\bibinfo {author} {\bibfnamefont {V.~N.}\ \bibnamefont
  {Glazkov}}, \bibinfo {author} {\bibfnamefont {A.~I.}\ \bibnamefont
  {Smirnov}}, \bibinfo {author} {\bibfnamefont {H.}~\bibnamefont {Tanaka}}, \
  and\ \bibinfo {author} {\bibfnamefont {A.}~\bibnamefont {Oosawa}},\ }\bibinfo
  {title} {{Spin-resonance modes of the spin-gap magnet
  ${\mathrm{TlCuCl}}_{3}$}},\ \href {\doibase 10.1103/PhysRevB.69.184410}
  {\bibfield  {journal} {\bibinfo  {journal} {Phys. Rev. B}\ }\textbf {\bibinfo
  {volume} {69}},\ \bibinfo {pages} {184410} (\bibinfo {year}
  {2004})}\BibitemShut {NoStop}%
\bibitem [{\citenamefont {Mermin}\ and\ \citenamefont
  {Wagner}(1966)}]{Mermin1966}%
  \BibitemOpen
  \bibfield  {author} {\bibinfo {author} {\bibfnamefont {N.~D.}\ \bibnamefont
  {Mermin}}\ and\ \bibinfo {author} {\bibfnamefont {H.}~\bibnamefont
  {Wagner}},\ }\bibinfo {title} {{Absence of Ferromagnetism or
  Antiferromagnetism in One- or Two-Dimensional Isotropic Heisenberg Models}},\
  \href {\doibase 10.1103/PhysRevLett.17.1133} {\bibfield  {journal} {\bibinfo
  {journal} {Phys. Rev. Lett.}\ }\textbf {\bibinfo {volume} {17}},\ \bibinfo
  {pages} {1133} (\bibinfo {year} {1966})}\BibitemShut {NoStop}%
\bibitem [{\citenamefont {Sachdev}(1994)}]{Sachdev1994}%
  \BibitemOpen
  \bibfield  {author} {\bibinfo {author} {\bibfnamefont {S.}~\bibnamefont
  {Sachdev}},\ }\bibinfo {title} {{Quantum phase transitions and conserved
  charges}},\ \href {\doibase 10.1007/BF01317409} {\bibfield  {journal}
  {\bibinfo  {journal} {Z. Phys. B}\ }\textbf {\bibinfo {volume} {94}},\
  \bibinfo {pages} {469} (\bibinfo {year} {1994})}\BibitemShut {NoStop}%
\bibitem [{\citenamefont {Giamarchi}\ and\ \citenamefont
  {Tsvelik}(1999)}]{Giamarchi1999}%
  \BibitemOpen
  \bibfield  {author} {\bibinfo {author} {\bibfnamefont {T.}~\bibnamefont
  {Giamarchi}}\ and\ \bibinfo {author} {\bibfnamefont {A.~M.}\ \bibnamefont
  {Tsvelik}},\ }\bibinfo {title} {{Coupled ladders in a magnetic field}},\
  \href {\doibase 10.1103/PhysRevB.59.11398} {\bibfield  {journal} {\bibinfo
  {journal} {Phys. Rev. B}\ }\textbf {\bibinfo {volume} {59}},\ \bibinfo
  {pages} {11398} (\bibinfo {year} {1999})}\BibitemShut {NoStop}%
\bibitem [{\citenamefont {Giamarchi}(2003)}]{Giamarchi2003}%
  \BibitemOpen
  \bibfield  {author} {\bibinfo {author} {\bibfnamefont {T.}~\bibnamefont
  {Giamarchi}},\ }\href@noop {} {\emph {\bibinfo {title} {{Quantum Physics in
  One Dimension}}}},\ International Series of Monographs on Physics\ (\bibinfo
  {publisher} {Clarendon Press, Oxford},\ \bibinfo {year} {2003})\BibitemShut
  {NoStop}%
\bibitem [{\citenamefont {{Berezinski{\v{i}}}}(1971)}]{Berezinskii1971}%
  \BibitemOpen
  \bibfield  {author} {\bibinfo {author} {\bibfnamefont {V.~L.}\ \bibnamefont
  {{Berezinski{\v{i}}}}},\ }\bibinfo {title} {{Destruction of Long-range Order
  in One-dimensional and Two-dimensional Systems having a Continuous Symmetry
  Group I. Classical Systems}},\ \href
  {https://ui.adsabs.harvard.edu/abs/1971JETP...32..493B} {\bibfield  {journal}
  {\bibinfo  {journal} {Sov. J. Exp. Theor. Phys.}\ }\textbf {\bibinfo {volume}
  {32}},\ \bibinfo {pages} {493} (\bibinfo {year} {1971})}\BibitemShut
  {NoStop}%
\bibitem [{\citenamefont {Kosterlitz}\ and\ \citenamefont
  {Thouless}(1973)}]{Kosterlitz1973}%
  \BibitemOpen
  \bibfield  {author} {\bibinfo {author} {\bibfnamefont {J.~M.}\ \bibnamefont
  {Kosterlitz}}\ and\ \bibinfo {author} {\bibfnamefont {D.~J.}\ \bibnamefont
  {Thouless}},\ }\bibinfo {title} {{Ordering, metastability and phase
  transitions in two-dimensional systems}},\ \href {\doibase
  10.1088/0022-3719/6/7/010} {\bibfield  {journal} {\bibinfo  {journal} {J.
  Phys. C: Solid State Phys.}\ }\textbf {\bibinfo {volume} {6}},\ \bibinfo
  {pages} {1181} (\bibinfo {year} {1973})}\BibitemShut {NoStop}%
\bibitem [{\citenamefont {Furuya}\ \emph {et~al.}(2016)\citenamefont {Furuya},
  \citenamefont {Dupont}, \citenamefont {Capponi}, \citenamefont
  {Laflorencie},\ and\ \citenamefont {Giamarchi}}]{Furuya2016}%
  \BibitemOpen
  \bibfield  {author} {\bibinfo {author} {\bibfnamefont {S.~C.}\ \bibnamefont
  {Furuya}}, \bibinfo {author} {\bibfnamefont {M.}~\bibnamefont {Dupont}},
  \bibinfo {author} {\bibfnamefont {S.}~\bibnamefont {Capponi}}, \bibinfo
  {author} {\bibfnamefont {N.}~\bibnamefont {Laflorencie}}, \ and\ \bibinfo
  {author} {\bibfnamefont {T.}~\bibnamefont {Giamarchi}},\ }\bibinfo {title}
  {{Dimensional modulation of spontaneous magnetic order in
  quasi-two-dimensional quantum antiferromagnets}},\ \href {\doibase
  10.1103/PhysRevB.94.144403} {\bibfield  {journal} {\bibinfo  {journal} {Phys.
  Rev. B.}\ }\textbf {\bibinfo {volume} {94}},\ \bibinfo {pages} {144403}
  (\bibinfo {year} {2016})}\BibitemShut {NoStop}%
\bibitem [{\citenamefont {Maeda}\ \emph {et~al.}(2007)\citenamefont {Maeda},
  \citenamefont {Hotta},\ and\ \citenamefont {Oshikawa}}]{Maeda2007}%
  \BibitemOpen
  \bibfield  {author} {\bibinfo {author} {\bibfnamefont {Y.}~\bibnamefont
  {Maeda}}, \bibinfo {author} {\bibfnamefont {C.}~\bibnamefont {Hotta}}, \ and\
  \bibinfo {author} {\bibfnamefont {M.}~\bibnamefont {Oshikawa}},\ }\bibinfo
  {title} {{Universal Temperature Dependence of the Magnetization of Gapped
  Spin Chains}},\ \href {\doibase 10.1103/PhysRevLett.99.057205} {\bibfield
  {journal} {\bibinfo  {journal} {Phys. Rev. Lett.}\ }\textbf {\bibinfo
  {volume} {99}},\ \bibinfo {pages} {057205} (\bibinfo {year}
  {2007})}\BibitemShut {NoStop}%
\bibitem [{\citenamefont {Klanj\ifmmode~\check{s}\else \v{s}\fi{}ek}\ \emph
  {et~al.}(2008)\citenamefont {Klanj\ifmmode~\check{s}\else \v{s}\fi{}ek},
  \citenamefont {Mayaffre}, \citenamefont {Berthier}, \citenamefont
  {Horvati\ifmmode~\acute{c}\else \'{c}\fi{}}, \citenamefont {Chiari},
  \citenamefont {Piovesana}, \citenamefont {Bouillot}, \citenamefont {Kollath},
  \citenamefont {Orignac}, \citenamefont {Citro},\ and\ \citenamefont
  {Giamarchi}}]{Klanjsek2008}%
  \BibitemOpen
  \bibfield  {author} {\bibinfo {author} {\bibfnamefont {M.}~\bibnamefont
  {Klanj\ifmmode~\check{s}\else \v{s}\fi{}ek}}, \bibinfo {author}
  {\bibfnamefont {H.}~\bibnamefont {Mayaffre}}, \bibinfo {author}
  {\bibfnamefont {C.}~\bibnamefont {Berthier}}, \bibinfo {author}
  {\bibfnamefont {M.}~\bibnamefont {Horvati\ifmmode~\acute{c}\else
  \'{c}\fi{}}}, \bibinfo {author} {\bibfnamefont {B.}~\bibnamefont {Chiari}},
  \bibinfo {author} {\bibfnamefont {O.}~\bibnamefont {Piovesana}}, \bibinfo
  {author} {\bibfnamefont {P.}~\bibnamefont {Bouillot}}, \bibinfo {author}
  {\bibfnamefont {C.}~\bibnamefont {Kollath}}, \bibinfo {author} {\bibfnamefont
  {E.}~\bibnamefont {Orignac}}, \bibinfo {author} {\bibfnamefont
  {R.}~\bibnamefont {Citro}}, \ and\ \bibinfo {author} {\bibfnamefont
  {T.}~\bibnamefont {Giamarchi}},\ }\bibinfo {title} {{Controlling Luttinger
  Liquid Physics in Spin Ladders under a Magnetic Field}},\ \href {\doibase
  10.1103/PhysRevLett.101.137207} {\bibfield  {journal} {\bibinfo  {journal}
  {Phys. Rev. Lett.}\ }\textbf {\bibinfo {volume} {101}},\ \bibinfo {pages}
  {137207} (\bibinfo {year} {2008})}\BibitemShut {NoStop}%
\bibitem [{\citenamefont {Lorenz}\ \emph {et~al.}(2008)\citenamefont {Lorenz},
  \citenamefont {Heyer}, \citenamefont {Garst}, \citenamefont {Anfuso},
  \citenamefont {Rosch}, \citenamefont {R\"uegg},\ and\ \citenamefont
  {Kr\"amer}}]{Lorenz2008}%
  \BibitemOpen
  \bibfield  {author} {\bibinfo {author} {\bibfnamefont {T.}~\bibnamefont
  {Lorenz}}, \bibinfo {author} {\bibfnamefont {O.}~\bibnamefont {Heyer}},
  \bibinfo {author} {\bibfnamefont {M.}~\bibnamefont {Garst}}, \bibinfo
  {author} {\bibfnamefont {F.}~\bibnamefont {Anfuso}}, \bibinfo {author}
  {\bibfnamefont {A.}~\bibnamefont {Rosch}}, \bibinfo {author} {\bibfnamefont
  {C.}~\bibnamefont {R\"uegg}}, \ and\ \bibinfo {author} {\bibfnamefont
  {K.}~\bibnamefont {Kr\"amer}},\ }\bibinfo {title} {{Diverging Thermal
  Expansion of the Spin-Ladder System
  $({\mathrm{C}}_{5}{\mathrm{H}}_{12}\mathrm{N}{)}_{2}{\mathrm{CuBr}}_{4}$}},\
  \href {\doibase 10.1103/PhysRevLett.100.067208} {\bibfield  {journal}
  {\bibinfo  {journal} {Phys. Rev. Lett.}\ }\textbf {\bibinfo {volume} {100}},\
  \bibinfo {pages} {067208} (\bibinfo {year} {2008})}\BibitemShut {NoStop}%
\bibitem [{\citenamefont {Thielemann}\ \emph {et~al.}(2009)\citenamefont
  {Thielemann}, \citenamefont {R\"uegg}, \citenamefont {Kiefer}, \citenamefont
  {R\o{}nnow}, \citenamefont {Normand}, \citenamefont {Bouillot}, \citenamefont
  {Kollath}, \citenamefont {Orignac}, \citenamefont {Citro}, \citenamefont
  {Giamarchi}, \citenamefont {L\"auchli}, \citenamefont {Biner}, \citenamefont
  {Kr\"amer}, \citenamefont {Wolff-Fabris}, \citenamefont {Zapf}, \citenamefont
  {Jaime}, \citenamefont {Stahn}, \citenamefont {Christensen}, \citenamefont
  {Grenier}, \citenamefont {McMorrow},\ and\ \citenamefont
  {Mesot}}]{Thielemann2009}%
  \BibitemOpen
  \bibfield  {author} {\bibinfo {author} {\bibfnamefont {B.}~\bibnamefont
  {Thielemann}}, \bibinfo {author} {\bibfnamefont {C.}~\bibnamefont {R\"uegg}},
  \bibinfo {author} {\bibfnamefont {K.}~\bibnamefont {Kiefer}}, \bibinfo
  {author} {\bibfnamefont {H.~M.}\ \bibnamefont {R\o{}nnow}}, \bibinfo {author}
  {\bibfnamefont {B.}~\bibnamefont {Normand}}, \bibinfo {author} {\bibfnamefont
  {P.}~\bibnamefont {Bouillot}}, \bibinfo {author} {\bibfnamefont
  {C.}~\bibnamefont {Kollath}}, \bibinfo {author} {\bibfnamefont
  {E.}~\bibnamefont {Orignac}}, \bibinfo {author} {\bibfnamefont
  {R.}~\bibnamefont {Citro}}, \bibinfo {author} {\bibfnamefont
  {T.}~\bibnamefont {Giamarchi}}, \bibinfo {author} {\bibfnamefont {A.~M.}\
  \bibnamefont {L\"auchli}}, \bibinfo {author} {\bibfnamefont {D.}~\bibnamefont
  {Biner}}, \bibinfo {author} {\bibfnamefont {K.~W.}\ \bibnamefont {Kr\"amer}},
  \bibinfo {author} {\bibfnamefont {F.}~\bibnamefont {Wolff-Fabris}}, \bibinfo
  {author} {\bibfnamefont {V.~S.}\ \bibnamefont {Zapf}}, \bibinfo {author}
  {\bibfnamefont {M.}~\bibnamefont {Jaime}}, \bibinfo {author} {\bibfnamefont
  {J.}~\bibnamefont {Stahn}}, \bibinfo {author} {\bibfnamefont {N.~B.}\
  \bibnamefont {Christensen}}, \bibinfo {author} {\bibfnamefont
  {B.}~\bibnamefont {Grenier}}, \bibinfo {author} {\bibfnamefont {D.~F.}\
  \bibnamefont {McMorrow}}, \ and\ \bibinfo {author} {\bibfnamefont
  {J.}~\bibnamefont {Mesot}},\ }\bibinfo {title} {{Field-controlled magnetic
  order in the quantum spin-ladder system
  ${(\text{Hpip})}_{2}{\text{CuBr}}_{4}$}},\ \href {\doibase
  10.1103/PhysRevB.79.020408} {\bibfield  {journal} {\bibinfo  {journal} {Phys.
  Rev. B}\ }\textbf {\bibinfo {volume} {79}},\ \bibinfo {pages} {020408}
  (\bibinfo {year} {2009})}\BibitemShut {NoStop}%
\bibitem [{\citenamefont {Schmidiger}\ \emph {et~al.}(2012)\citenamefont
  {Schmidiger}, \citenamefont {Bouillot}, \citenamefont {M\"uhlbauer},
  \citenamefont {Gvasaliya}, \citenamefont {Kollath}, \citenamefont
  {Giamarchi},\ and\ \citenamefont {Zheludev}}]{Schmidiger2012}%
  \BibitemOpen
  \bibfield  {author} {\bibinfo {author} {\bibfnamefont {D.}~\bibnamefont
  {Schmidiger}}, \bibinfo {author} {\bibfnamefont {P.}~\bibnamefont
  {Bouillot}}, \bibinfo {author} {\bibfnamefont {S.}~\bibnamefont
  {M\"uhlbauer}}, \bibinfo {author} {\bibfnamefont {S.}~\bibnamefont
  {Gvasaliya}}, \bibinfo {author} {\bibfnamefont {C.}~\bibnamefont {Kollath}},
  \bibinfo {author} {\bibfnamefont {T.}~\bibnamefont {Giamarchi}}, \ and\
  \bibinfo {author} {\bibfnamefont {A.}~\bibnamefont {Zheludev}},\ }\bibinfo
  {title} {{Spectral and Thermodynamic Properties of a Strong-Leg Quantum Spin
  Ladder}},\ \href {\doibase 10.1103/PhysRevLett.108.167201} {\bibfield
  {journal} {\bibinfo  {journal} {Phys. Rev. Lett.}\ }\textbf {\bibinfo
  {volume} {108}},\ \bibinfo {pages} {167201} (\bibinfo {year}
  {2012})}\BibitemShut {NoStop}%
\bibitem [{\citenamefont {Ninios}\ \emph {et~al.}(2012)\citenamefont {Ninios},
  \citenamefont {Hong}, \citenamefont {Manabe}, \citenamefont {Hotta},
  \citenamefont {Herringer}, \citenamefont {Turnbull}, \citenamefont {Landee},
  \citenamefont {Takano},\ and\ \citenamefont {Chan}}]{Ninios2012}%
  \BibitemOpen
  \bibfield  {author} {\bibinfo {author} {\bibfnamefont {K.}~\bibnamefont
  {Ninios}}, \bibinfo {author} {\bibfnamefont {T.}~\bibnamefont {Hong}},
  \bibinfo {author} {\bibfnamefont {T.}~\bibnamefont {Manabe}}, \bibinfo
  {author} {\bibfnamefont {C.}~\bibnamefont {Hotta}}, \bibinfo {author}
  {\bibfnamefont {S.~N.}\ \bibnamefont {Herringer}}, \bibinfo {author}
  {\bibfnamefont {M.~M.}\ \bibnamefont {Turnbull}}, \bibinfo {author}
  {\bibfnamefont {C.~P.}\ \bibnamefont {Landee}}, \bibinfo {author}
  {\bibfnamefont {Y.}~\bibnamefont {Takano}}, \ and\ \bibinfo {author}
  {\bibfnamefont {H.~B.}\ \bibnamefont {Chan}},\ }\bibinfo {title} {{Wilson
  Ratio of a Tomonaga-Luttinger Liquid in a Spin-$1/2$ Heisenberg Ladder}},\
  \href {\doibase 10.1103/PhysRevLett.108.097201} {\bibfield  {journal}
  {\bibinfo  {journal} {Phys. Rev. Lett.}\ }\textbf {\bibinfo {volume} {108}},\
  \bibinfo {pages} {097201} (\bibinfo {year} {2012})}\BibitemShut {NoStop}%
\bibitem [{\citenamefont {Allenspach}\ \emph {et~al.}(2021)\citenamefont
  {Allenspach}, \citenamefont {Puphal}, \citenamefont {Link}, \citenamefont
  {Heinmaa}, \citenamefont {Pomjakushina}, \citenamefont {Krellner},
  \citenamefont {Lass}, \citenamefont {Tucker}, \citenamefont {Niedermayer},
  \citenamefont {Imajo}, \citenamefont {Kohama}, \citenamefont {Kindo},
  \citenamefont {Kr\"amer}, \citenamefont {Horvati\ifmmode~\acute{c}\else
  \'{c}\fi{}}, \citenamefont {Jaime}, \citenamefont {Madsen}, \citenamefont
  {Mira}, \citenamefont {Laflorencie}, \citenamefont {Mila}, \citenamefont
  {Normand}, \citenamefont {R\"uegg}, \citenamefont {Stern},\ and\
  \citenamefont {Weickert}}]{Allenspach2021}%
  \BibitemOpen
  \bibfield  {author} {\bibinfo {author} {\bibfnamefont {S.}~\bibnamefont
  {Allenspach}}, \bibinfo {author} {\bibfnamefont {P.}~\bibnamefont {Puphal}},
  \bibinfo {author} {\bibfnamefont {J.}~\bibnamefont {Link}}, \bibinfo {author}
  {\bibfnamefont {I.}~\bibnamefont {Heinmaa}}, \bibinfo {author} {\bibfnamefont
  {E.}~\bibnamefont {Pomjakushina}}, \bibinfo {author} {\bibfnamefont
  {C.}~\bibnamefont {Krellner}}, \bibinfo {author} {\bibfnamefont
  {J.}~\bibnamefont {Lass}}, \bibinfo {author} {\bibfnamefont {G.~S.}\
  \bibnamefont {Tucker}}, \bibinfo {author} {\bibfnamefont {C.}~\bibnamefont
  {Niedermayer}}, \bibinfo {author} {\bibfnamefont {S.}~\bibnamefont {Imajo}},
  \bibinfo {author} {\bibfnamefont {Y.}~\bibnamefont {Kohama}}, \bibinfo
  {author} {\bibfnamefont {K.}~\bibnamefont {Kindo}}, \bibinfo {author}
  {\bibfnamefont {S.}~\bibnamefont {Kr\"amer}}, \bibinfo {author}
  {\bibfnamefont {M.}~\bibnamefont {Horvati\ifmmode~\acute{c}\else
  \'{c}\fi{}}}, \bibinfo {author} {\bibfnamefont {M.}~\bibnamefont {Jaime}},
  \bibinfo {author} {\bibfnamefont {A.}~\bibnamefont {Madsen}}, \bibinfo
  {author} {\bibfnamefont {A.}~\bibnamefont {Mira}}, \bibinfo {author}
  {\bibfnamefont {N.}~\bibnamefont {Laflorencie}}, \bibinfo {author}
  {\bibfnamefont {F.}~\bibnamefont {Mila}}, \bibinfo {author} {\bibfnamefont
  {B.}~\bibnamefont {Normand}}, \bibinfo {author} {\bibfnamefont
  {C.}~\bibnamefont {R\"uegg}}, \bibinfo {author} {\bibfnamefont
  {R.}~\bibnamefont {Stern}}, \ and\ \bibinfo {author} {\bibfnamefont
  {F.}~\bibnamefont {Weickert}},\ }\bibinfo {title} {{Revealing
  three-dimensional quantum criticality by Sr substitution in Han purple}},\
  \href {\doibase 10.1103/PhysRevResearch.3.023177} {\bibfield  {journal}
  {\bibinfo  {journal} {Phys. Rev. Res.}\ }\textbf {\bibinfo {volume} {3}},\
  \bibinfo {pages} {023177} (\bibinfo {year} {2021})}\BibitemShut {NoStop}%
\bibitem [{\citenamefont {Kageyama}\ \emph {et~al.}(1999)\citenamefont
  {Kageyama}, \citenamefont {Yoshimura}, \citenamefont {Stern}, \citenamefont
  {Mushnikov}, \citenamefont {Onizuka}, \citenamefont {Kato}, \citenamefont
  {Kosuge}, \citenamefont {Slichter}, \citenamefont {Goto},\ and\ \citenamefont
  {Ueda}}]{Kageyama1999}%
  \BibitemOpen
  \bibfield  {author} {\bibinfo {author} {\bibfnamefont {H.}~\bibnamefont
  {Kageyama}}, \bibinfo {author} {\bibfnamefont {K.}~\bibnamefont {Yoshimura}},
  \bibinfo {author} {\bibfnamefont {R.}~\bibnamefont {Stern}}, \bibinfo
  {author} {\bibfnamefont {N.~V.}\ \bibnamefont {Mushnikov}}, \bibinfo {author}
  {\bibfnamefont {K.}~\bibnamefont {Onizuka}}, \bibinfo {author} {\bibfnamefont
  {M.}~\bibnamefont {Kato}}, \bibinfo {author} {\bibfnamefont {K.}~\bibnamefont
  {Kosuge}}, \bibinfo {author} {\bibfnamefont {C.~P.}\ \bibnamefont
  {Slichter}}, \bibinfo {author} {\bibfnamefont {T.}~\bibnamefont {Goto}}, \
  and\ \bibinfo {author} {\bibfnamefont {Y.}~\bibnamefont {Ueda}},\ }\bibinfo
  {title} {{Exact dimer ground state and quantized magnetization plateaus in
  the two-dimemsional spin system SrCu$_2$(BO$_3$)$_2$}},\ \href {\doibase
  10.1103/PhysRevLett.82.3168} {\bibfield  {journal} {\bibinfo  {journal}
  {Phys. Rev. Lett.}\ }\textbf {\bibinfo {volume} {82}},\ \bibinfo {pages}
  {3168} (\bibinfo {year} {1999})}\BibitemShut {NoStop}%
\bibitem [{\citenamefont {Stone}\ \emph {et~al.}(2001)\citenamefont {Stone},
  \citenamefont {Zaliznyak}, \citenamefont {Reich},\ and\ \citenamefont
  {Broholm}}]{Stone2001}%
  \BibitemOpen
  \bibfield  {author} {\bibinfo {author} {\bibfnamefont {M.~B.}\ \bibnamefont
  {Stone}}, \bibinfo {author} {\bibfnamefont {I.}~\bibnamefont {Zaliznyak}},
  \bibinfo {author} {\bibfnamefont {D.~H.}\ \bibnamefont {Reich}}, \ and\
  \bibinfo {author} {\bibfnamefont {C.}~\bibnamefont {Broholm}},\ }\bibinfo
  {title} {{Frustration-induced two-dimensional quantum disordered phase in
  piperazinium hexachlorodicuprate}},\ \href {\doibase
  10.1103/PhysRevB.64.144405} {\bibfield  {journal} {\bibinfo  {journal} {Phys.
  Rev. B}\ }\textbf {\bibinfo {volume} {64}},\ \bibinfo {pages} {144405}
  (\bibinfo {year} {2001})}\BibitemShut {NoStop}%
\bibitem [{\citenamefont {Stone}\ \emph {et~al.}(2006)\citenamefont {Stone},
  \citenamefont {Zaliznyak}, \citenamefont {Hong}, \citenamefont {Broholm},\
  and\ \citenamefont {Reich}}]{Stone2006}%
  \BibitemOpen
  \bibfield  {author} {\bibinfo {author} {\bibfnamefont {M.~B.}\ \bibnamefont
  {Stone}}, \bibinfo {author} {\bibfnamefont {I.~A.}\ \bibnamefont
  {Zaliznyak}}, \bibinfo {author} {\bibfnamefont {T.}~\bibnamefont {Hong}},
  \bibinfo {author} {\bibfnamefont {C.~L.}\ \bibnamefont {Broholm}}, \ and\
  \bibinfo {author} {\bibfnamefont {D.~H.}\ \bibnamefont {Reich}},\ }\bibinfo
  {title} {{Quasiparticle breakdown in a quantum spin liquid}},\ \href
  {\doibase 10.1038/nature04593} {\bibfield  {journal} {\bibinfo  {journal}
  {Nature}\ }\textbf {\bibinfo {volume} {440}},\ \bibinfo {pages} {187}
  (\bibinfo {year} {2006})}\BibitemShut {NoStop}%
\bibitem [{\citenamefont {Nakajima}\ \emph {et~al.}(2006)\citenamefont
  {Nakajima}, \citenamefont {Mitamura},\ and\ \citenamefont
  {Ueda}}]{Nakajima2006}%
  \BibitemOpen
  \bibfield  {author} {\bibinfo {author} {\bibfnamefont {T.}~\bibnamefont
  {Nakajima}}, \bibinfo {author} {\bibfnamefont {H.}~\bibnamefont {Mitamura}},
  \ and\ \bibinfo {author} {\bibfnamefont {Y.}~\bibnamefont {Ueda}},\ }\bibinfo
  {title} {{Singlet Ground State and Magnetic Interactions in New Spin Dimer
  System Ba$_3$Cr$_2$O$_8$}},\ \href {\doibase 10.1143/JPSJ.75.054706}
  {\bibfield  {journal} {\bibinfo  {journal} {J. Phys. Soc. Jpn.}\ }\textbf
  {\bibinfo {volume} {75}},\ \bibinfo {pages} {054706} (\bibinfo {year}
  {2006})}\BibitemShut {NoStop}%
\bibitem [{\citenamefont {Kofu}\ \emph {et~al.}(2009)\citenamefont {Kofu},
  \citenamefont {Kim}, \citenamefont {Ji}, \citenamefont {Lee}, \citenamefont
  {Ueda}, \citenamefont {Qiu}, \citenamefont {Kang}, \citenamefont {Green},\
  and\ \citenamefont {Ueda}}]{Kofu2009}%
  \BibitemOpen
  \bibfield  {author} {\bibinfo {author} {\bibfnamefont {M.}~\bibnamefont
  {Kofu}}, \bibinfo {author} {\bibfnamefont {J.-H.}\ \bibnamefont {Kim}},
  \bibinfo {author} {\bibfnamefont {S.}~\bibnamefont {Ji}}, \bibinfo {author}
  {\bibfnamefont {S.-H.}\ \bibnamefont {Lee}}, \bibinfo {author} {\bibfnamefont
  {H.}~\bibnamefont {Ueda}}, \bibinfo {author} {\bibfnamefont {Y.}~\bibnamefont
  {Qiu}}, \bibinfo {author} {\bibfnamefont {H.-J.}\ \bibnamefont {Kang}},
  \bibinfo {author} {\bibfnamefont {M.~A.}\ \bibnamefont {Green}}, \ and\
  \bibinfo {author} {\bibfnamefont {Y.}~\bibnamefont {Ueda}},\ }\bibinfo
  {title} {{Weakly Coupled $s = 1/2$ Quantum Spin Singlets in
  Ba$_3$Cr$_2$O$_8$}},\ \href {\doibase 10.1103/PhysRevLett.102.037206}
  {\bibfield  {journal} {\bibinfo  {journal} {Phys. Rev. Lett.}\ }\textbf
  {\bibinfo {volume} {102}},\ \bibinfo {pages} {037206} (\bibinfo {year}
  {2009})}\BibitemShut {NoStop}%
\bibitem [{\citenamefont {Aczel}\ \emph
  {et~al.}(2009{\natexlab{a}})\citenamefont {Aczel}, \citenamefont {Kohama},
  \citenamefont {Jaime}, \citenamefont {Balicas}, \citenamefont {Ninios},
  \citenamefont {Chan}, \citenamefont {Dabkowska},\ and\ \citenamefont
  {Luke}}]{Aczel2009a}%
  \BibitemOpen
  \bibfield  {author} {\bibinfo {author} {\bibfnamefont {A.~A.}\ \bibnamefont
  {Aczel}}, \bibinfo {author} {\bibfnamefont {Y.}~\bibnamefont {Kohama}},
  \bibinfo {author} {\bibfnamefont {M.}~\bibnamefont {Jaime}}, \bibinfo
  {author} {\bibfnamefont {L.}~\bibnamefont {Balicas}}, \bibinfo {author}
  {\bibfnamefont {K.}~\bibnamefont {Ninios}}, \bibinfo {author} {\bibfnamefont
  {H.~B.}\ \bibnamefont {Chan}}, \bibinfo {author} {\bibfnamefont {H.~A.}\
  \bibnamefont {Dabkowska}}, \ and\ \bibinfo {author} {\bibfnamefont {G.~M.}\
  \bibnamefont {Luke}},\ }\bibinfo {title} {{Bose-Einstein condensation of
  triplons in Ba$_3$Cr$_2$O$_8$}},\ \href {\doibase 10.1103/PhysRevB.79.100409}
  {\bibfield  {journal} {\bibinfo  {journal} {Phys. Rev. B}\ }\textbf {\bibinfo
  {volume} {79}},\ \bibinfo {pages} {100409} (\bibinfo {year}
  {2009}{\natexlab{a}})}\BibitemShut {NoStop}%
\bibitem [{\citenamefont {Aczel}\ \emph
  {et~al.}(2009{\natexlab{b}})\citenamefont {Aczel}, \citenamefont {Kohama},
  \citenamefont {Marcenat}, \citenamefont {Weickert}, \citenamefont
  {Ayala-Valenzuela}, \citenamefont {Jaime}, \citenamefont {McDonald},
  \citenamefont {Selesnic}, \citenamefont {Dabkowska},\ and\ \citenamefont
  {Luke}}]{Aczel2009b}%
  \BibitemOpen
  \bibfield  {author} {\bibinfo {author} {\bibfnamefont {A.~A.}\ \bibnamefont
  {Aczel}}, \bibinfo {author} {\bibfnamefont {Y.}~\bibnamefont {Kohama}},
  \bibinfo {author} {\bibfnamefont {C.}~\bibnamefont {Marcenat}}, \bibinfo
  {author} {\bibfnamefont {F.}~\bibnamefont {Weickert}}, \bibinfo {author}
  {\bibfnamefont {O.~E.}\ \bibnamefont {Ayala-Valenzuela}}, \bibinfo {author}
  {\bibfnamefont {M.}~\bibnamefont {Jaime}}, \bibinfo {author} {\bibfnamefont
  {R.~D.}\ \bibnamefont {McDonald}}, \bibinfo {author} {\bibfnamefont {S.~D.}\
  \bibnamefont {Selesnic}}, \bibinfo {author} {\bibfnamefont {H.~A.}\
  \bibnamefont {Dabkowska}}, \ and\ \bibinfo {author} {\bibfnamefont {G.~M.}\
  \bibnamefont {Luke}},\ }\bibinfo {title} {{Field-Induced Bose-Einstein
  Condensation of Triplons up to 8 K in Sr$_3$Cr$_2$O$_8$}},\ \href {\doibase
  10.1103/PhysRevLett.103.207203} {\bibfield  {journal} {\bibinfo  {journal}
  {Phys. Rev. Lett.}\ }\textbf {\bibinfo {volume} {103}},\ \bibinfo {pages}
  {207203} (\bibinfo {year} {2009}{\natexlab{b}})}\BibitemShut {NoStop}%
\bibitem [{\citenamefont {Islam}\ \emph {et~al.}(2010)\citenamefont {Islam},
  \citenamefont {Quintero-Castro}, \citenamefont {Lake}, \citenamefont
  {Siemensmeyer}, \citenamefont {Kiefer}, \citenamefont {Skourski},\ and\
  \citenamefont {Herrmannsdorfer}}]{Islam2010}%
  \BibitemOpen
  \bibfield  {author} {\bibinfo {author} {\bibfnamefont {A.~T. M.~N.}\
  \bibnamefont {Islam}}, \bibinfo {author} {\bibfnamefont {D.}~\bibnamefont
  {Quintero-Castro}}, \bibinfo {author} {\bibfnamefont {B.}~\bibnamefont
  {Lake}}, \bibinfo {author} {\bibfnamefont {K.}~\bibnamefont {Siemensmeyer}},
  \bibinfo {author} {\bibfnamefont {K.}~\bibnamefont {Kiefer}}, \bibinfo
  {author} {\bibfnamefont {Y.}~\bibnamefont {Skourski}}, \ and\ \bibinfo
  {author} {\bibfnamefont {T.}~\bibnamefont {Herrmannsdorfer}},\ }\bibinfo
  {title} {{Optical Floating-Zone Growth of Large Single Crystal of Spin Half
  Dimer Sr$_3$Cr$_2$O$_8$}},\ \href {\doibase 10.1021/cg9010339} {\bibfield
  {journal} {\bibinfo  {journal} {Crystal Growth \& Design}\ }\textbf {\bibinfo
  {volume} {10}},\ \bibinfo {pages} {465} (\bibinfo {year} {2010})}\BibitemShut
  {NoStop}%
\bibitem [{\citenamefont {Nomura}\ \emph {et~al.}(2020)\citenamefont {Nomura},
  \citenamefont {Skourski}, \citenamefont {Quintero-Castro}, \citenamefont
  {Zvyagin}, \citenamefont {Suslov}, \citenamefont {Gorbunov}, \citenamefont
  {Yasin}, \citenamefont {Wosnitza}, \citenamefont {Kindo}, \citenamefont
  {Islam}, \citenamefont {Lake}, \citenamefont {Kohama}, \citenamefont
  {Zherlitsyn},\ and\ \citenamefont {Jaime}}]{Nomura2020}%
  \BibitemOpen
  \bibfield  {author} {\bibinfo {author} {\bibfnamefont {T.}~\bibnamefont
  {Nomura}}, \bibinfo {author} {\bibfnamefont {Y.}~\bibnamefont {Skourski}},
  \bibinfo {author} {\bibfnamefont {D.~L.}\ \bibnamefont {Quintero-Castro}},
  \bibinfo {author} {\bibfnamefont {A.~A.}\ \bibnamefont {Zvyagin}}, \bibinfo
  {author} {\bibfnamefont {A.~V.}\ \bibnamefont {Suslov}}, \bibinfo {author}
  {\bibfnamefont {D.}~\bibnamefont {Gorbunov}}, \bibinfo {author}
  {\bibfnamefont {S.}~\bibnamefont {Yasin}}, \bibinfo {author} {\bibfnamefont
  {J.}~\bibnamefont {Wosnitza}}, \bibinfo {author} {\bibfnamefont
  {K.}~\bibnamefont {Kindo}}, \bibinfo {author} {\bibfnamefont {A.~T. M.~N.}\
  \bibnamefont {Islam}}, \bibinfo {author} {\bibfnamefont {B.}~\bibnamefont
  {Lake}}, \bibinfo {author} {\bibfnamefont {Y.}~\bibnamefont {Kohama}},
  \bibinfo {author} {\bibfnamefont {S.}~\bibnamefont {Zherlitsyn}}, \ and\
  \bibinfo {author} {\bibfnamefont {M.}~\bibnamefont {Jaime}},\ }\bibinfo
  {title} {{Enhanced spin correlations in the Bose-Einstein condensate compound
  Sr$_3$Cr$_2$O$_8$}},\ \href {\doibase 10.1103/PhysRevB.102.165144} {\bibfield
   {journal} {\bibinfo  {journal} {Phys. Rev. B}\ }\textbf {\bibinfo {volume}
  {102}},\ \bibinfo {pages} {165144} (\bibinfo {year} {2020})}\BibitemShut
  {NoStop}%
\bibitem [{\citenamefont {Tutsch}\ \emph {et~al.}(2014)\citenamefont {Tutsch},
  \citenamefont {Wolf}, \citenamefont {Wessel}, \citenamefont {Postulka},
  \citenamefont {Tsui}, \citenamefont {Jeschke}, \citenamefont {Opahle},
  \citenamefont {Saha-Dasgupta}, \citenamefont {Valent{\'i}}, \citenamefont
  {Br{\"u}hl}, \citenamefont {Removi{\'c}-Langer}, \citenamefont {Kretz},
  \citenamefont {Lerner}, \citenamefont {Wagner},\ and\ \citenamefont
  {Lang}}]{Tutsch2014}%
  \BibitemOpen
  \bibfield  {author} {\bibinfo {author} {\bibfnamefont {U.}~\bibnamefont
  {Tutsch}}, \bibinfo {author} {\bibfnamefont {B.}~\bibnamefont {Wolf}},
  \bibinfo {author} {\bibfnamefont {S.}~\bibnamefont {Wessel}}, \bibinfo
  {author} {\bibfnamefont {L.}~\bibnamefont {Postulka}}, \bibinfo {author}
  {\bibfnamefont {Y.}~\bibnamefont {Tsui}}, \bibinfo {author} {\bibfnamefont
  {H.}~\bibnamefont {Jeschke}}, \bibinfo {author} {\bibfnamefont
  {I.}~\bibnamefont {Opahle}}, \bibinfo {author} {\bibfnamefont
  {T.}~\bibnamefont {Saha-Dasgupta}}, \bibinfo {author} {\bibfnamefont
  {R.}~\bibnamefont {Valent{\'i}}}, \bibinfo {author} {\bibfnamefont
  {A.}~\bibnamefont {Br{\"u}hl}}, \bibinfo {author} {\bibfnamefont
  {K.}~\bibnamefont {Removi{\'c}-Langer}}, \bibinfo {author} {\bibfnamefont
  {T.}~\bibnamefont {Kretz}}, \bibinfo {author} {\bibfnamefont
  {H.}~\bibnamefont {Lerner}}, \bibinfo {author} {\bibfnamefont
  {M.}~\bibnamefont {Wagner}}, \ and\ \bibinfo {author} {\bibfnamefont
  {M.}~\bibnamefont {Lang}},\ }\bibinfo {title} {{Evidence of a field-induced
  Berezinskii-Kosterlitz-Thouless scenario in a two-dimensional spin-dimer
  system}},\ \href {\doibase 10.1038/ncomms6169} {\bibfield  {journal}
  {\bibinfo  {journal} {Nat. Commun.}\ }\textbf {\bibinfo {volume} {5}},\
  \bibinfo {pages} {5169} (\bibinfo {year} {2014})}\BibitemShut {NoStop}%
\bibitem [{\citenamefont {FitzHugh}\ and\ \citenamefont
  {Zycherman}(1992)}]{FitzHugh1992}%
  \BibitemOpen
  \bibfield  {author} {\bibinfo {author} {\bibfnamefont {E.~W.}\ \bibnamefont
  {FitzHugh}}\ and\ \bibinfo {author} {\bibfnamefont {L.~A.}\ \bibnamefont
  {Zycherman}},\ }\bibinfo {title} {{A Purple Barium Copper Silicate Pigment
  from Early China}},\ \href {\doibase
  https://doi.org/10.1179/sic.1992.37.3.145} {\bibfield  {journal} {\bibinfo
  {journal} {Stud. Conserv.}\ }\textbf {\bibinfo {volume} {37}},\ \bibinfo
  {pages} {145} (\bibinfo {year} {1992})}\BibitemShut {NoStop}%
\bibitem [{\citenamefont {Sebastian}\ \emph {et~al.}(2006)\citenamefont
  {Sebastian}, \citenamefont {Harrison}, \citenamefont {Batista}, \citenamefont
  {Balicas}, \citenamefont {Jaime}, \citenamefont {Sharma}, \citenamefont
  {Kawashima},\ and\ \citenamefont {Fisher}}]{Sebastian2006}%
  \BibitemOpen
  \bibfield  {author} {\bibinfo {author} {\bibfnamefont {S.~E.}\ \bibnamefont
  {Sebastian}}, \bibinfo {author} {\bibfnamefont {N.}~\bibnamefont {Harrison}},
  \bibinfo {author} {\bibfnamefont {C.~D.}\ \bibnamefont {Batista}}, \bibinfo
  {author} {\bibfnamefont {L.}~\bibnamefont {Balicas}}, \bibinfo {author}
  {\bibfnamefont {M.}~\bibnamefont {Jaime}}, \bibinfo {author} {\bibfnamefont
  {P.~A.}\ \bibnamefont {Sharma}}, \bibinfo {author} {\bibfnamefont
  {N.}~\bibnamefont {Kawashima}}, \ and\ \bibinfo {author} {\bibfnamefont
  {I.~R.}\ \bibnamefont {Fisher}},\ }\bibinfo {title} {Dimensional reduction at
  a quantum critical point},\ \href {\doibase 10.1038/nature04732} {\bibfield
  {journal} {\bibinfo  {journal} {Nature}\ }\textbf {\bibinfo {volume} {441}},\
  \bibinfo {pages} {617} (\bibinfo {year} {2006})}\BibitemShut {NoStop}%
\bibitem [{\citenamefont {Samulon}\ \emph {et~al.}(2006)\citenamefont
  {Samulon}, \citenamefont {Islam}, \citenamefont {Sebastian}, \citenamefont
  {Brooks}, \citenamefont {McCourt}, \citenamefont {Ilavsky},\ and\
  \citenamefont {Fisher}}]{Samulon2006}%
  \BibitemOpen
  \bibfield  {author} {\bibinfo {author} {\bibfnamefont {E.~C.}\ \bibnamefont
  {Samulon}}, \bibinfo {author} {\bibfnamefont {Z.}~\bibnamefont {Islam}},
  \bibinfo {author} {\bibfnamefont {S.~E.}\ \bibnamefont {Sebastian}}, \bibinfo
  {author} {\bibfnamefont {P.~B.}\ \bibnamefont {Brooks}}, \bibinfo {author}
  {\bibfnamefont {M.~K.}\ \bibnamefont {McCourt}}, \bibinfo {author}
  {\bibfnamefont {J.}~\bibnamefont {Ilavsky}}, \ and\ \bibinfo {author}
  {\bibfnamefont {I.~R.}\ \bibnamefont {Fisher}},\ }\bibinfo {title}
  {{Low-temperature structural phase transition and incommensurate lattice
  modulation in the spin-gap compound BaCuSi$_2$O$_6$}},\ \href {\doibase
  10.1103/PhysRevB.73.100407} {\bibfield  {journal} {\bibinfo  {journal} {Phys.
  Rev. B}\ }\textbf {\bibinfo {volume} {73}},\ \bibinfo {pages} {100407}
  (\bibinfo {year} {2006})}\BibitemShut {NoStop}%
\bibitem [{\citenamefont {Sheptyakov}\ \emph {et~al.}(2012)\citenamefont
  {Sheptyakov}, \citenamefont {Pomjakushin}, \citenamefont {Stern},
  \citenamefont {Heinmaa}, \citenamefont {Nakamura},\ and\ \citenamefont
  {Kimura}}]{Sheptyakov2012}%
  \BibitemOpen
  \bibfield  {author} {\bibinfo {author} {\bibfnamefont {D.~V.}\ \bibnamefont
  {Sheptyakov}}, \bibinfo {author} {\bibfnamefont {V.~Y.}\ \bibnamefont
  {Pomjakushin}}, \bibinfo {author} {\bibfnamefont {R.}~\bibnamefont {Stern}},
  \bibinfo {author} {\bibfnamefont {I.}~\bibnamefont {Heinmaa}}, \bibinfo
  {author} {\bibfnamefont {H.}~\bibnamefont {Nakamura}}, \ and\ \bibinfo
  {author} {\bibfnamefont {T.}~\bibnamefont {Kimura}},\ }\bibinfo {title} {{Two
  types of adjacent dimer layers in the low-temperature phase of
  BaCuSi$_2$O$_6$}},\ \href {\doibase 10.1103/physrevb.86.014433} {\bibfield
  {journal} {\bibinfo  {journal} {Phys. Rev. B}\ }\textbf {\bibinfo {volume}
  {86}},\ \bibinfo {pages} {014433} (\bibinfo {year} {2012})}\BibitemShut
  {NoStop}%
\bibitem [{\citenamefont {R\"{u}egg}\ \emph {et~al.}(2007)\citenamefont
  {R\"{u}egg}, \citenamefont {McMorrow}, \citenamefont {Normand}, \citenamefont
  {R{\o}nnow}, \citenamefont {Sebastian}, \citenamefont {Fisher}, \citenamefont
  {Batista}, \citenamefont {Gvasaliya}, \citenamefont {Niedermayer},\ and\
  \citenamefont {Stahn}}]{Rueegg2007}%
  \BibitemOpen
  \bibfield  {author} {\bibinfo {author} {\bibfnamefont {C.}~\bibnamefont
  {R\"{u}egg}}, \bibinfo {author} {\bibfnamefont {D.~F.}\ \bibnamefont
  {McMorrow}}, \bibinfo {author} {\bibfnamefont {B.}~\bibnamefont {Normand}},
  \bibinfo {author} {\bibfnamefont {H.~M.}\ \bibnamefont {R{\o}nnow}}, \bibinfo
  {author} {\bibfnamefont {S.~E.}\ \bibnamefont {Sebastian}}, \bibinfo {author}
  {\bibfnamefont {I.~R.}\ \bibnamefont {Fisher}}, \bibinfo {author}
  {\bibfnamefont {C.~D.}\ \bibnamefont {Batista}}, \bibinfo {author}
  {\bibfnamefont {S.~N.}\ \bibnamefont {Gvasaliya}}, \bibinfo {author}
  {\bibfnamefont {C.}~\bibnamefont {Niedermayer}}, \ and\ \bibinfo {author}
  {\bibfnamefont {J.}~\bibnamefont {Stahn}},\ }\bibinfo {title} {{Multiple
  Magnon Modes and Consequences for the Bose-Einstein Condensed Phase in
  BaCuSi$_2$O$_6$}},\ \href {\doibase 10.1103/physrevlett.98.017202} {\bibfield
   {journal} {\bibinfo  {journal} {Phys. Rev. Lett.}\ }\textbf {\bibinfo
  {volume} {98}},\ \bibinfo {pages} {017202} (\bibinfo {year}
  {2007})}\BibitemShut {NoStop}%
\bibitem [{\citenamefont {Kr\"{a}mer}\ \emph {et~al.}(2007)\citenamefont
  {Kr\"{a}mer}, \citenamefont {Stern}, \citenamefont {Horvati{\'{c}}},
  \citenamefont {Berthier}, \citenamefont {Kimura},\ and\ \citenamefont
  {Fisher}}]{Kraemer2007}%
  \BibitemOpen
  \bibfield  {author} {\bibinfo {author} {\bibfnamefont {S.}~\bibnamefont
  {Kr\"{a}mer}}, \bibinfo {author} {\bibfnamefont {R.}~\bibnamefont {Stern}},
  \bibinfo {author} {\bibfnamefont {M.}~\bibnamefont {Horvati{\'{c}}}},
  \bibinfo {author} {\bibfnamefont {C.}~\bibnamefont {Berthier}}, \bibinfo
  {author} {\bibfnamefont {T.}~\bibnamefont {Kimura}}, \ and\ \bibinfo {author}
  {\bibfnamefont {I.~R.}\ \bibnamefont {Fisher}},\ }\bibinfo {title} {{Nuclear
  magnetic resonance evidence for a strong modulation of the Bose-Einstein
  condensate in BaCuSi$_2$O$_6$}},\ \href {\doibase 10.1103/physrevb.76.100406}
  {\bibfield  {journal} {\bibinfo  {journal} {Phys. Rev. B}\ }\textbf {\bibinfo
  {volume} {76}},\ \bibinfo {pages} {100406} (\bibinfo {year}
  {2007})}\BibitemShut {NoStop}%
\bibitem [{\citenamefont {Allenspach}\ \emph {et~al.}(2020)\citenamefont
  {Allenspach}, \citenamefont {Biffin}, \citenamefont {Stuhr}, \citenamefont
  {Tucker}, \citenamefont {Ohira-Kawamura}, \citenamefont {Kofu}, \citenamefont
  {Voneshen}, \citenamefont {Boehm}, \citenamefont {Normand}, \citenamefont
  {Laflorencie}, \citenamefont {Mila},\ and\ \citenamefont
  {R\"{u}egg}}]{Allenspach2020}%
  \BibitemOpen
  \bibfield  {author} {\bibinfo {author} {\bibfnamefont {S.}~\bibnamefont
  {Allenspach}}, \bibinfo {author} {\bibfnamefont {A.}~\bibnamefont {Biffin}},
  \bibinfo {author} {\bibfnamefont {U.}~\bibnamefont {Stuhr}}, \bibinfo
  {author} {\bibfnamefont {G.~S.}\ \bibnamefont {Tucker}}, \bibinfo {author}
  {\bibfnamefont {S.}~\bibnamefont {Ohira-Kawamura}}, \bibinfo {author}
  {\bibfnamefont {M.}~\bibnamefont {Kofu}}, \bibinfo {author} {\bibfnamefont
  {D.~J.}\ \bibnamefont {Voneshen}}, \bibinfo {author} {\bibfnamefont
  {M.}~\bibnamefont {Boehm}}, \bibinfo {author} {\bibfnamefont
  {B.}~\bibnamefont {Normand}}, \bibinfo {author} {\bibfnamefont
  {N.}~\bibnamefont {Laflorencie}}, \bibinfo {author} {\bibfnamefont
  {F.}~\bibnamefont {Mila}}, \ and\ \bibinfo {author} {\bibfnamefont
  {C.}~\bibnamefont {R\"{u}egg}},\ }\bibinfo {title} {{Multiple Magnetic
  Bilayers and Unconventional Criticality without Frustration in
  BaCuSi$_2$O$_6$}},\ \href {\doibase 10.1103/PhysRevLett.124.177205}
  {\bibfield  {journal} {\bibinfo  {journal} {Phys. Rev. Lett.}\ }\textbf
  {\bibinfo {volume} {124}},\ \bibinfo {pages} {177205} (\bibinfo {year}
  {2020})}\BibitemShut {NoStop}%
\bibitem [{\citenamefont {Sparta}\ \emph {et~al.}(2006)\citenamefont {Sparta},
  \citenamefont {Merz}, \citenamefont {Roth}, \citenamefont {Stern},
  \citenamefont {Cerny},\ and\ \citenamefont {Kimura}}]{Sparta2006}%
  \BibitemOpen
  \bibfield  {author} {\bibinfo {author} {\bibfnamefont {K.}~\bibnamefont
  {Sparta}}, \bibinfo {author} {\bibfnamefont {M.}~\bibnamefont {Merz}},
  \bibinfo {author} {\bibfnamefont {G.}~\bibnamefont {Roth}}, \bibinfo {author}
  {\bibfnamefont {R.}~\bibnamefont {Stern}}, \bibinfo {author} {\bibfnamefont
  {R.}~\bibnamefont {Cerny}}, \ and\ \bibinfo {author} {\bibfnamefont
  {T.}~\bibnamefont {Kimura}},\ }\bibinfo {title} {{Low temperature phase
  transition in BaCuSi$_2$O$_6$}},\ \href {\doibase 10.1107/S0108767306096061}
  {\bibfield  {journal} {\bibinfo  {journal} {Acta Cryst. A}\ }\textbf
  {\bibinfo {volume} {62}},\ \bibinfo {pages} {S197} (\bibinfo {year}
  {2006})}\BibitemShut {NoStop}%
\bibitem [{\citenamefont {Kr\"{a}mer}\ \emph {et~al.}(2013)\citenamefont
  {Kr\"{a}mer}, \citenamefont {Laflorencie}, \citenamefont {Stern},
  \citenamefont {Horvati{\'{c}}}, \citenamefont {Berthier}, \citenamefont
  {Nakamura}, \citenamefont {Kimura},\ and\ \citenamefont
  {Mila}}]{Kraemer2013}%
  \BibitemOpen
  \bibfield  {author} {\bibinfo {author} {\bibfnamefont {S.}~\bibnamefont
  {Kr\"{a}mer}}, \bibinfo {author} {\bibfnamefont {N.}~\bibnamefont
  {Laflorencie}}, \bibinfo {author} {\bibfnamefont {R.}~\bibnamefont {Stern}},
  \bibinfo {author} {\bibfnamefont {M.}~\bibnamefont {Horvati{\'{c}}}},
  \bibinfo {author} {\bibfnamefont {C.}~\bibnamefont {Berthier}}, \bibinfo
  {author} {\bibfnamefont {H.}~\bibnamefont {Nakamura}}, \bibinfo {author}
  {\bibfnamefont {T.}~\bibnamefont {Kimura}}, \ and\ \bibinfo {author}
  {\bibfnamefont {F.}~\bibnamefont {Mila}},\ }\bibinfo {title} {{Spatially
  resolved magnetization in the Bose-Einstein condensed state of
  BaCuSi$_2$O$_6$: Evidence for imperfect frustration}},\ \href {\doibase
  10.1103/PhysRevB.87.180405} {\bibfield  {journal} {\bibinfo  {journal} {Phys.
  Rev. B}\ }\textbf {\bibinfo {volume} {87}},\ \bibinfo {pages} {180405}
  (\bibinfo {year} {2013})}\BibitemShut {NoStop}%
\bibitem [{\citenamefont {Jaime}\ \emph {et~al.}(2004)\citenamefont {Jaime},
  \citenamefont {Correa}, \citenamefont {Harrison}, \citenamefont {Batista},
  \citenamefont {Kawashima}, \citenamefont {Kazuma}, \citenamefont {Jorge},
  \citenamefont {Stern}, \citenamefont {Heinmaa}, \citenamefont {Zvyagin},
  \citenamefont {Sasago},\ and\ \citenamefont {Uchinokura}}]{Jaime2004}%
  \BibitemOpen
  \bibfield  {author} {\bibinfo {author} {\bibfnamefont {M.}~\bibnamefont
  {Jaime}}, \bibinfo {author} {\bibfnamefont {V.~F.}\ \bibnamefont {Correa}},
  \bibinfo {author} {\bibfnamefont {N.}~\bibnamefont {Harrison}}, \bibinfo
  {author} {\bibfnamefont {C.~D.}\ \bibnamefont {Batista}}, \bibinfo {author}
  {\bibfnamefont {N.}~\bibnamefont {Kawashima}}, \bibinfo {author}
  {\bibfnamefont {Y.}~\bibnamefont {Kazuma}}, \bibinfo {author} {\bibfnamefont
  {G.~A.}\ \bibnamefont {Jorge}}, \bibinfo {author} {\bibfnamefont
  {R.}~\bibnamefont {Stern}}, \bibinfo {author} {\bibfnamefont
  {I.}~\bibnamefont {Heinmaa}}, \bibinfo {author} {\bibfnamefont {S.~A.}\
  \bibnamefont {Zvyagin}}, \bibinfo {author} {\bibfnamefont {Y.}~\bibnamefont
  {Sasago}}, \ and\ \bibinfo {author} {\bibfnamefont {K.}~\bibnamefont
  {Uchinokura}},\ }\bibinfo {title} {{Magnetic-Field-Induced Condensation of
  Triplons in Han Purple Pigment BaCuSi$_2$O$_6$}},\ \href {\doibase
  10.1103/physrevlett.93.087203} {\bibfield  {journal} {\bibinfo  {journal}
  {Phys. Rev. Lett.}\ }\textbf {\bibinfo {volume} {93}},\ \bibinfo {pages}
  {087203} (\bibinfo {year} {2004})}\BibitemShut {NoStop}%
\bibitem [{\citenamefont {Sebastian}\ \emph {et~al.}(2005)\citenamefont
  {Sebastian}, \citenamefont {Sharma}, \citenamefont {Jaime}, \citenamefont
  {Harrison}, \citenamefont {Correa}, \citenamefont {Balicas}, \citenamefont
  {Kawashima}, \citenamefont {Batista},\ and\ \citenamefont
  {Fisher}}]{Sebastian2005}%
  \BibitemOpen
  \bibfield  {author} {\bibinfo {author} {\bibfnamefont {S.~E.}\ \bibnamefont
  {Sebastian}}, \bibinfo {author} {\bibfnamefont {P.~A.}\ \bibnamefont
  {Sharma}}, \bibinfo {author} {\bibfnamefont {M.}~\bibnamefont {Jaime}},
  \bibinfo {author} {\bibfnamefont {N.}~\bibnamefont {Harrison}}, \bibinfo
  {author} {\bibfnamefont {V.}~\bibnamefont {Correa}}, \bibinfo {author}
  {\bibfnamefont {L.}~\bibnamefont {Balicas}}, \bibinfo {author} {\bibfnamefont
  {N.}~\bibnamefont {Kawashima}}, \bibinfo {author} {\bibfnamefont {C.~D.}\
  \bibnamefont {Batista}}, \ and\ \bibinfo {author} {\bibfnamefont {I.~R.}\
  \bibnamefont {Fisher}},\ }\bibinfo {title} {{Characteristic Bose-Einstein
  condensation scaling close to a quantum critical point in BaCuSi$_2$O$_6$}},\
  \href {\doibase 10.1103/physrevb.72.100404} {\bibfield  {journal} {\bibinfo
  {journal} {Phys. Rev. B}\ }\textbf {\bibinfo {volume} {72}},\ \bibinfo
  {pages} {100404} (\bibinfo {year} {2005})}\BibitemShut {NoStop}%
\bibitem [{\citenamefont {Batista}\ \emph {et~al.}(2007)\citenamefont
  {Batista}, \citenamefont {Schmalian}, \citenamefont {Kawashima},
  \citenamefont {Sengupta}, \citenamefont {Sebastian}, \citenamefont
  {Harrison}, \citenamefont {Jaime},\ and\ \citenamefont
  {Fisher}}]{Batista2007}%
  \BibitemOpen
  \bibfield  {author} {\bibinfo {author} {\bibfnamefont {C.~D.}\ \bibnamefont
  {Batista}}, \bibinfo {author} {\bibfnamefont {J.}~\bibnamefont {Schmalian}},
  \bibinfo {author} {\bibfnamefont {N.}~\bibnamefont {Kawashima}}, \bibinfo
  {author} {\bibfnamefont {P.}~\bibnamefont {Sengupta}}, \bibinfo {author}
  {\bibfnamefont {S.~E.}\ \bibnamefont {Sebastian}}, \bibinfo {author}
  {\bibfnamefont {N.}~\bibnamefont {Harrison}}, \bibinfo {author}
  {\bibfnamefont {M.}~\bibnamefont {Jaime}}, \ and\ \bibinfo {author}
  {\bibfnamefont {I.~R.}\ \bibnamefont {Fisher}},\ }\bibinfo {title}
  {{Geometric Frustration and Dimensional Reduction at a Quantum Critical
  Point}},\ \href {\doibase 10.1103/PhysRevLett.98.257201} {\bibfield
  {journal} {\bibinfo  {journal} {Phys. Rev. Lett.}\ }\textbf {\bibinfo
  {volume} {98}},\ \bibinfo {pages} {257201} (\bibinfo {year}
  {2007})}\BibitemShut {NoStop}%
\bibitem [{\citenamefont {Schmalian}\ and\ \citenamefont
  {Batista}(2008)}]{Schmalian2008}%
  \BibitemOpen
  \bibfield  {author} {\bibinfo {author} {\bibfnamefont {J.}~\bibnamefont
  {Schmalian}}\ and\ \bibinfo {author} {\bibfnamefont {C.~D.}\ \bibnamefont
  {Batista}},\ }\bibinfo {title} {{Emergent symmetry and dimensional reduction
  at a quantum critical point}},\ \href {\doibase 10.1103/PhysRevB.77.094406}
  {\bibfield  {journal} {\bibinfo  {journal} {Phys. Rev. B}\ }\textbf {\bibinfo
  {volume} {77}},\ \bibinfo {pages} {094406} (\bibinfo {year}
  {2008})}\BibitemShut {NoStop}%
\bibitem [{\citenamefont {Maltseva}\ and\ \citenamefont
  {Coleman}(2005)}]{Maltseva2005}%
  \BibitemOpen
  \bibfield  {author} {\bibinfo {author} {\bibfnamefont {M.}~\bibnamefont
  {Maltseva}}\ and\ \bibinfo {author} {\bibfnamefont {P.}~\bibnamefont
  {Coleman}},\ }\bibinfo {title} {{Failure of geometric frustration to preserve
  a quasi-two-dimensional spin fluid}},\ \href {\doibase
  10.1103/PhysRevB.72.174415} {\bibfield  {journal} {\bibinfo  {journal} {Phys.
  Rev. B}\ }\textbf {\bibinfo {volume} {72}},\ \bibinfo {pages} {174415}
  (\bibinfo {year} {2005})}\BibitemShut {NoStop}%
\bibitem [{\citenamefont {R\"osch}\ and\ \citenamefont
  {Vojta}(2007{\natexlab{a}})}]{Roesch2007a}%
  \BibitemOpen
  \bibfield  {author} {\bibinfo {author} {\bibfnamefont {O.}~\bibnamefont
  {R\"osch}}\ and\ \bibinfo {author} {\bibfnamefont {M.}~\bibnamefont
  {Vojta}},\ }\bibinfo {title} {{Quantum phase transitions and dimensional
  reduction in antiferromagnets with interlayer frustration}},\ \href {\doibase
  10.1103/PhysRevB.76.180401} {\bibfield  {journal} {\bibinfo  {journal} {Phys.
  Rev. B}\ }\textbf {\bibinfo {volume} {76}},\ \bibinfo {pages} {180401}
  (\bibinfo {year} {2007}{\natexlab{a}})}\BibitemShut {NoStop}%
\bibitem [{\citenamefont {R\"osch}\ and\ \citenamefont
  {Vojta}(2007{\natexlab{b}})}]{Roesch2007b}%
  \BibitemOpen
  \bibfield  {author} {\bibinfo {author} {\bibfnamefont {O.}~\bibnamefont
  {R\"osch}}\ and\ \bibinfo {author} {\bibfnamefont {M.}~\bibnamefont
  {Vojta}},\ }\bibinfo {title} {{Reduced dimensionality in layered quantum
  dimer magnets: Frustration vs. inhomogeneous condensates}},\ \href {\doibase
  10.1103/PhysRevB.76.224408} {\bibfield  {journal} {\bibinfo  {journal} {Phys.
  Rev. B}\ }\textbf {\bibinfo {volume} {76}},\ \bibinfo {pages} {224408}
  (\bibinfo {year} {2007}{\natexlab{b}})}\BibitemShut {NoStop}%
\bibitem [{\citenamefont {Mazurenko}\ \emph {et~al.}(2014)\citenamefont
  {Mazurenko}, \citenamefont {Valentyuk}, \citenamefont {Stern},\ and\
  \citenamefont {Tsirlin}}]{Mazurenko2014}%
  \BibitemOpen
  \bibfield  {author} {\bibinfo {author} {\bibfnamefont {V.~V.}\ \bibnamefont
  {Mazurenko}}, \bibinfo {author} {\bibfnamefont {M.~V.}\ \bibnamefont
  {Valentyuk}}, \bibinfo {author} {\bibfnamefont {R.}~\bibnamefont {Stern}}, \
  and\ \bibinfo {author} {\bibfnamefont {A.~A.}\ \bibnamefont {Tsirlin}},\
  }\bibinfo {title} {{Nonfrustrated Interlayer Order and its Relevance to the
  Bose-Einstein Condensation of Magnons in BaCuSi$_2$O$_6$}},\ \href {\doibase
  10.1103/physrevlett.112.107202} {\bibfield  {journal} {\bibinfo  {journal}
  {Phys. Rev. Lett.}\ }\textbf {\bibinfo {volume} {112}},\ \bibinfo {pages}
  {107202} (\bibinfo {year} {2014})}\BibitemShut {NoStop}%
\bibitem [{\citenamefont {Puphal}\ \emph {et~al.}(2016)\citenamefont {Puphal},
  \citenamefont {Sheptyakov}, \citenamefont {van Well}, \citenamefont
  {Postulka}, \citenamefont {Heinmaa}, \citenamefont {Ritter}, \citenamefont
  {Assmus}, \citenamefont {Wolf}, \citenamefont {Lang}, \citenamefont
  {Jeschke}, \citenamefont {Valent{\'{\i}}}, \citenamefont {Stern},
  \citenamefont {R\"{u}egg},\ and\ \citenamefont {Krellner}}]{Puphal2016}%
  \BibitemOpen
  \bibfield  {author} {\bibinfo {author} {\bibfnamefont {P.}~\bibnamefont
  {Puphal}}, \bibinfo {author} {\bibfnamefont {D.}~\bibnamefont {Sheptyakov}},
  \bibinfo {author} {\bibfnamefont {N.}~\bibnamefont {van Well}}, \bibinfo
  {author} {\bibfnamefont {L.}~\bibnamefont {Postulka}}, \bibinfo {author}
  {\bibfnamefont {I.}~\bibnamefont {Heinmaa}}, \bibinfo {author} {\bibfnamefont
  {F.}~\bibnamefont {Ritter}}, \bibinfo {author} {\bibfnamefont
  {W.}~\bibnamefont {Assmus}}, \bibinfo {author} {\bibfnamefont
  {B.}~\bibnamefont {Wolf}}, \bibinfo {author} {\bibfnamefont {M.}~\bibnamefont
  {Lang}}, \bibinfo {author} {\bibfnamefont {H.~O.}\ \bibnamefont {Jeschke}},
  \bibinfo {author} {\bibfnamefont {R.}~\bibnamefont {Valent{\'{\i}}}},
  \bibinfo {author} {\bibfnamefont {R.}~\bibnamefont {Stern}}, \bibinfo
  {author} {\bibfnamefont {C.}~\bibnamefont {R\"{u}egg}}, \ and\ \bibinfo
  {author} {\bibfnamefont {C.}~\bibnamefont {Krellner}},\ }\bibinfo {title}
  {{Stabilization of the tetragonal structure in
  (Ba$_{1-x}$Sr$_x$)CuSi$_2$O$_6$}},\ \href {\doibase
  10.1103/physrevb.93.174121} {\bibfield  {journal} {\bibinfo  {journal} {Phys.
  Rev. B}\ }\textbf {\bibinfo {volume} {93}},\ \bibinfo {pages} {174121}
  (\bibinfo {year} {2016})}\BibitemShut {NoStop}%
\bibitem [{\citenamefont {Gibertini}\ \emph {et~al.}(2019)\citenamefont
  {Gibertini}, \citenamefont {Koperski}, \citenamefont {Morpurgo},\ and\
  \citenamefont {Novoselov}}]{Gibertini2019}%
  \BibitemOpen
  \bibfield  {author} {\bibinfo {author} {\bibfnamefont {M.}~\bibnamefont
  {Gibertini}}, \bibinfo {author} {\bibfnamefont {M.}~\bibnamefont {Koperski}},
  \bibinfo {author} {\bibfnamefont {A.~F.}\ \bibnamefont {Morpurgo}}, \ and\
  \bibinfo {author} {\bibfnamefont {K.~S.}\ \bibnamefont {Novoselov}},\
  }\bibinfo {title} {{Magnetic {2D} materials and heterostructures}},\ \href
  {\doibase 10.1038/s41565-019-0438-6} {\bibfield  {journal} {\bibinfo
  {journal} {Nat. Nanotechnol.}\ }\textbf {\bibinfo {volume} {14}},\ \bibinfo
  {pages} {408} (\bibinfo {year} {2019})}\BibitemShut {NoStop}%
\bibitem [{\citenamefont {Klein}\ \emph {et~al.}(2019)\citenamefont {Klein},
  \citenamefont {MacNeill}, \citenamefont {Song}, \citenamefont {Larson},
  \citenamefont {Fang}, \citenamefont {Xu}, \citenamefont {Ribeiro},
  \citenamefont {Canfield}, \citenamefont {Kaxiras}, \citenamefont {Comin},\
  and\ \citenamefont {Jarillo-Herrero}}]{Klein2019}%
  \BibitemOpen
  \bibfield  {author} {\bibinfo {author} {\bibfnamefont {D.~R.}\ \bibnamefont
  {Klein}}, \bibinfo {author} {\bibfnamefont {D.}~\bibnamefont {MacNeill}},
  \bibinfo {author} {\bibfnamefont {Q.}~\bibnamefont {Song}}, \bibinfo {author}
  {\bibfnamefont {D.~T.}\ \bibnamefont {Larson}}, \bibinfo {author}
  {\bibfnamefont {S.}~\bibnamefont {Fang}}, \bibinfo {author} {\bibfnamefont
  {M.}~\bibnamefont {Xu}}, \bibinfo {author} {\bibfnamefont {R.~A.}\
  \bibnamefont {Ribeiro}}, \bibinfo {author} {\bibfnamefont {P.~C.}\
  \bibnamefont {Canfield}}, \bibinfo {author} {\bibfnamefont {E.}~\bibnamefont
  {Kaxiras}}, \bibinfo {author} {\bibfnamefont {R.}~\bibnamefont {Comin}}, \
  and\ \bibinfo {author} {\bibfnamefont {P.}~\bibnamefont {Jarillo-Herrero}},\
  }\bibinfo {title} {{Enhancement of interlayer exchange in an ultrathin {2D}
  magnet}},\ \href {\doibase 10.1038/s41567-019-0651-0} {\bibfield  {journal}
  {\bibinfo  {journal} {Nat. Phys.}\ }\textbf {\bibinfo {volume} {15}},\
  \bibinfo {pages} {1255} (\bibinfo {year} {2019})}\BibitemShut {NoStop}%
\bibitem [{\citenamefont {Ubrig}\ \emph {et~al.}(2019)\citenamefont {Ubrig},
  \citenamefont {Wang}, \citenamefont {Teyssier}, \citenamefont {Taniguchi},
  \citenamefont {Watanabe}, \citenamefont {Giannini}, \citenamefont
  {Morpurgo},\ and\ \citenamefont {Gibertini}}]{Ubrig2019}%
  \BibitemOpen
  \bibfield  {author} {\bibinfo {author} {\bibfnamefont {N.}~\bibnamefont
  {Ubrig}}, \bibinfo {author} {\bibfnamefont {Z.}~\bibnamefont {Wang}},
  \bibinfo {author} {\bibfnamefont {J.}~\bibnamefont {Teyssier}}, \bibinfo
  {author} {\bibfnamefont {T.}~\bibnamefont {Taniguchi}}, \bibinfo {author}
  {\bibfnamefont {K.}~\bibnamefont {Watanabe}}, \bibinfo {author}
  {\bibfnamefont {E.}~\bibnamefont {Giannini}}, \bibinfo {author}
  {\bibfnamefont {A.~F.}\ \bibnamefont {Morpurgo}}, \ and\ \bibinfo {author}
  {\bibfnamefont {M.}~\bibnamefont {Gibertini}},\ }\bibinfo {title}
  {{Low-temperature monoclinic layer stacking in atomically thin {CrI$_3$}
  crystals}},\ \href {\doibase 10.1088/2053-1583/ab4c64} {\bibfield  {journal}
  {\bibinfo  {journal} {2D Mater.}\ }\textbf {\bibinfo {volume} {7}},\ \bibinfo
  {pages} {015007} (\bibinfo {year} {2019})}\BibitemShut {NoStop}%
\bibitem [{\citenamefont {Cai}\ \emph {et~al.}(2019)\citenamefont {Cai},
  \citenamefont {Song}, \citenamefont {Wilson}, \citenamefont {Clark},
  \citenamefont {He}, \citenamefont {Zhang}, \citenamefont {Taniguchi},
  \citenamefont {Watanabe}, \citenamefont {Yao}, \citenamefont {Xiao},
  \citenamefont {McGuire}, \citenamefont {Cobden},\ and\ \citenamefont
  {Xu}}]{Cao2019}%
  \BibitemOpen
  \bibfield  {author} {\bibinfo {author} {\bibfnamefont {X.}~\bibnamefont
  {Cai}}, \bibinfo {author} {\bibfnamefont {T.}~\bibnamefont {Song}}, \bibinfo
  {author} {\bibfnamefont {N.~P.}\ \bibnamefont {Wilson}}, \bibinfo {author}
  {\bibfnamefont {G.}~\bibnamefont {Clark}}, \bibinfo {author} {\bibfnamefont
  {M.}~\bibnamefont {He}}, \bibinfo {author} {\bibfnamefont {X.}~\bibnamefont
  {Zhang}}, \bibinfo {author} {\bibfnamefont {T.}~\bibnamefont {Taniguchi}},
  \bibinfo {author} {\bibfnamefont {K.}~\bibnamefont {Watanabe}}, \bibinfo
  {author} {\bibfnamefont {W.}~\bibnamefont {Yao}}, \bibinfo {author}
  {\bibfnamefont {D.}~\bibnamefont {Xiao}}, \bibinfo {author} {\bibfnamefont
  {M.~A.}\ \bibnamefont {McGuire}}, \bibinfo {author} {\bibfnamefont {D.~H.}\
  \bibnamefont {Cobden}}, \ and\ \bibinfo {author} {\bibfnamefont
  {X.}~\bibnamefont {Xu}},\ }\bibinfo {title} {{Atomically Thin CrCl$_3$: An
  In-Plane Layered Antiferromagnetic Insulator}},\ \href {\doibase
  10.1021/acs.nanolett.9b01317} {\bibfield  {journal} {\bibinfo  {journal}
  {Nano Lett.}\ }\textbf {\bibinfo {volume} {19}},\ \bibinfo {pages} {3993}
  (\bibinfo {year} {2019})}\BibitemShut {NoStop}%
\bibitem [{\citenamefont {Wang}\ \emph {et~al.}(2019)\citenamefont {Wang},
  \citenamefont {Gibertini}, \citenamefont {Dumcenco}, \citenamefont
  {Taniguchi}, \citenamefont {Watanabe}, \citenamefont {Giannini},\ and\
  \citenamefont {Morpurgo}}]{Wang2019}%
  \BibitemOpen
  \bibfield  {author} {\bibinfo {author} {\bibfnamefont {Z.}~\bibnamefont
  {Wang}}, \bibinfo {author} {\bibfnamefont {M.}~\bibnamefont {Gibertini}},
  \bibinfo {author} {\bibfnamefont {D.}~\bibnamefont {Dumcenco}}, \bibinfo
  {author} {\bibfnamefont {T.}~\bibnamefont {Taniguchi}}, \bibinfo {author}
  {\bibfnamefont {K.}~\bibnamefont {Watanabe}}, \bibinfo {author}
  {\bibfnamefont {E.}~\bibnamefont {Giannini}}, \ and\ \bibinfo {author}
  {\bibfnamefont {A.~F.}\ \bibnamefont {Morpurgo}},\ }\bibinfo {title}
  {{Determining the phase diagram of the atomically thin layered
  antiferromagnet {CrCl$_3$}}},\ \href {\doibase 10.1038/s41565-019-0565-0}
  {\bibfield  {journal} {\bibinfo  {journal} {Nat. Nanotechnol.}\ }\textbf
  {\bibinfo {volume} {14}},\ \bibinfo {pages} {1116} (\bibinfo {year}
  {2019})}\BibitemShut {NoStop}%
\bibitem [{\citenamefont {Kim}\ \emph {et~al.}(2019)\citenamefont {Kim},
  \citenamefont {Yang}, \citenamefont {Li}, \citenamefont {Jiang},
  \citenamefont {Jin}, \citenamefont {Tao}, \citenamefont {Nichols},
  \citenamefont {Sfigakis}, \citenamefont {Zhong}, \citenamefont {Li},
  \citenamefont {Tian}, \citenamefont {Cory}, \citenamefont {Miao},
  \citenamefont {Shan}, \citenamefont {Mak}, \citenamefont {Lei}, \citenamefont
  {Sun}, \citenamefont {Zhao},\ and\ \citenamefont {Tsen}}]{Kim2019}%
  \BibitemOpen
  \bibfield  {author} {\bibinfo {author} {\bibfnamefont {H.~H.}\ \bibnamefont
  {Kim}}, \bibinfo {author} {\bibfnamefont {B.}~\bibnamefont {Yang}}, \bibinfo
  {author} {\bibfnamefont {S.}~\bibnamefont {Li}}, \bibinfo {author}
  {\bibfnamefont {S.}~\bibnamefont {Jiang}}, \bibinfo {author} {\bibfnamefont
  {C.}~\bibnamefont {Jin}}, \bibinfo {author} {\bibfnamefont {Z.}~\bibnamefont
  {Tao}}, \bibinfo {author} {\bibfnamefont {G.}~\bibnamefont {Nichols}},
  \bibinfo {author} {\bibfnamefont {F.}~\bibnamefont {Sfigakis}}, \bibinfo
  {author} {\bibfnamefont {S.}~\bibnamefont {Zhong}}, \bibinfo {author}
  {\bibfnamefont {C.}~\bibnamefont {Li}}, \bibinfo {author} {\bibfnamefont
  {S.}~\bibnamefont {Tian}}, \bibinfo {author} {\bibfnamefont {D.~G.}\
  \bibnamefont {Cory}}, \bibinfo {author} {\bibfnamefont {G.-X.}\ \bibnamefont
  {Miao}}, \bibinfo {author} {\bibfnamefont {J.}~\bibnamefont {Shan}}, \bibinfo
  {author} {\bibfnamefont {K.~F.}\ \bibnamefont {Mak}}, \bibinfo {author}
  {\bibfnamefont {H.}~\bibnamefont {Lei}}, \bibinfo {author} {\bibfnamefont
  {K.}~\bibnamefont {Sun}}, \bibinfo {author} {\bibfnamefont {L.}~\bibnamefont
  {Zhao}}, \ and\ \bibinfo {author} {\bibfnamefont {A.~W.}\ \bibnamefont
  {Tsen}},\ }\bibinfo {title} {{Evolution of interlayer and intralayer
  magnetism in three atomically thin chromium trihalides}},\ \href {\doibase
  10.1073/pnas.1902100116} {\bibfield  {journal} {\bibinfo  {journal} {Proc.
  Natl. Acad. Sci. U.S.A.}\ }\textbf {\bibinfo {volume} {116}},\ \bibinfo
  {pages} {11131} (\bibinfo {year} {2019})}\BibitemShut {NoStop}%
\bibitem [{\citenamefont {Prokhnenko}\ \emph {et~al.}(2017)\citenamefont
  {Prokhnenko}, \citenamefont {Smeibidl}, \citenamefont {Stein}, \citenamefont
  {Bartkowiak},\ and\ \citenamefont {St\"{u}sser}}]{Prokhnenko2017}%
  \BibitemOpen
  \bibfield  {author} {\bibinfo {author} {\bibfnamefont {O.}~\bibnamefont
  {Prokhnenko}}, \bibinfo {author} {\bibfnamefont {P.}~\bibnamefont
  {Smeibidl}}, \bibinfo {author} {\bibfnamefont {W.-D.}\ \bibnamefont {Stein}},
  \bibinfo {author} {\bibfnamefont {M.}~\bibnamefont {Bartkowiak}}, \ and\
  \bibinfo {author} {\bibfnamefont {N.}~\bibnamefont {St\"{u}sser}},\ }\bibinfo
  {title} {{HFM/EXED: The High Magnetic Field Facility for Neutron Scattering
  at BER II}},\ \href {\doibase 10.17815/jlsrf-3-111} {\bibfield  {journal}
  {\bibinfo  {journal} {JLSRF}\ }\textbf {\bibinfo {volume} {3}},\ \bibinfo
  {pages} {A115} (\bibinfo {year} {2017})}\BibitemShut {NoStop}%
\bibitem [{\citenamefont {Smeibidl}\ \emph {et~al.}(2016)\citenamefont
  {Smeibidl}, \citenamefont {Bird}, \citenamefont {Ehmler}, \citenamefont
  {Dixon}, \citenamefont {Heinrich}, \citenamefont {Hoffmann}, \citenamefont
  {Kempfer}, \citenamefont {Bole}, \citenamefont {Toth}, \citenamefont
  {Prokhnenko},\ and\ \citenamefont {Lake}}]{Smeibidl2016}%
  \BibitemOpen
  \bibfield  {author} {\bibinfo {author} {\bibfnamefont {P.}~\bibnamefont
  {Smeibidl}}, \bibinfo {author} {\bibfnamefont {M.}~\bibnamefont {Bird}},
  \bibinfo {author} {\bibfnamefont {H.}~\bibnamefont {Ehmler}}, \bibinfo
  {author} {\bibfnamefont {I.}~\bibnamefont {Dixon}}, \bibinfo {author}
  {\bibfnamefont {J.}~\bibnamefont {Heinrich}}, \bibinfo {author}
  {\bibfnamefont {M.}~\bibnamefont {Hoffmann}}, \bibinfo {author}
  {\bibfnamefont {S.}~\bibnamefont {Kempfer}}, \bibinfo {author} {\bibfnamefont
  {S.}~\bibnamefont {Bole}}, \bibinfo {author} {\bibfnamefont {J.}~\bibnamefont
  {Toth}}, \bibinfo {author} {\bibfnamefont {O.}~\bibnamefont {Prokhnenko}}, \
  and\ \bibinfo {author} {\bibfnamefont {B.}~\bibnamefont {Lake}},\ }\bibinfo
  {title} {{First Hybrid Magnet for Neutron Scattering at Helmholtz-Zentrum
  Berlin}},\ \href {\doibase 10.1109/TASC.2016.2525773} {\bibfield  {journal}
  {\bibinfo  {journal} {IEEE Trans. Appl. Supercond.}\ }\textbf {\bibinfo
  {volume} {26}},\ \bibinfo {pages} {1} (\bibinfo {year} {2016})}\BibitemShut
  {NoStop}%
\bibitem [{\citenamefont {Bartkowiak}\ \emph {et~al.}(2015)\citenamefont
  {Bartkowiak}, \citenamefont {St\"{u}sser},\ and\ \citenamefont
  {Prokhnenko}}]{Bartkowiak2015}%
  \BibitemOpen
  \bibfield  {author} {\bibinfo {author} {\bibfnamefont {M.}~\bibnamefont
  {Bartkowiak}}, \bibinfo {author} {\bibfnamefont {N.}~\bibnamefont
  {St\"{u}sser}}, \ and\ \bibinfo {author} {\bibfnamefont {O.}~\bibnamefont
  {Prokhnenko}},\ }\bibinfo {title} {{The design of the inelastic neutron
  scattering mode for the Extreme Environment Diffractometer with the 26T High
  Field Magnet}},\ \href {\doibase https://doi.org/10.1016/j.nima.2015.06.028}
  {\bibfield  {journal} {\bibinfo  {journal} {Nucl. Instrum. Methods Phys. Res.
  A}\ }\textbf {\bibinfo {volume} {797}},\ \bibinfo {pages} {121} (\bibinfo
  {year} {2015})}\BibitemShut {NoStop}%
\bibitem [{\citenamefont {Prokhnenko}\ \emph {et~al.}(2015)\citenamefont
  {Prokhnenko}, \citenamefont {Stein}, \citenamefont {Bleif}, \citenamefont
  {Fromme}, \citenamefont {Bartkowiak},\ and\ \citenamefont
  {Wilpert}}]{Prokhnenko2015}%
  \BibitemOpen
  \bibfield  {author} {\bibinfo {author} {\bibfnamefont {O.}~\bibnamefont
  {Prokhnenko}}, \bibinfo {author} {\bibfnamefont {W.-D.}\ \bibnamefont
  {Stein}}, \bibinfo {author} {\bibfnamefont {H.-J.}\ \bibnamefont {Bleif}},
  \bibinfo {author} {\bibfnamefont {M.}~\bibnamefont {Fromme}}, \bibinfo
  {author} {\bibfnamefont {M.}~\bibnamefont {Bartkowiak}}, \ and\ \bibinfo
  {author} {\bibfnamefont {T.}~\bibnamefont {Wilpert}},\ }\bibinfo {title}
  {{Time-of-flight Extreme Environment Diffractometer at the Helmholtz-Zentrum
  Berlin}},\ \href {\doibase 10.1063/1.4913656} {\bibfield  {journal} {\bibinfo
   {journal} {Rev. Sci. Instr.}\ }\textbf {\bibinfo {volume} {86}},\ \bibinfo
  {pages} {033102} (\bibinfo {year} {2015})}\BibitemShut {NoStop}%
\bibitem [{\citenamefont {Berthier}\ \emph {et~al.}(2017)\citenamefont
  {Berthier}, \citenamefont {Horvati{\'{c}}}, \citenamefont {Julien},
  \citenamefont {Mayaffre},\ and\ \citenamefont {Kr\"{a}mer}}]{Berthier2017}%
  \BibitemOpen
  \bibfield  {author} {\bibinfo {author} {\bibfnamefont {C.}~\bibnamefont
  {Berthier}}, \bibinfo {author} {\bibfnamefont {M.}~\bibnamefont
  {Horvati{\'{c}}}}, \bibinfo {author} {\bibfnamefont {M.-H.}\ \bibnamefont
  {Julien}}, \bibinfo {author} {\bibfnamefont {H.}~\bibnamefont {Mayaffre}}, \
  and\ \bibinfo {author} {\bibfnamefont {S.}~\bibnamefont {Kr\"{a}mer}},\
  }\bibinfo {title} {{Nuclear magnetic resonance in high magnetic field:
  Application to condensed matter physics}},\ \href {\doibase
  https://doi.org/10.1016/j.crhy.2017.09.009} {\bibfield  {journal} {\bibinfo
  {journal} {C. R. Phys.}\ }\textbf {\bibinfo {volume} {18}},\ \bibinfo {pages}
  {331} (\bibinfo {year} {2017})}\BibitemShut {NoStop}%
\bibitem [{\citenamefont {Stern}\ \emph {et~al.}(2014)\citenamefont {Stern},
  \citenamefont {Heinmaa}, \citenamefont {Joon}, \citenamefont {Tsirlin},
  \citenamefont {Nakamura},\ and\ \citenamefont {Kimura}}]{Stern2014}%
  \BibitemOpen
  \bibfield  {author} {\bibinfo {author} {\bibfnamefont {R.}~\bibnamefont
  {Stern}}, \bibinfo {author} {\bibfnamefont {I.}~\bibnamefont {Heinmaa}},
  \bibinfo {author} {\bibfnamefont {E.}~\bibnamefont {Joon}}, \bibinfo {author}
  {\bibfnamefont {A.~A.}\ \bibnamefont {Tsirlin}}, \bibinfo {author}
  {\bibfnamefont {H.}~\bibnamefont {Nakamura}}, \ and\ \bibinfo {author}
  {\bibfnamefont {T.}~\bibnamefont {Kimura}},\ }\bibinfo {title}
  {{Low-Temperature High-Resolution Solid-State (cryoMAS) NMR of Han Purple
  BaCuSi$_2$O$_6$}},\ \href {\doibase 10.1007/s00723-014-0597-4} {\bibfield
  {journal} {\bibinfo  {journal} {Appl. Magn. Reson.}\ }\textbf {\bibinfo
  {volume} {45}},\ \bibinfo {pages} {1253} (\bibinfo {year}
  {2014})}\BibitemShut {NoStop}%
\bibitem [{\citenamefont {Zvyagin}\ \emph {et~al.}(2006)\citenamefont
  {Zvyagin}, \citenamefont {Wosnitza}, \citenamefont {Krzystek}, \citenamefont
  {Stern}, \citenamefont {Jaime}, \citenamefont {Sasago},\ and\ \citenamefont
  {Uchinokura}}]{Zvyagin2006}%
  \BibitemOpen
  \bibfield  {author} {\bibinfo {author} {\bibfnamefont {S.~A.}\ \bibnamefont
  {Zvyagin}}, \bibinfo {author} {\bibfnamefont {J.}~\bibnamefont {Wosnitza}},
  \bibinfo {author} {\bibfnamefont {J.}~\bibnamefont {Krzystek}}, \bibinfo
  {author} {\bibfnamefont {R.}~\bibnamefont {Stern}}, \bibinfo {author}
  {\bibfnamefont {M.}~\bibnamefont {Jaime}}, \bibinfo {author} {\bibfnamefont
  {Y.}~\bibnamefont {Sasago}}, \ and\ \bibinfo {author} {\bibfnamefont
  {K.}~\bibnamefont {Uchinokura}},\ }\bibinfo {title} {{Spin-triplet excitons
  in the $S = 1/2$ gapped antiferromagnet BaCuSi$_2$O$_6$: Electron
  paramagnetic resonance studies}},\ \href {\doibase
  10.1103/PhysRevB.73.094446} {\bibfield  {journal} {\bibinfo  {journal} {Phys.
  Rev. B}\ }\textbf {\bibinfo {volume} {73}},\ \bibinfo {pages} {094446}
  (\bibinfo {year} {2006})}\BibitemShut {NoStop}%
\bibitem [{\citenamefont {Bartkowiak}\ \emph {et~al.}(2020)\citenamefont
  {Bartkowiak}, \citenamefont {Proke\v{s}}, \citenamefont {Fromme},
  \citenamefont {Budack}, \citenamefont {Dirlick},\ and\ \citenamefont
  {Prokhnenko}}]{Bartkowiak2020}%
  \BibitemOpen
  \bibfield  {author} {\bibinfo {author} {\bibfnamefont {M.}~\bibnamefont
  {Bartkowiak}}, \bibinfo {author} {\bibfnamefont {K.}~\bibnamefont
  {Proke\v{s}}}, \bibinfo {author} {\bibfnamefont {M.}~\bibnamefont {Fromme}},
  \bibinfo {author} {\bibfnamefont {A.}~\bibnamefont {Budack}}, \bibinfo
  {author} {\bibfnamefont {J.}~\bibnamefont {Dirlick}}, \ and\ \bibinfo
  {author} {\bibfnamefont {O.}~\bibnamefont {Prokhnenko}},\ }\bibinfo {title}
  {{\it EXEQ} and {\it InEXEQ}: software tools for experiment planning at the
  Extreme Environment Diffractometer},\ \href {\doibase
  10.1107/S1600576720011942} {\bibfield  {journal} {\bibinfo  {journal} {J.
  Appl. Crystallogr.}\ }\textbf {\bibinfo {volume} {53}},\ \bibinfo {pages}
  {1613} (\bibinfo {year} {2020})}\BibitemShut {NoStop}%
\bibitem [{\citenamefont {Proke\ifmmode~\check{s}\else \v{s}\fi{}}\ \emph
  {et~al.}(2017)\citenamefont {Proke\ifmmode~\check{s}\else \v{s}\fi{}},
  \citenamefont {Bartkowiak}, \citenamefont {Rivin}, \citenamefont
  {Prokhnenko}, \citenamefont {F\"orster}, \citenamefont {Gerischer},
  \citenamefont {Wahle}, \citenamefont {Huang},\ and\ \citenamefont
  {Mydosh}}]{Prokes2017}%
  \BibitemOpen
  \bibfield  {author} {\bibinfo {author} {\bibfnamefont {K.}~\bibnamefont
  {Proke\ifmmode~\check{s}\else \v{s}\fi{}}}, \bibinfo {author} {\bibfnamefont
  {M.}~\bibnamefont {Bartkowiak}}, \bibinfo {author} {\bibfnamefont
  {O.}~\bibnamefont {Rivin}}, \bibinfo {author} {\bibfnamefont
  {O.}~\bibnamefont {Prokhnenko}}, \bibinfo {author} {\bibfnamefont
  {T.}~\bibnamefont {F\"orster}}, \bibinfo {author} {\bibfnamefont
  {S.}~\bibnamefont {Gerischer}}, \bibinfo {author} {\bibfnamefont
  {R.}~\bibnamefont {Wahle}}, \bibinfo {author} {\bibfnamefont {Y.-K.}\
  \bibnamefont {Huang}}, \ and\ \bibinfo {author} {\bibfnamefont {J.~A.}\
  \bibnamefont {Mydosh}},\ }\bibinfo {title} {{Magnetic structure in a
  $\mathrm{U}{({\mathrm{Ru}}_{0.92}{\mathrm{Rh}}_{0.08})}_{2}{\mathrm{Si}}_{2}$
  single crystal studied by neutron diffraction in static magnetic fields up to
  24 T}},\ \href {\doibase 10.1103/PhysRevB.96.121117} {\bibfield  {journal}
  {\bibinfo  {journal} {Phys. Rev. B}\ }\textbf {\bibinfo {volume} {96}},\
  \bibinfo {pages} {121117} (\bibinfo {year} {2017})}\BibitemShut {NoStop}%
\bibitem [{\citenamefont {Fogh}\ \emph {et~al.}(2020)\citenamefont {Fogh},
  \citenamefont {Kihara}, \citenamefont {Toft-Petersen}, \citenamefont
  {Bartkowiak}, \citenamefont {Narumi}, \citenamefont {Prokhnenko},
  \citenamefont {Miyake}, \citenamefont {Tokunaga}, \citenamefont {Oikawa},
  \citenamefont {S\o{}rensen}, \citenamefont {Dyrnum}, \citenamefont {Grimmer},
  \citenamefont {Nojiri},\ and\ \citenamefont {Christensen}}]{Fogh2020}%
  \BibitemOpen
  \bibfield  {author} {\bibinfo {author} {\bibfnamefont {E.}~\bibnamefont
  {Fogh}}, \bibinfo {author} {\bibfnamefont {T.}~\bibnamefont {Kihara}},
  \bibinfo {author} {\bibfnamefont {R.}~\bibnamefont {Toft-Petersen}}, \bibinfo
  {author} {\bibfnamefont {M.}~\bibnamefont {Bartkowiak}}, \bibinfo {author}
  {\bibfnamefont {Y.}~\bibnamefont {Narumi}}, \bibinfo {author} {\bibfnamefont
  {O.}~\bibnamefont {Prokhnenko}}, \bibinfo {author} {\bibfnamefont
  {A.}~\bibnamefont {Miyake}}, \bibinfo {author} {\bibfnamefont
  {M.}~\bibnamefont {Tokunaga}}, \bibinfo {author} {\bibfnamefont
  {K.}~\bibnamefont {Oikawa}}, \bibinfo {author} {\bibfnamefont {M.~K.}\
  \bibnamefont {S\o{}rensen}}, \bibinfo {author} {\bibfnamefont {J.~C.}\
  \bibnamefont {Dyrnum}}, \bibinfo {author} {\bibfnamefont {H.}~\bibnamefont
  {Grimmer}}, \bibinfo {author} {\bibfnamefont {H.}~\bibnamefont {Nojiri}}, \
  and\ \bibinfo {author} {\bibfnamefont {N.~B.}\ \bibnamefont {Christensen}},\
  }\bibinfo {title} {{Magnetic structures and quadratic magnetoelectric effect
  in ${\mathrm{LiNiPO}}_{4}$ beyond 30 T}},\ \href {\doibase
  10.1103/PhysRevB.101.024403} {\bibfield  {journal} {\bibinfo  {journal}
  {Phys. Rev. B}\ }\textbf {\bibinfo {volume} {101}},\ \bibinfo {pages}
  {024403} (\bibinfo {year} {2020})}\BibitemShut {NoStop}%
\bibitem [{\citenamefont {Proke\ifmmode~\check{s}\else \v{s}\fi{}}\ \emph
  {et~al.}(2020)\citenamefont {Proke\ifmmode~\check{s}\else \v{s}\fi{}},
  \citenamefont {Bartkowiak}, \citenamefont {Gorbunov}, \citenamefont
  {Prokhnenko}, \citenamefont {Rivin},\ and\ \citenamefont
  {Smeibidl}}]{Prokes2020}%
  \BibitemOpen
  \bibfield  {author} {\bibinfo {author} {\bibfnamefont {K.}~\bibnamefont
  {Proke\ifmmode~\check{s}\else \v{s}\fi{}}}, \bibinfo {author} {\bibfnamefont
  {M.}~\bibnamefont {Bartkowiak}}, \bibinfo {author} {\bibfnamefont {D.~I.}\
  \bibnamefont {Gorbunov}}, \bibinfo {author} {\bibfnamefont {O.}~\bibnamefont
  {Prokhnenko}}, \bibinfo {author} {\bibfnamefont {O.}~\bibnamefont {Rivin}}, \
  and\ \bibinfo {author} {\bibfnamefont {P.}~\bibnamefont {Smeibidl}},\
  }\bibinfo {title} {{Noncollinear magnetic structure in
  ${\mathrm{U}}_{2}{\mathrm{Pd}}_{2}\mathrm{In}$ at high magnetic fields}},\
  \href {\doibase 10.1103/PhysRevResearch.2.013137} {\bibfield  {journal}
  {\bibinfo  {journal} {Phys. Rev. Res.}\ }\textbf {\bibinfo {volume} {2}},\
  \bibinfo {pages} {013137} (\bibinfo {year} {2020})}\BibitemShut {NoStop}%
\bibitem [{\citenamefont {Arnold}\ \emph {et~al.}(2014)\citenamefont {Arnold},
  \citenamefont {Bilheux}, \citenamefont {Borreguero}, \citenamefont {Buts},
  \citenamefont {Campbell}, \citenamefont {Chapon}, \citenamefont {Doucet},
  \citenamefont {Draper}, \citenamefont {Leal}, \citenamefont {Gigg},
  \citenamefont {Lynch}, \citenamefont {Markvardsen}, \citenamefont
  {Mikkelson}, \citenamefont {Mikkelson}, \citenamefont {Miller}, \citenamefont
  {Palmen}, \citenamefont {Parker}, \citenamefont {Passos}, \citenamefont
  {Perring}, \citenamefont {Peterson}, \citenamefont {Ren}, \citenamefont
  {Reuter}, \citenamefont {Savici}, \citenamefont {Taylor}, \citenamefont
  {Taylor}, \citenamefont {Tolchenov}, \citenamefont {Zhou},\ and\
  \citenamefont {Zikovsky}}]{Arnold2014}%
  \BibitemOpen
  \bibfield  {author} {\bibinfo {author} {\bibfnamefont {O.}~\bibnamefont
  {Arnold}}, \bibinfo {author} {\bibfnamefont {J.~C.}\ \bibnamefont {Bilheux}},
  \bibinfo {author} {\bibfnamefont {J.~M.}\ \bibnamefont {Borreguero}},
  \bibinfo {author} {\bibfnamefont {A.}~\bibnamefont {Buts}}, \bibinfo {author}
  {\bibfnamefont {S.~I.}\ \bibnamefont {Campbell}}, \bibinfo {author}
  {\bibfnamefont {L.}~\bibnamefont {Chapon}}, \bibinfo {author} {\bibfnamefont
  {M.}~\bibnamefont {Doucet}}, \bibinfo {author} {\bibfnamefont
  {N.}~\bibnamefont {Draper}}, \bibinfo {author} {\bibfnamefont {R.~F.}\
  \bibnamefont {Leal}}, \bibinfo {author} {\bibfnamefont {M.~A.}\ \bibnamefont
  {Gigg}}, \bibinfo {author} {\bibfnamefont {V.~E.}\ \bibnamefont {Lynch}},
  \bibinfo {author} {\bibfnamefont {A.}~\bibnamefont {Markvardsen}}, \bibinfo
  {author} {\bibfnamefont {D.~J.}\ \bibnamefont {Mikkelson}}, \bibinfo {author}
  {\bibfnamefont {R.~L.}\ \bibnamefont {Mikkelson}}, \bibinfo {author}
  {\bibfnamefont {R.}~\bibnamefont {Miller}}, \bibinfo {author} {\bibfnamefont
  {K.}~\bibnamefont {Palmen}}, \bibinfo {author} {\bibfnamefont
  {P.}~\bibnamefont {Parker}}, \bibinfo {author} {\bibfnamefont
  {G.}~\bibnamefont {Passos}}, \bibinfo {author} {\bibfnamefont {T.~G.}\
  \bibnamefont {Perring}}, \bibinfo {author} {\bibfnamefont {P.~F.}\
  \bibnamefont {Peterson}}, \bibinfo {author} {\bibfnamefont {S.}~\bibnamefont
  {Ren}}, \bibinfo {author} {\bibfnamefont {M.~A.}\ \bibnamefont {Reuter}},
  \bibinfo {author} {\bibfnamefont {A.~T.}\ \bibnamefont {Savici}}, \bibinfo
  {author} {\bibfnamefont {J.~W.}\ \bibnamefont {Taylor}}, \bibinfo {author}
  {\bibfnamefont {R.~J.}\ \bibnamefont {Taylor}}, \bibinfo {author}
  {\bibfnamefont {R.}~\bibnamefont {Tolchenov}}, \bibinfo {author}
  {\bibfnamefont {W.}~\bibnamefont {Zhou}}, \ and\ \bibinfo {author}
  {\bibfnamefont {J.}~\bibnamefont {Zikovsky}},\ }\bibinfo {title} {{Mantid --
  Data analysis and visualization package for neutron scattering and $\mu$SR
  experiments}},\ \href {\doibase https://doi.org/10.1016/j.nima.2014.07.029}
  {\bibfield  {journal} {\bibinfo  {journal} {Nucl. Instrum. Methods Phys.
  Res., Sect. A}\ }\textbf {\bibinfo {volume} {764}},\ \bibinfo {pages} {156}
  (\bibinfo {year} {2014})}\BibitemShut {NoStop}%
\bibitem [{\citenamefont {{Debye}}(1913)}]{Debye1913}%
  \BibitemOpen
  \bibfield  {author} {\bibinfo {author} {\bibfnamefont {P.}~\bibnamefont
  {{Debye}}},\ }\bibinfo {title} {{Interferenz von R{\"o}ntgenstrahlen und
  W{\"a}rmebewegung}},\ \href {\doibase 10.1002/andp.19133480105} {\bibfield
  {journal} {\bibinfo  {journal} {Ann. Physik}\ }\textbf {\bibinfo {volume}
  {348}},\ \bibinfo {pages} {49} (\bibinfo {year} {1913})}\BibitemShut
  {NoStop}%
\bibitem [{\citenamefont {Sivia}(1996)}]{Sivia1996}%
  \BibitemOpen
  \bibfield  {author} {\bibinfo {author} {\bibfnamefont {D.~S.}\ \bibnamefont
  {Sivia}},\ }\href@noop {} {\emph {\bibinfo {title} {Data Analysis: A Bayesian
  Tutorial}}}\ (\bibinfo  {publisher} {Clarendon Press, Oxford},\ \bibinfo
  {year} {1996})\BibitemShut {NoStop}%
\bibitem [{\citenamefont {Bishop}(2006)}]{Bishop2006}%
  \BibitemOpen
  \bibfield  {author} {\bibinfo {author} {\bibfnamefont {C.~M.}\ \bibnamefont
  {Bishop}},\ }\href@noop {} {\emph {\bibinfo {title} {{Pattern Recognition and
  Machine Learning (Information Science and Statistics)}}}}\ (\bibinfo
  {publisher} {Springer-Verlag},\ \bibinfo {address} {Berlin, Heidelberg},\
  \bibinfo {year} {2006})\BibitemShut {NoStop}%
\bibitem [{\citenamefont {Campostrini}\ \emph {et~al.}(2002)\citenamefont
  {Campostrini}, \citenamefont {Hasenbusch}, \citenamefont {Pelissetto},
  \citenamefont {Rossi},\ and\ \citenamefont {Vicari}}]{Campostrini2002}%
  \BibitemOpen
  \bibfield  {author} {\bibinfo {author} {\bibfnamefont {M.}~\bibnamefont
  {Campostrini}}, \bibinfo {author} {\bibfnamefont {M.}~\bibnamefont
  {Hasenbusch}}, \bibinfo {author} {\bibfnamefont {A.}~\bibnamefont
  {Pelissetto}}, \bibinfo {author} {\bibfnamefont {P.}~\bibnamefont {Rossi}}, \
  and\ \bibinfo {author} {\bibfnamefont {E.}~\bibnamefont {Vicari}},\ }\bibinfo
  {title} {{Critical exponents and equation of state of the three-dimensional
  Heisenberg universality class}},\ \href {\doibase 10.1103/PhysRevB.65.144520}
  {\bibfield  {journal} {\bibinfo  {journal} {Phys. Rev. B}\ }\textbf {\bibinfo
  {volume} {65}},\ \bibinfo {pages} {144520} (\bibinfo {year}
  {2002})}\BibitemShut {NoStop}%
\bibitem [{\citenamefont {Sasago}\ \emph {et~al.}(1997)\citenamefont {Sasago},
  \citenamefont {Uchinokura}, \citenamefont {Zheludev},\ and\ \citenamefont
  {Shirane}}]{Sasago1997}%
  \BibitemOpen
  \bibfield  {author} {\bibinfo {author} {\bibfnamefont {Y.}~\bibnamefont
  {Sasago}}, \bibinfo {author} {\bibfnamefont {K.}~\bibnamefont {Uchinokura}},
  \bibinfo {author} {\bibfnamefont {A.}~\bibnamefont {Zheludev}}, \ and\
  \bibinfo {author} {\bibfnamefont {G.}~\bibnamefont {Shirane}},\ }\bibinfo
  {title} {{Temperature-dependent spin gap and singlet ground state in
  ${\mathrm{BaCuSi}}_{2}$${\mathrm{O}}_{6}$}},\ \href {\doibase
  10.1103/PhysRevB.55.8357} {\bibfield  {journal} {\bibinfo  {journal} {Phys.
  Rev. B}\ }\textbf {\bibinfo {volume} {55}},\ \bibinfo {pages} {8357}
  (\bibinfo {year} {1997})}\BibitemShut {NoStop}%
\bibitem [{\citenamefont {Nayak}\ \emph {et~al.}(2020)\citenamefont {Nayak},
  \citenamefont {Blosser}, \citenamefont {Zheludev},\ and\ \citenamefont
  {Mila}}]{Nayak2020}%
  \BibitemOpen
  \bibfield  {author} {\bibinfo {author} {\bibfnamefont {M.}~\bibnamefont
  {Nayak}}, \bibinfo {author} {\bibfnamefont {D.}~\bibnamefont {Blosser}},
  \bibinfo {author} {\bibfnamefont {A.}~\bibnamefont {Zheludev}}, \ and\
  \bibinfo {author} {\bibfnamefont {F.}~\bibnamefont {Mila}},\ }\bibinfo
  {title} {{Magnetic-Field-Induced Bound States in Spin-$\frac12$ Ladders}},\
  \href {\doibase 10.1103/physrevlett.124.087203} {\bibfield  {journal}
  {\bibinfo  {journal} {Phys. Rev. Lett.}\ }\textbf {\bibinfo {volume} {124}},\
  \bibinfo {pages} {087203} (\bibinfo {year} {2020})}\BibitemShut {NoStop}%
\bibitem [{\citenamefont {Gelman}\ \emph {et~al.}(2004)\citenamefont {Gelman},
  \citenamefont {Carlin}, \citenamefont {Stern},\ and\ \citenamefont
  {Rubin}}]{Gelman2004}%
  \BibitemOpen
  \bibfield  {author} {\bibinfo {author} {\bibfnamefont {A.}~\bibnamefont
  {Gelman}}, \bibinfo {author} {\bibfnamefont {J.}~\bibnamefont {Carlin}},
  \bibinfo {author} {\bibfnamefont {H.}~\bibnamefont {Stern}}, \ and\ \bibinfo
  {author} {\bibfnamefont {D.}~\bibnamefont {Rubin}},\ }\href@noop {} {\emph
  {\bibinfo {title} {{Bayesian Data Analysis}}}}\ (\bibinfo  {publisher}
  {Chapman \& Hall/CRC, Boca Raton},\ \bibinfo {year} {2004})\BibitemShut
  {NoStop}%
\bibitem [{\citenamefont {Hoffman}\ and\ \citenamefont
  {Gelman}(2014)}]{Hoffman2014}%
  \BibitemOpen
  \bibfield  {author} {\bibinfo {author} {\bibfnamefont {M.~D.}\ \bibnamefont
  {Hoffman}}\ and\ \bibinfo {author} {\bibfnamefont {A.}~\bibnamefont
  {Gelman}},\ }\bibinfo {title} {{The No-U-Turn Sampler: Adaptively Setting
  Path Lengths in Hamiltonian Monte Carlo}},\ \href
  {http://jmlr.org/papers/v15/hoffman14a.html} {\bibfield  {journal} {\bibinfo
  {journal} {J. Mach. Learn. Res.}\ }\textbf {\bibinfo {volume} {15}},\
  \bibinfo {pages} {1593} (\bibinfo {year} {2014})}\BibitemShut {NoStop}%
\bibitem [{\citenamefont {Salvatier}\ \emph {et~al.}(2016)\citenamefont
  {Salvatier}, \citenamefont {Wiecki},\ and\ \citenamefont
  {Fonnesbeck}}]{Salvatier2016}%
  \BibitemOpen
  \bibfield  {author} {\bibinfo {author} {\bibfnamefont {J.}~\bibnamefont
  {Salvatier}}, \bibinfo {author} {\bibfnamefont {T.}~\bibnamefont {Wiecki}}, \
  and\ \bibinfo {author} {\bibfnamefont {C.}~\bibnamefont {Fonnesbeck}},\
  }\bibinfo {title} {{Probabilistic programming in Python using PyMC3}},\ \href
  {\doibase 10.7717/peerj-cs.55} {\bibfield  {journal} {\bibinfo  {journal}
  {PeerJ Comp. Sci.}\ }\textbf {\bibinfo {volume} {2}},\ \bibinfo {pages} {e55}
  (\bibinfo {year} {2016})}\BibitemShut {NoStop}%
\end{thebibliography}%


\begin{thebibliography}{10} 
\end{thebibliography}

\end{document}